\documentclass[prd,a4paper,showpacs,onecolumn,nofootinbib,preprintnumbers]{revtex4}

\usepackage[top=2.8cm, bottom=2.8cm, left=2.4cm, right=2.4cm]{geometry}
\usepackage[utf8]{inputenc}  
\usepackage[T1]{fontenc}     
\usepackage{amsmath,amssymb,lmodern} 
\usepackage{graphicx}
\usepackage{diagbox}
\usepackage{hyperref}
\usepackage{color}
\usepackage{lipsum}
\usepackage{bm}
\usepackage{url}

\usepackage{xcolor}
\definecolor{blue}{rgb}{0.19,0.64,0.54}
\definecolor{reddish}{rgb}{0.65, 0.2, 0.2}
\definecolor{red}{rgb}{0.7,0.3,0.3}
\definecolor{darkgreen}{rgb}{0.2,0.7,0.3}
\definecolor{darkblue}{rgb}{0.3,0.40,0.48}
\definecolor{gray}{rgb}{.8,.8,.8}

\hypersetup{,
    unicode=false,          % non-Latin characters in Acrobat‚Äôs bookmarks
    pdftoolbar=true,        % show Acrobat‚Äôs toolbar?
    pdfmenubar=true,        % show Acrobat‚Äôs menu?
    pdffitwindow=false,     % window fit to page when opened
    pdfstartview={FitH},    % fits the width of the page to the window
    pdftitle={Generalized Proca SU(2)},    % title
    pdfauthor={E. Allys, P. Peter and Y. Rodriguez},     % author
    pdfsubject={Subject},   % subject of the document
    pdfcreator={Creator},   % creator of the document
    pdfproducer={Producer}, % producer of the document
    pdfkeywords={keyword1} {key2} {key3}, % list of keywords
    pdfnewwindow=true,      % links in new window
    colorlinks=true,       % false: boxed links; true: colored links
    linkcolor=red,          % color of internal links
    citecolor=darkblue,        % color of links to bibliography
    filecolor=magenta,      % color of file links
    urlcolor=darkblue,           % color of external links
    linktocpage=true
}

\newcommand{\dd}{\mbox{d}}

\renewcommand{\a}{{\color{red}a\color{black}}}
\renewcommand{\b}{{\color{red}b\color{black}}}
\renewcommand{\c}{{\color{red}c\color{black}}}
\renewcommand{\d}{{\color{red}d\color{black}}}
\newcommand{\e}{{\color{red}e\color{black}}}
\newcommand{\al}{{\color{red}d\color{black}}}
\newcommand{\be}{{\color{red}e\color{black}}}

\begin{document}

\preprint{PI/UAN-2016-597FT}

\title{Generalized SU(2) Proca Theory}

\author{Erwan Allys,}
\email{allys@iap.fr}
\affiliation{Institut d’Astrophysique de Paris, UMR 7095, \\
UPMC Universit\'e Paris 6 et CNRS,\\
98 bis boulevard Arago, 75014 Paris, France}

\author{Patrick Peter}
\email{peter@iap.fr}
\affiliation{Institut d’Astrophysique de Paris, UMR 7095, \\
UPMC Universit\'e Paris 6 et CNRS,\\
98 bis boulevard Arago, 75014 Paris, France}
\affiliation{Institut Lagrange de Paris,\\
UPMC Universit\'e Paris 6 et CNRS,\\
Sorbonne Universit\'es, 75014 Paris, France}

\author{Yeinzon Rodr\'iguez}
\email{yeinzon.rodriguez@uan.edu.co}
\affiliation{Centro de Investigaciones en Ciencias B\'asicas y Aplicadas, Universidad Antonio Nari\~no, \\ Cra 3 Este \# 47A-15, Bogot\'a D.C. 110231, Colombia}
\affiliation{Escuela  de  F\'isica,  Universidad  Industrial  de  Santander, \\ Ciudad  Universitaria,  Bucaramanga  680002,  Colombia}
\affiliation{Simons Associate at The Abdus Salam International Centre for Theoretical Physics, \\ Strada Costiera 11, I-34151, Trieste, Italy}

\date{\today}

\begin{abstract}
Following previous works on generalized Abelian Proca theory, also
called vector Galileon, we investigate the massive extension of an SU(2)
gauge theory, i.e., the generalized SU(2) Proca model, which could be
dubbed non-Abelian vector Galileon. This particular symmetry group
permits fruitful applications in cosmology such as inflation driven by
gauge fields. Our approach consists in building, in an exhaustive way, all
the Lagrangians containing up to six contracted Lorentz indices. For
this purpose, and after identifying by group theoretical considerations
all the independent Lagrangians which can be written at these orders, we
consider the only linear combinations propagating three degrees of
freedom and having healthy dynamics for their longitudinal mode, i.e.,
whose pure Stückelberg contribution turns into the SU(2) multi-Galileon
dynamics. Finally, and after having considered the curved space-time
expansion of these Lagrangians, we discuss the form of the theory at all
subsequent orders.
\end{abstract}


\maketitle

\section{Introduction}

In the search for well-motivated theories that describe the primordial
universe, several attempts have been made to obtain inflationary
descriptions from particle physics (the Standard Model, Supersymmetry,
Grand Unified Theories, etc.; see, e.g., Refs.~\cite{Bezrukov:2012sa,Bezrukov:2007ep,Mazumdar:2010sa,Lyth:1998xn,Hertzberg:2014sza}),
 or from quantum theories of gravity such as Supergravity, String
 Theory, and Loop Quantum Gravity (see, e.g.,
 Refs.~\cite{Ferrara:2013rsa,Olive:1989nu,Baumann:2014nda,Ashtekar:2009mm,Barrau:2010nd}).
This top-down approach has been very fruitful, providing new ways to
understand the structure of the high energy theories necessary to
reproduce the observable properties of the Universe, ranging from the
Cosmic Microwave Background Radiation (CMB) to the Large-Scale Structure
(LSS). However, little is known from the observational point of view for
many of these theories (those whose characteristic energy scale is
much higher than the electroweak one), the CMB and LSS being, at present, the
only situations in which they would have had observable consequences and
would thus leave testable signatures. Since the power of the current and
proposed accelerators is not going to increase as much as would be
needed to directly test these theories in the foreseeable future, we
need to devise another approach to the fundamental theory that describes
nature.

Such an approach already exists, and it boils down to the question of
whether there is any choice in formulating the fundamental theory. This
bottom-up approach consists in finding an action completely free of
pathologies, the first of them being the Ostrogradski instability
\cite{ostro} (the Hamiltonian could be unbounded from below), and
satisfying a given set of assumptions, e.g., symmetry requirements.  One
then needs to define the material content of the universe (scalar
fields, vector fields, ...), although, in principle, the construction itself
and the stability requirements constrain some content and allow others
so that the material content is, once the conditions are applied,
somehow redefined. This very ambitious program is just beginning to be
implemented, and interesting works have been carried out in which the
extra material content (on top of gravity) is composed of one or many
scalar fields. It was Horndeski \cite{Horndeski:1974wa} who found, for
the first time, the most general action for a scalar field and gravity
that produces second-order equations of motion. In general, if the
Lagrangian is nondegenerate, having equations of motion of second order
at most is a necessary requirement to avoid the Ostrogradsky instability
\cite{Woodard:2006nt,Woodard:2015zca}. By pursuing this goal, an action
is found that, however, still requires a Hamiltonian analysis in order
to guarantee that the instability is not present.

Horndeski's construction was rediscovered in the context of what is
nowadays called Galileons \cite{Nicolis:2008in}.  The Galileons are the
scalar fields whose action, in flat spacetime, leads to equations of
motion that involve only second-order derivatives.  The idea has been
extended by finding the so-called Generalized Galileons, by allowing for
lower-order derivatives in the equations of motion
\cite{Deffayet:2011gz,Deffayet:2013lga}.  The background space-time
geometry where these Generalized Galileons live can be promoted to a
curved one by replacing the ordinary derivatives with covariant ones and
adding some counterterms that involve nonminimal couplings to the
curvature \cite{Deffayet:2009wt,Deffayet:2009mn}.  The latter guarantees
the equations of motion for both geometry and matter are still
second order, so the Galileons, both Generalized and Covariantized,
are found.  This procedure is equivalent to that proposed by Horndeski
for one scalar field \cite{Kobayashi:2011nu}, but it loses some interesting
terms when more than one scalar field is present
\cite{Kobayashi:2013ina}. The Galileon approach for scalar fields has
found multiple applications in cosmology, ranging from inflation (see,
e.g., Refs.
\cite{Creminelli:2010ba,Kobayashi:2010cm,Mizuno:2010ag,Burrage:2010cu,Creminelli:2010qf,Kamada:2010qe,Libanov:2016kfc,Banerjee:2016hom,Hirano:2016gmv,Brandenberger:2016vhg,Nishi:2016wty}) to dark energy (see, e.g., Refs. \cite{Chow:2009fm,Silva:2009km,Kobayashi:2010wa,Gannouji:2010au,Tsujikawa:2010zza,DeFelice:2010pv,Ali:2010gr,Padilla:2010tj,DeFelice:2010nf,Mota:2010bs,Nesseris:2010pc,Gabadadze:2016llq,Neveu:2016gxp,Salvatelli:2016mgy,Shahalam:2016kkg,Minamitsuji:2016qyc,Saridakis:2016ahq,Biswas:2016bwq}).

The original proposal was based on the requirement of second-order
equations of motion for all the additional degrees of freedom to gravity, all of them therefore being 
dynamical so that the system is nondegenerate. The generalization
to the so-called extended Horndeski theories also includes nonphysical
degrees of freedom and thus considers degenerate theories
\cite{Gleyzes:2014dya,Langlois:2015cwa,Langlois:2015skt,Motohashi:2016ftl,Crisostomi:2016czh}.
%
%Being the non-degeneracy condition the base on which the requirement of,
%at most, second-order equations of motion stands on, extended Horndeski
%theories have been looked for by considering degenerate theories
%\cite{Gleyzes:2014dya,Langlois:2015cwa,Langlois:2015skt,Motohashi:2016ftl,Crisostomi:2016czh}.
Such a construction is by now well understood, and some cosmological
applications have also been considered 
\cite{Harko:2016xip,Babichev:2016rlq,Sakstein:2016ggl,Kobayashi:2016xpl,Lagos:2016wyv,Crisostomi:2016czh,Frusciante:2016xoj,Qiu:2015aha,Akita:2015mho}.

However, scalar fields are not the only possibilities as the matter
content of the universe.  Horndeski indeed wondered some fourty years
ago what the action would be for an Abelian vector field in curved
spacetime \cite{Horndeski:1976gi}.  Working with curvature is a way to
bypass the no-go theorem presented in Ref. \cite{Deffayet:2013tca}, which
states that the only possible action for an Abelian vector field in flat
spacetime that leads to second-order equations of motion is the
Maxwell-type one.  Relaxing the gauge invariance allows for a nontrivial action in flat spacetime, in this way generalizing the Proca action
\cite{Heisenberg:2014rta,Tasinato:2014eka}.  The construction of the
resulting vector Galileon action has been well investigated and discussed, so
there is already a consensus about the number and type of terms in the
action, even in the covariantized version
\cite{Allys:2015sht,Jimenez:2016isa,Allys:2016jaq}.  Moreover, the
analogous extended Horndeski theories have been built for a vector field
\cite{Heisenberg:2016eld,Kimura:2016rzw}, and the corresponding
cosmological applications have been explored
\cite{Tasinato:2014eka,Tasinato:2014mia,Hull:2014bga,DeFelice:2016cri,DeFelice:2016yws,DeFelice:2016uil,Heisenberg:2016wtr}.

Some cosmological applications of vector fields have been investigated,
and interesting scenarios, such as the $fF^2$ model
\cite{Watanabe:2009ct} and the vector curvaton
\cite{Dimopoulos:2006ms,Dimopoulos:2009am}, have been devised. There is,
however, an obstacle when dealing with vector fields in cosmology:  they
produce too much anisotropy, both at the background and at the perturbation
levels, well above the observable limits, unless one implements some
dilution mechanism or considers only the
temporal component of the vector field (which is, however, usually nondynamical). In the $fF^2$ model, the potentially huge anisotropy is
addressed by coupling the vector field to a scalar that dominates the
energy density of the universe and, therefore, dilutes the anisotropy;
in contrast, in the vector curvaton scenario, the anisotropy is diluted
by the very rapid oscillations of the vector curvaton around the minimum
of its potential.  Another dilution mechanism is to consider many
randomly oriented vector fields \cite{Golovnev:2008cf}; however, this requires
a large number of them, indeed hundreds, so it is difficult to
justify it from a particle physics point of view.  There is,
nevertheless, another possibility, the so-called ``cosmic triad''
\cite{Golovnev:2008cf,ArmendarizPicon:2004pm}, a situation in which three
vector fields orthogonal to each other and of the same norm can give
rise to a rich phenomenology while making the background and
perturbations completely isotropic \cite{Rodriguez:2015xra}.  A couple
of very interesting models, gauge-flation
\cite{Maleknejad:2011jw,Maleknejad:2011sq} and chromo-natural inflation
\cite{Adshead:2012kp}, have implemented this idea by embedding it in a
non-Abelian framework and exploiting the local isomorphism between the
SO(3) and SU(2) groups of transformations.  At first sight, the cosmic
triad configuration looks very unnatural, but dynamical system studies
have shown that it represents an attractor configuration
\cite{Maleknejad:2011jr}.  Unfortunately, although the background
dynamics of these two models is successful, their perturbative dynamics
makes them incompatible with the latest Planck observations
\cite{Namba:2013kia,Adshead:2013nka}.  Despite this failure, such models
have shown the applicability that non-Abelian gauge fields can have in
cosmological scenarios.

Having in mind the above motivations, the purpose of this paper is to
build the first-order terms of the generalized SU(2) Proca theory and
to discuss the general form of the complete theory. For the most part,
we focus on those Lagrangians containing up to six contracted Lorentz indices,
which we obtain exhaustively. To ensure that we do not forget some terms, we first
construct from group theoretical considerations all possible Lagrangians
at these orders, before imposing the standard dynamical condition,
i.e., that only three degrees of freedom propagate. Then, after
identifying all the Lagrangians that imply the same dynamics, e.g.,
those related by a conserved current, we verify that the pure
Stückelberg part of the Lagrangians is healthy, i.e., that it implies the SU(2)
multi-Galileon dynamics. To this end, it is useful to derive all the
equivalent formulations of the SU(2) adjoint multi-Galileon model, which
we provide in the Appendix. Then, after computing the relevant
curved space-time extension of our Lagrangians, we conclude about the status
of the complete formulation of the theory, i.e., that containing the higher
order terms we did not consider in this work.

The layout of this paper is the following.  In Section  \ref{gnapc}, the
generalized non-Abelian Proca theory is introduced, and some technical
aspects needed for later sections are laid out;  the procedure to build
the theory is also described.  In Section \ref{PartSinglet}, the
building blocks of the Lagrangian are systematically obtained.  Section
\ref{cht} deals with the right number of propagating degrees of freedom
and the consistency of the obtained Lagrangian with the scalar Galileon
nature of its longitudinal part.  The covariantization of the theory is
performed in Section \ref{covproc} and the final model, together with a
discussion and comparison with the Abelian case, is presented in Section
\ref{FinalModel}.  The appendix presents the construction of the
multi-Galileon scalar Lagrangian in the 3-dimensional representation of
SU(2) and its equivalent formulations.  Throughout this paper, we have
employed the mostly plus signature, i.e., $\eta_{\mu\nu} =
\rm{diag}\,\left( -, +,+,+\right)$, and set $\hbar=c=1$.

\section{Generalized non-Abelian Proca Theory} \label{gnapc}

Our aim is to generalize the non-Abelian Proca theory, described below,
to include all possible second-order ghost-free terms propagating only
three degrees of freedom. After discussing the general symmetry case, we
concentrate on the SU(2) symmetry, which is particularly interesting in
a cosmological perspective, as discussed in the Introduction, and we
roughly present the procedure, which will be thoroughly explained below.

\subsection{Non-Abelian Proca Theory}

Let us first present the nowadays standard non-Abelian Proca theory.
Also called a massive Yang-Mills model, this theory had been extensively
studied in the past, such as, e.g., in
Refs.~\cite{Boulware:1970zc,Shizuya:1975ek,GrosseKnetter:1993nu,Banerjee:1994pp,Su:2002qj},
with a Hamiltonian formulation detailed in
Refs.~\cite{Senjanovic:1976br,Banerjee:1997sf}. Our starting point
Lagrangian, including the mass term, reads
\begin{equation}
\label{EqLagProcaNonAbelian}
\mathcal{L} = -\frac{1}{4} F^{\mu\nu}_{\a} F^{\a}_{\mu\nu} +\frac{1}{2}
m^2 A^{\mu}_{\a} A_{\mu}^{\a},
\end{equation}
with the non-Abelian Faraday tensor given by
\begin{equation}
\label{EqFaradayNonAbelian}
F_{\mu \nu}^{\a} = \partial_\mu A_\nu^{\a}  - \partial_\nu A_\nu^{\a} +
g f^{\a}{}_{\b \c} A_\mu^{\b} A_\nu ^{\c},
\end{equation}
with $g$ being the coupling constant and $f^{\a}{}_{\b \c}$ the structure
constants of the symmetry group under consideration. This can be
considered as the limit of a valid particle physics model based on a
Higgs condensate whose corresponding degree of freedom is assumed
to be frozen, hence breaking the relevant
symmetry~\cite{Hull:2014bga,Tasinato:2014mia}.

Let us emphasize a technical point at this stage: one could work with
the vector field assumed as an operator, namely,
\begin{equation}
A_\mu (x) = A_\mu ^{\a} (x) T_{\a},
\end{equation}
with $T_{\a}$ representing the operators associated with corresponding
elements of the underlying group in a given representation. We then
have, by definition of the algebra, the commutation relations
\begin{equation}
\left[T_{\a}, T_{\b} \right] = i f_{\a\b}{}^{\c} T_{\c}.
\label{TaTb}
\end{equation}
Since this work concentrates on the vector fields
themselves and not on their action on other fields, it is simpler to
restrict our attention to the fields themselves, i.e.,
\begin{equation}
A_\mu (x)=\left\{ A_\mu^{\a} (x) \right\},
\end{equation}
which are in the Lie algebra of the symmetry group under consideration.
These two ways of writing the field operators are, of course, strictly
equivalent, but the latter formalism, with group indices attached to the
vectors themselves, merely does not need the introduction of the algebra
operators themselves and is thus more appropriate for our purpose.

Any action needs to be a scalar, and this includes not only the Lorentz
group but also any internal symmetry, such as that stemming from the
algebra in Eq. \eqref{TaTb}. If the relevant symmetry is of the local type,
and for an infinitesimal transformation,
the vectors transform through
\begin{equation}
\delta A_\mu ^{\a} = -\frac{1}{g}  \partial_\mu \alpha^{\a}(x)+
f^{\a}{}_{\b\c}\alpha^{\b}(x) A_\mu ^{\c},
\label{dAloc}
\end{equation}
which leaves invariant only the kinetic term $F_{\mu\nu}^{\a} F^{\mu\nu}_{\a}$,
but of course not even a mass term $A_{\mu}^{\a} A^{\mu}_{\a}$, much less any
extension such as those we want to consider below. This is merely a restatement of
the well-known fact that mass breaks gauge symmetry. We therefore restrict our
attention to global transformations of the kind
\begin{equation}
\partial_\mu \alpha^{\a}=0\ \ \ \ \Longrightarrow \ \ \ \ \
\delta A_\mu ^{\a} = f^{\a}{}_{\b\c}\alpha^{\b} A_\mu ^{\c};
\label{dAglo}
\end{equation}
i.e., we assume the vector field itself transforms as the adjoint
representation, with dimension equal to that of the symmetry group
itself. It is also profitable, and maybe more enlightening, to look at
the effect of a finite local transformation of the group, still
described by a set of parameters $\alpha^{\a}(x)$. Under this
transformation, the vector field transforms as
\begin{equation}
\label{EqGaugeTransfoInt}
A_\mu (x)=A_\mu^\a (x) T_\a \mapsto U\left[\alpha^{\a}(x)\right]
\left[-\frac1{g}\partial_\mu + A_\mu (x) \right]
U^{-1}\left[\alpha^{\a}(x)\right],
\end{equation}
where $U\left[\alpha^{\a}(x)\right]$ describes the action of the group
element labeled by $\alpha^{\a}(x)$. This allows us to emphasize that in the
case where the symmetry becomes global, i.e., where $\alpha^{\a}(x)$ no longer depends on the space-time point, the vector field transforms
exactly as the adjoint representation of the symmetry group. This is
indeed the symmetry assumed for the non-Abelian Proca (massive
Yang-Mills) field. In the Abelian case, this transformation is trivial
because the action of the group commutes with the vector field, and the
transformation in Eq. \eqref{EqGaugeTransfoInt} thus reduces to the identity in the
global symmetry case. In the non-Abelian case, however, one needs to
specify how the extra indices are to be summed over in order to produce
a singlet with respect to this global symmetry transformation. To relate
the set of theories under considerations here with the more usual ones
in particle physics involving a local symmetry broken by means of a
Higgs field, one can envisage our transformation in Eq. \eqref{dAglo} as the
limit of that in Eq. \eqref{dAloc}.

With these motivating considerations, we now move on to evaluating the
most general theory with a massive vector field transforming according
to the adjoint representation of a given global symmetry group.

\subsection{Restricting Attention to the SU(2) Case}

In view of the potentially relevant cosmological consequences, from now on we restrict our attention to the case for which the relevant
symmetry group is SU(2), with dimension equal to $3$, and therefore
consider a vector field also of dimension $3$. Since SU(2) is locally
isomorphic to SO(3), one can then simply use a vector representation
with group indices varying from $1$ to $3$ in $A_{\mu}^{\a}$; i.e., we
restrict our attention to the fundamental representation of SO(3).

The set of SU(2) structure constants is identical to the 3-dimensional Levi-Civita
tensor $\epsilon^{\a}_{\ \b\c}$, whereas the group metric $g_{\a\b}$, given by
$g_{\a\b} = -f_{\a\d}{}^{\e} f_{\b\e}{}^{\d}$, is simply the flat metric
$2 \delta_{\a\b}$. The only primitive invariants are $\epsilon_{\a\b\c}$
and $\delta_{\a\b}$ \cite{Slansky:1981yr,Fuchs:1997jv,Padilla:2010ir},
and one can therefore write all possible contractions by merely
contracting fields with contravariant indices with all appropriate
combinations of those two primitive invariants written with covariant indices.
Recall also the further simplification induced by the fact that
contractions among structure constants (Levi-Civita symbols in the
case at hand) leaving one, two or three free indices will, respectively,
lead to a vanishing result, or terms proportional to $\delta_{\a\b}$ and
$\epsilon_{\a\b\c}$ \cite{Metha:1983mng}; it is therefore often
unnecessary to use multiple contractions.

As already alluded to earlier, choosing SU(2) is not innocuous as we aim
at cosmological applications, in view, in particular, of implementing
inflation driven by gauge fields (see, e.g., Refs.
\cite{Maleknejad:2011jw,Maleknejad:2011sq,Namba:2013kia,Maeda:2012eg,Adshead:2012qe,Nieto:2016gnp,Davydov:2015epx,Alexander:2014uza,Sharif:2014aaa,Darabi:2014pua,Maleknejad:2013npa,Maeda:2013daa,Setare:2013dra,Maleknejad:2011jr,Maleknejad:2012fw,Cembranos:2012ng,Adshead:2012kp,Dimastrogiovanni:2012ew,Adshead:2013nka,Ghalee:2012gg}):
since its adjoint representation is 3-dimensional, SU(2) permits us to
generate configurations for which all three vectors are nonvanishing
while ensuring isotropy.

\subsection{Generalization}

What follows is very similar to the generalized Abelian Proca case as
discussed, e.g., in
Refs.~\cite{Heisenberg:2014rta,Tasinato:2014eka,Li:2015vwa,Allys:2015sht,Jimenez:2016isa,Allys:2016jaq} 
(see also Refs.~\cite{deRham:2011by,Jimenez:2013qsa,Hull:2015uwa}
for the equivalent curved space-time construction). In brief, we 
construct the most general action generalizing that of Proca for a
massive SU(2) vector field, i.e.,
\begin{equation}
\mathcal{S}_\mathrm{Proca}  = \int \mathcal{L}_\mathrm{Proca}\,\dd^4x =
\int \left( -\frac14 F^{\a}_{\mu\nu} F_{\a}^{\mu\nu} + \frac12 m_A^2 X
\right)\,\dd^4x,
\label{Proca}
\end{equation}
where $X\equiv A_\mu^{\a} A^\mu_{\a}$. To the above action
(\ref{Proca}), we add all possible terms containing not only
functions of $X$ but also derivative self-interactions. These terms will
have to fulfill some conditions for the corresponding theory to make
sense. We first split the vector into a scalar-pure vector decomposition
\begin{equation}
A_\mu^{\a}=\partial_\mu \pi^\a + \bar{A}_\mu^\a,
\end{equation}
 where $\pi^\a$ is a scalar multiplet in the $\bm{3}$ representation of
 SU(2), i.e., the St\"uckelberg field generalized to the non-Abelian
 case, and $\bar{A}_\mu^\a$ is a divergence-free vector ($\partial_\mu
 \bar{A}^{\mu\a}=0$), containing the curl part of the field, i.e., that
 for which the Abelian form of the Faraday tensor is nonvanishing. The conditions one then
 must  impose on the theory in order for it to make (classical) sense are
\begin{itemize}
\item[a)] the equations of motion for all physical degrees of freedom,
i.e.,, for both $\bar{A}_\mu^\a$ and $\pi^\a$, and hence $A_\mu^\a$ and
$\pi^\a$, must be at most second order, thus ensuring stability
\cite{ostro,Woodard:2006nt,Woodard:2015zca},
\item[b)] the action may contain at most second-order derivative terms
in $\pi^\a$ and first-order derivatives for $A_\mu^\a$,
\item[c)] each component of the SU(2) multiplet propagates only three
degrees of freedom, the zeroth component being nondynamical.
\end{itemize} 
In what follows, we apply these conditions and restrict our attention
to the theories involving terms with up to six Lorentz indices
contracted. From the cosmological perspective, such theories are
expected to allow for a richer phenomenology since this is what happens
for the Abelian Proca case \cite{Tasinato:2014eka,Tasinato:2014mia,Hull:2014bga,DeFelice:2016cri,DeFelice:2016yws,DeFelice:2016uil,Heisenberg:2016wtr}

\subsection{Procedure}
\label{PartProcedure}

We now proceed along the lines of Ref.~\cite{Allys:2015sht}; i.e.,
we build, in Sec.~ \ref{PartSinglet}, a complete basis of linearly
independent test Lagrangians describing all possible Lagrangians
containing a given number of vector fields and their derivatives; the
detailed prescription is given in Sec.~\ref{PartProcedureSinglet}.
Next we demand only three degrees of freedom per multiplet component of
the vector field, which translates into a condition on the
Hessian~\cite{Heisenberg:2014rta,Allys:2015sht}, the latter being
defined by
\begin{equation}
\label{EqDefHessian}
\mathcal{H}^{\mu\nu \d \e} = \frac{\partial}{\partial (\partial_0
A_{\mu\d}) } \frac{\partial}{\partial (\partial_0 A_{\nu \e} )}
\mathcal{L},
\end{equation}
for a given Lagrangian $\mathcal{L}$. This functional over the fields is
symmetric under the index exchange $\left(\mu,\d \right) \leftrightarrow
\left( \nu,\e \right)$.

In order for $\mathcal{H}^{\mu\nu \d \e}$ to have three vanishing
eigenvalues, one for each timelike component of the three vectors
$A_{\mu\d}$, and since all the terms it is built of are a priori
independent (up to symmetries), a necessary condition is that we demand
$\mathcal{H}^{0 \nu \d \e}=0$; this requirement will be explicitly
checked in Sec.~\ref{PartHessianCondition} for each test Lagrangian.

The above condition is, however, not sufficient, for it does not exhaust
all the constraints and thus does not count the effectively propagating
degrees of freedom. For instance, some terms inducing no dynamics for
the time components of the vector fields may also yield no dynamics for
some other component, or even for the overall vector field. The required
analysis is tedious and must be followed step by
step~\cite{Senjanovic:1976br,Banerjee:1997sf}.

As the final step of the above analysis, we consider the scalar
part associated with those linear combinations of test Lagrangians
verifying the Hessian condition. One must check, which is done in
Sec.~\ref{PartScalarContribution}, that they are of two kinds: either
they have no dynamics at all, being vanishing or given by a total
derivative, or their dynamics is second order in the equations of motion
of the scalar field; i.e., they belong to the class of generalized
Galileons
\cite{Deffayet:2010zh,Padilla:2010ir,Padilla:2010de,Hinterbichler:2010xn,Padilla:2012dx,Sivanesan:2013tba}. 
This will provide the most general terms that verify the requirements we
demand, formulated in terms of the non-Abelian Faraday tensor; see
Sec.~\ref{PartSwitchFaradayYM}.

Before moving on, we mention that even though the
procedure discussed above and applied below is allegedly tedious, it
guarantees an exhaustive list of all possible terms at each
order, and, in particular, all those specific to the non-Abelian case.
Those terms might have been obtained by some quicker method, but we prefer to
be able to produce all the theoretically acceptable terms rather than
constructing a few. In view of possible cosmological applications, there
is indeed no way to say which terms will be relevant and which ones will not.

\section{Construction of the test Lagrangians}
\label{PartSinglet}

As anticipated above, our method relies heavily on the construction of a
basis of test Lagrangians satisfying the symmetry requirement, on which
we later apply the Hessian condition. This is the purpose of this
section.

\subsection{Description of the Procedure}
\label{PartProcedureSinglet}

We now proceed to build the complete basis, in the sense of linear
algebra, of test Lagrangians, for a given number of fields and their
first derivatives. Since they are linearly independent, we will then be able to
write down the most general theory at the given order as a linear
combination of these Lagrangians.

In order to construct Lagrangians, i.e., scalars, we need to consider the
Lorentz and group indices. The former spacetime indices run from $0$
to $3$ and are denoted by small Greek letters, while the latter group
indices run from $1$ to $3$, since we assume the adjoint 3-dimensional
SU(2) representation, and are represented by small Latin letters from
the beginning of the alphabet. We first write down all the Lorentz
scalar quantities that may be formed with a given number of fields and
first derivatives, and then consider all the SU(2) index combinations
leading to SU(2) scalars of these Lorentz scalars.

For the sake of simplicity, beginning with the Lorentz sector, we 
dismiss the group indices altogether, keeping in mind, however, that their
presence might spoil some symmetry properties: contractions between
symmetric and anti-symmetric (with respect to Lorentz indices only)
tensors will not necessarily vanish when group indices are included,
as exemplified by the starred equations in the next section. The Lorentz scalars,
once formed, will then subsequently be assigned SU(2) indices following
simple alphabetical order, leaving as many free SU(2) indices as there
are fields in the term, to then be contracted with a relevant pure SU(2)
tensor. For instance, a term like $A^{\mu}A^{\nu}\left(\partial_{\mu}
A^{\nu} \right)$ will be indexed as
$A^{\mu\a}A^{\nu\b}\left(\partial_{\mu} A^{\nu\c} \right)$, demanding
contraction with a structure constant $\epsilon_{\a\b\c}$ to form a
Lorentz and SU(2) scalar.
This procedure can seem rather tedious, and it most
definitely is, but it ensures that we construct a complete basis.

For simplicity, we restrict our attention to those Lagrangians
containing up to 6 Lorentz indices contracted as they should be to form a
scalar.

\subsection{Lorentz Sector}  \label{LSSection}

An easy way to classify the Lorentz scalars that one can form with a given
number of 4-vectors consists in using the local equivalence, at the
Lie-algebraic level, between SO(3,1) and SU(2)$\times$SU(2) (see, e.g.,
Ref. \cite{Ramond:2010zz}). One obtains the following
table~\cite{Slansky:1981yr,Feger:2012bs}:

\begin{center}
\begin{tabular}{|l|c|c|c|c|c|c|c|c|}
\hline
$\#$ of vector fields $A^{\mu} $ & 1 & 2 & 3 & 4 & 5 & 6 & 7 & 8 \\
\hline
$\# $ of Lorentz scalars & 0  & 1 & 0 & 4 & 0 & 25 & 0 & 196 \\
\hline
\end{tabular}
\end{center}

These scalars can be written in terms of the primitive invariants,
namely, $g_{\mu\nu}$ and $\epsilon_{\mu\nu\rho\sigma}$. As shown in the
table, an odd number of vector fields is impossible, as is obvious from
the fact that one cannot form primitive Lorentz invariants with an odd number of
indices. For two fields, the only contracting possibility is $g_{\mu\nu}$,
while for four free Lorentz indices, the contractions 
with a term of the form $A^\mu B^\nu C^\rho D^\sigma$ 
can be performed with any member of
the list
\begin{equation}
\left\{
\begin{array}{l}
g_{\mu\nu} g_{\rho\sigma},\\
g_{\mu\rho}g_{\nu\sigma},\\
g_{\mu\sigma}g_{\nu\rho},\\
\epsilon_{\mu\nu\rho\sigma}.
\end{array}
\right.
\end{equation}
For the case with six free indices 
of the form $A^\mu B^\nu C^\rho D^\sigma E^\delta F^\epsilon$, 
one finds the fifteen independent
possibilities of combining three metrics, i.e., $g_{\mu\nu}g_{\rho\sigma}
g_{\delta\epsilon}$ and the nonequivalent permutations of indices, as
well as fifteen combinations of a metric and a Levi-Civita tensor, of which
only ten are independent, which we choose to be
\begin{equation}
\left\{
\begin{array}{l}
g_{\nu\rho} \epsilon_{\mu\sigma\delta\epsilon},\\
g_{\nu\sigma} \epsilon_{\mu\rho\delta\epsilon},\\
g_{\nu\delta} \epsilon_{\mu\rho\sigma\epsilon},\\
g_{\nu\epsilon} \epsilon_{\mu\rho\sigma\delta},\\
g_{\rho\sigma} \epsilon_{\mu\nu\delta\epsilon},\\
g_{\rho\delta}\epsilon_{\mu\nu\sigma\epsilon},\\
g_{\rho\epsilon}\epsilon_{\mu\nu\sigma\delta},\\
g_{\sigma\delta}\epsilon_{\mu\nu\rho\epsilon},\\
g_{\sigma\epsilon}\epsilon_{\mu\nu\rho\delta},\\
g_{\epsilon\delta}\epsilon_{\mu\nu\rho\sigma}.
\end{array}
\right.
\end{equation}

Now, one needs to take into account that when only one vector $A^\mu$
and its gradient are plugged into these expressions, some terms are
identical and can thus be simplified. The following table sums up the
number of independent terms that can be built for a given product of vectors and gradients. Numbers in parentheses indicate those terms that would vanish
if it were not for the group index; in our listings of all available
Lagrangians below, we indicate these contractions with a star.
Given the above discussion, we are sure that all the possible terms have
been found, and they are all linearly independent.
\begin{center}
\begin{tabular}{|c|c|c|c|}
\hline
\backslashbox{$\# (\partial^\mu A^\nu)$}{$\# A^\rho A^\sigma$} & 0 & 1 & 2 \\
\hline
1 & 1 (0)  & 3 (1) & 6 (4)  \\
\hline
2 & 4 (0)  & 13 (3)  & 34 (23) \\
\hline
3 & 9 (2)  & 52 (22)  &  \\
\hline
\end{tabular}
\end{center}
We now discuss each case separately.

For a single derivative and no additional field, one gets the simplest
combination, namely, $\left(\partial \cdot A \right)$. With two
additional fields, one gets
\begin{equation}
\left\{ \begin{array}{l}
\left(\partial \cdot A \right) \left(A \cdot A \right),\\
\left[\left(\partial^\mu A^\nu \right) A_\mu A_\nu \right],\\
\left[ \epsilon_{\mu \nu \rho \sigma} \left(\partial^\mu A^\nu \right) 
A^\rho A^\sigma \right], ~~~~ ~~~~ (*)
\end{array} \right.
\end{equation}
and with four additional fields, one obtains
\begin{equation}
\left\{ \begin{array}{l}
\left(\partial \cdot A \right) \left(A \cdot A \right)\left(A \cdot A
\right),\\
\left[\left(\partial^\mu A^\nu \right) A_\mu A_\nu \right]\left(A \cdot
A \right),\\
\left[ \epsilon_{\mu \nu \rho \sigma} \left(\partial^\mu A^\nu \right) 
A^\rho A^\sigma \right]\left(A \cdot A \right), ~~~~ ~~~~ (*) \\
\left[\epsilon_{\mu \nu \rho \sigma}\left(\partial^\mu A^\alpha \right)
A^\nu A^\rho A^\sigma A_\alpha    \right], ~~~~ ~~~~ (*)  \\
\left[\epsilon_{\mu \nu \rho \sigma}\left(\partial^\alpha A^\mu \right)
A^\nu A^\rho A^\sigma A_\alpha    \right],  ~~~~ ~~~~ (*) \\
\left(\partial \cdot A \right) \left[\epsilon_{\mu \nu \rho \sigma}
A^\mu A^\nu A^\rho A^\sigma  \right]. ~~~~ ~~~~ (*)
\end{array} \right.
\end{equation}

With two derivatives and no additional field, one then finds
\begin{equation}
\left\{ \begin{array}{l}
\left(\partial \cdot A \right)\left(\partial \cdot A \right),\\
\left[\left(\partial^\mu A^\nu \right) \left(\partial_\mu A_\nu \right)
\right],\\
\left[\left(\partial^\mu A^\nu \right) \left(\partial_\nu A_\mu \right)
\right],\\
\left[ \epsilon_{\mu \nu \rho \sigma} \left(\partial^\mu A^\nu \right)
\left(\partial^\rho A^\sigma \right)  \right],
\end{array} \right.
\end{equation}
whereas with two additional fields, one finds\footnote{As an example of the fact that not
every reshuffling of indices is independent, let us consider the term $
\epsilon_{\mu \nu \rho \sigma} A^\mu A^\nu \left(\partial^\alpha A^\rho
\right)   \left(\partial_\alpha A^\sigma \right)$, which could, in
principle, have appeared in the list in Eq. \eqref{17}. It is indeed not
necessary because the property
\begin{equation}
g_{\mu\rho} \epsilon_{\nu\sigma\delta\epsilon} = g_{\nu\rho}
\epsilon_{\mu\sigma\delta\epsilon} - g_{\rho\sigma}
\epsilon_{\mu\nu\delta\epsilon} + g_{\rho\delta}
\epsilon_{\mu\nu\sigma\epsilon} -
g_{\rho\epsilon}\epsilon_{\mu\nu\sigma\delta}
\end{equation}
allows us to write it as a linear combination of the terms in
Eq.~\eqref{17}.
}

\begin{equation}
\left\{ \begin{array}{l}
\left(\partial \cdot A \right)\left(\partial \cdot A \right)\left(A
\cdot A \right),\\
\left[\left(\partial^\mu A^\nu \right) \left(\partial_\mu A_\nu \right)
\right]\left(A \cdot A \right),\\
\left[\left(\partial^\mu A^\nu \right) \left(\partial_\nu A_\mu \right)
\right]\left(A \cdot A \right),\\
\left[ \epsilon_{\mu \nu \rho \sigma} \left(\partial^\mu A^\nu \right)
\left(\partial^\rho A^\sigma \right)  \right]\left(A \cdot A \right),\\
\left[\left(\partial^\mu A^\nu \right) A_\mu A_\nu \right]\left(\partial
\cdot A \right),\\
\left[ \epsilon_{\mu \nu \rho \sigma} \left(\partial^\mu A^\nu \right) 
A^\rho A^\sigma \right]\left(\partial \cdot A \right),  ~~~~ ~~~~ (*) \\
\left[ A_\mu A_\nu  \left(\partial^\mu A^\alpha \right)
\left(\partial^\nu A_\alpha \right) \right],\\
\left[ A_\mu A_\nu  \left(\partial^\mu A^\alpha \right)
\left(\partial_\alpha A^\nu \right) \right],\\
\left[ A_\mu A_\nu  \left(\partial^\alpha A^\mu \right)
\left(\partial_\alpha A^\nu \right) \right],\\
\left[ \epsilon_{\mu \nu \rho \sigma} A^\mu A^\nu  \left(\partial^\rho
A^\alpha \right)  \left(\partial^\sigma A_\alpha \right)\right],  ~~~~
~~~~ (*) \\
\left[ \epsilon_{\mu \nu \rho \sigma} A^\mu A^\nu  \left(\partial^\rho
A^\alpha \right)  \left(\partial_\alpha A^\sigma \right)\right],  ~~~~
~~~~ (*) \\
\left[ \epsilon_{\mu \nu \rho \sigma} A^\mu A^\alpha  \left(\partial^\nu
A^\rho \right)  \left(\partial^\sigma A_\alpha \right)\right], \\
\left[ \epsilon_{\mu \nu \rho \sigma} A^\mu A^\alpha  \left(\partial^\nu
A^\rho \right)  \left(\partial_\alpha A^\sigma \right)\right].
\end{array} \right.
\label{17}
\end{equation}

Finally, demanding three gradients of the vector field and no vector
field itself, one obtains
\begin{equation}
\left\{ \begin{array}{l}
\left(\partial \cdot A \right)\left(\partial \cdot A \right)\left(
\partial \cdot A \right),\\
\left[\left(\partial^\mu A^\nu \right) \left(\partial_\mu A_\nu \right)
\right]\left( \partial \cdot A \right),\\
\left[\left(\partial^\mu A^\nu \right) \left(\partial_\nu A_\mu \right)
\right]\left( \partial \cdot A \right),\\
\left[ \epsilon_{\mu \nu \rho \sigma} \left(\partial^\mu A^\nu \right) 
\left(\partial^\rho A^\sigma \right)  \right] \left( \partial \cdot A
\right),  \\
\left[ \left(\partial^\mu A_\nu \right) \left(\partial^\nu A_\rho
\right) \left(\partial^\rho A_\mu \right) \right],\\
\left[ \left(\partial^\mu A_\nu \right) \left(\partial^\nu A_\rho
\right) \left(\partial_\mu A^\rho \right) \right],\\
\left[ \epsilon_{\mu \nu \rho \sigma} \left(\partial^\mu A^\alpha
\right) \left(\partial^\nu A_\alpha \right) \left(\partial^\rho A^\sigma
\right) \right], ~~~~ ~~~~ (*) \\
\left[ \epsilon_{\mu \nu \rho \sigma} \left(\partial^\mu A^\alpha
\right) \left(\partial_\alpha A^\nu \right) \left(\partial^\rho A^\sigma
\right) \right],\\
\left[ \epsilon_{\mu \nu \rho \sigma} \left(\partial^\alpha A^\mu
\right)\left(\partial_\alpha A^\nu \right)\left(\partial^\rho A^\sigma
\right) \right]. ~~~~ ~~~~ (*)
\end{array} \right.
\end{equation}

\subsection{Group Sector}

Let us now proceed with the similar procedure but now in the group sector.
Since we assumed that the vector fields transform according to the
representation of dimension $3$ of SU(2), one can safely use known
results from representation theory of compact Lie groups. The table
below summarizes the different possibilities to obtain an SU(2) singlet
as a function of the number of fields belonging to the $\mathbf{3}$
representation of SU(2) \cite{Slansky:1981yr,Feger:2012bs}:

\begin{center}
\begin{tabular}{|l|c|c|c|c|c|c|c|}
\hline
$\#$ vector fields in the $\mathbf{3}$ of SU(2) & 1 & 2 & 3 & 4 & 5 & 6
& 7\\
\hline
$\#$ of SU(2) singlets & 0  & 1 & 1 & 3 & 6 & 15 & 36 \\
\hline
\end{tabular}
\end{center}

We reproduce below the procedure explained in
Sec.~\ref{PartProcedureSinglet}, whereby one constructs the necessary
products of group metric coefficients $\delta_{ab}$ and structure
constants $\epsilon_{abc}$. Getting as many independent terms as
predicted by the representation theory (table above) ensures completeness of
the basis. Similar to the Lorentz invariance discussed in the previous
section, these two tensors are the only
primitive invariants of the group
\cite{Slansky:1981yr,Fuchs:1997jv,Padilla:2010ir}.

To contract with two or three free SU(2) indices, the only possible
choices are, respectively, $\delta_{ab}$ and $\epsilon_{abc}$. With four fields,
one can make use of the three combinations
\begin{equation}
\label{DabDcd}
\left\{ \begin{array}{l}
\delta_{ab}\delta_{cd},\\
\delta_{ac}\delta_{bd},\\
\delta_{ad}\delta_{bc},
\end{array} \right.
\end{equation}
while five fields demand the following six possibilities, namely
\begin{equation}
 \left\{ \begin{array}{l}
\delta_{ab}\epsilon_{cde},\\
\delta_{ac}\epsilon_{bde},\\
\delta_{ad}\epsilon_{bce},\\
\delta_{bc}\epsilon_{ade},\\
\delta_{bd}\epsilon_{ace},\\
\delta_{cd}\epsilon_{abe}.
\end{array} \right. 
\end{equation}

As in Sec.~\ref{LSSection}, one can devise other possible
formulations that apply, but they will always be expressible as linear
combinations of the above. For instance, relations between the structure
constants, such as
\begin{equation}
\epsilon_{ab}{}^{e}\epsilon_{cde}=\delta_{ac}\delta_{bd} -
\delta_{ad}\delta_{bc},
\end{equation}
imply that contracting a four-index term with two structure
constants is equivalent to a linear combination of the terms given in
Eq.~\eqref{DabDcd}.

\subsection{Final Test Lagrangians}

Gathering the results and applying the procedure of
Sec.~\ref{PartProcedureSinglet}, we are now in a position to write down
our test Lagrangians, scalars under both Lorentz and SU(2)
transformations. Some of these terms simplify through contractions, e.g.,
$\epsilon_{\a\b\c}\left(A^{\a} \cdot A^{\b} \right)\left(\partial \cdot
A^{\c}\right) = 0$, and we are left with fewer terms than the naive
multiplication of all singlet possibilities of each sector would have
otherwise suggested. This is fortunate because the number of terms to be
considered a priori is quickly increasing with the number of fields
involved, as shown in the table below:
\begin{center}
\begin{tabular}{|c|c|c|c|}
\hline
\backslashbox{$\# \partial^\mu A^{\nu\a}$}{$\# A^{\rho \b}$} & 0 & 2 & 4
\\
\hline
1 & 0  & 3 & 36  \\
\hline
2 & 4  & 42  & 510 \\
\hline
3 & 9  & 312  &  \\
\hline
\end{tabular}
\end{center}					

After simplifications, we find two terms (instead of three according to the
table) containing a single derivative term and two additional vector fields,
\begin{equation}
\left\{ \begin{array}{l}
\mathcal{L}_{1}=\epsilon_{\a \b \c}\left[\left(\partial^\mu A^{\a\nu}
\right) A^{\b}_\mu A^{\c}_\nu \right],\\
\mathcal{L}_{2}=\epsilon_{\a\b\c}\left[ \epsilon_{\mu \nu \rho \sigma}
\left(\partial^\mu A^{\a\nu} \right)  A^{\b\rho} A^{\c\sigma} \right]  ,
\end{array} \right.
\end{equation}
and eight with four such fields, namely,
\begin{equation}
\left\{ \begin{array}{l}
\mathcal{L}_{1}=\epsilon_{\a\b\c}\left[\left(\partial^\mu A^{\d\nu}
\right) A^{\a}_\mu A^{\b}_\nu \right]\left(A^{\c} \cdot A_{\d}
\right),\\
\mathcal{L}_{2}=\epsilon_{\a\b\c}\left[\left(\partial^\mu A^{\a\nu}
\right) A^{\d}_\mu A^{\b}_\nu \right]\left(A^{\c} \cdot A_{\d}
\right),\\
\mathcal{L}_{3}=\epsilon_{\a\b\c}\left[\left(\partial^\mu A^{\a\nu}
\right) A^{\b}_\mu A^{\d}_\nu \right]\left(A^{\c} \cdot A_{\d}
\right),\\
\mathcal{L}_{4}=\epsilon_{\a\b\c}\left[ \epsilon_{\mu \nu \rho \sigma}
\left(\partial^\mu A^{\d\nu} \right)  A^{\a\rho} A^{\b\sigma}
\right]\left(A^{\c} \cdot A_{\d} \right),  \\
\mathcal{L}_{5}=\epsilon_{\a\b\c}\left[ \epsilon_{\mu \nu \rho \sigma}
\left(\partial^\mu A^{\a\nu} \right)  A^{\d\rho} A^{\b\sigma}
\right]\left(A^{\c} \cdot A_{\d} \right),  \\
\mathcal{L}_{6}=\epsilon_{\a\b\c}\left[\epsilon_{\mu \nu \rho
\sigma}\left(\partial^\mu A^{\d\alpha} \right) A_{\d}^\nu A^{\a\rho}
A^{\b\sigma} A^{\c}_\alpha    \right],\\
\mathcal{L}_{7}=\epsilon_{\a\b\c}\left[\epsilon_{\mu \nu \rho
\sigma}\left(\partial^\alpha A^{\d\mu} \right) A_{\d}^\nu A^{\a\rho}
A^{\b\sigma} A^{\c}_\alpha    \right] ,\\
\mathcal{L}_{8}=\epsilon_{\a\b\c}\left[\epsilon_{\mu \nu \rho
\sigma}\left(\partial \cdot A_{\d} \right) A^{\d \mu} A^{\a\nu}
A^{\b\rho} A^{\c \sigma}    \right] .\\
\end{array} \right.
\end{equation}
Note that one cannot build a single derivative term without an additional
field, as it would otherwise belong to the $\mathbf{3}$ representation of SU(2).

For two first-order vector field derivatives without additional fields, one gets
\begin{equation}
\left\{ \begin{array}{l}
\mathcal{L}_1=\left(\partial \cdot A^{\a} \right)\left(\partial \cdot
A_{\a} \right),\\ \mathcal{L}_2=\left[\left(\partial^\mu A^\nu_{\a}
\right) \left(\partial_\mu A_\nu^{\a} \right) \right],\\
\mathcal{L}_3=\left[\left(\partial^\mu A^\nu_{\a} \right)
\left(\partial_\nu A_\mu^{\a} \right) \right],\\
\mathcal{L}_4=\left[ \epsilon_{\mu \nu \rho \sigma} \left(\partial^\mu
A^{\a\nu} \right)  \left(\partial^\rho A^\sigma_{\a} \right)  \right],
\end{array} \right.
\end{equation}
whereas with two additional vector fields, one gets 
\begin{equation}
\label{EqLagBeforeHessian}
\left\{ \begin{array}{l}
\mathcal{L}_1=\left(\partial \cdot A^{\a}\right)\left(\partial \cdot
A_{\a} \right)\left(A^{\b} \cdot A_{\b} \right),\\
\mathcal{L}_2=\left(\partial \cdot A^{\a} \right)\left(\partial \cdot
A^{\b} \right)\left(A_{\a} \cdot A_{\b} \right),\\
\mathcal{L}_3=\left[\left(\partial^\mu A^\nu_{\a} \right)
\left(\partial_\mu A_\nu^{\a} \right) \right]\left(A^{\b} \cdot A_{\b}
\right),\\
\mathcal{L}_4=\left[\left(\partial^\mu A^\nu_{\a} \right)
\left(\partial_\mu A_\nu^{\b} \right) \right]\left(A^{\a} \cdot A_{\b}
\right),\\
\mathcal{L}_5=\left[\left(\partial^\mu A^\nu_{\a} \right)
\left(\partial_\nu A_\mu^{\a} \right) \right]\left(A^{\b} \cdot A_{\b}
\right),\\
\mathcal{L}_6=\left[\left(\partial^\mu A^\nu_{\a} \right)
\left(\partial_\nu A_\mu^{\b} \right) \right]\left(A^{\a} \cdot A_{\b}
\right),\\
\mathcal{L}_7=\left[ \epsilon_{\mu \nu \rho \sigma} \left(\partial^\mu
A^{\a\nu} \right)  \left(\partial^\rho A^\sigma_{\a} \right) 
\right]\left(A^{\b} \cdot A_{\b} \right),\\
\mathcal{L}_8=\left[ \epsilon_{\mu \nu \rho \sigma} \left(\partial^\mu
A^{\a\nu} \right)  \left(\partial^\rho A^\sigma_{\b} \right) 
\right]\left(A_{\a} \cdot A^{\b} \right),\\
\mathcal{L}_9=\left[\left(\partial^\mu A^\nu_{\a} \right) A_\mu^{\a}
A_\nu^{\b} \right]\left(\partial \cdot A_{\b} \right),\\
\mathcal{L}_{10}=\left[\left(\partial^\mu A^\nu_{\a} \right) A_\mu^{\b}
A_\nu^{\a} \right]\left(\partial \cdot A_{\b} \right),\\
\mathcal{L}_{11}=\left[\left(\partial^\mu A^\nu_{\a} \right) A_\mu^{\b}
A_{\b\nu} \right]\left(\partial \cdot A^{\a} \right),\\
\mathcal{L}_{12}=\left[ \epsilon_{\mu \nu \rho \sigma}
\left(\partial^\mu A^{\a\nu} \right)  A^\rho_{\a} A^\sigma_{\b}
\right]\left(\partial \cdot A^{\b} \right), \\
\mathcal{L}_{13}=\left[ A_\mu^{\a} A_{\a\nu}  \left(\partial^\mu
A^\alpha_{\b} \right) \left(\partial^\nu A_{\alpha}^{\b} \right)
\right],\\
\mathcal{L}_{14}=\left[ A_\mu^{\a} A_{\nu}^{\b}  \left(\partial^\mu
A^\alpha_{\a} \right) \left(\partial^\nu A_{\b\alpha} \right) \right],\\
\mathcal{L}_{15}=\left[ A_\mu^{\a} A_{\a\nu}  \left(\partial^\mu
A^{\b\alpha} \right) \left(\partial_\alpha A^\nu_{\b} \right) \right],\\
\mathcal{L}_{16}=\left[ A_\mu^{\a} A_{\nu}^{\b}  \left(\partial^\mu
A^{\alpha}_{\a} \right) \left(\partial_\alpha A^\nu_{\b} \right)
\right],\\
\mathcal{L}_{17}=\left[ A_\mu^{\a} A_{\nu}^{\b}  \left(\partial^\mu
A^{\alpha}_{\b} \right) \left(\partial_\alpha A^\nu_{\a} \right)
\right],\\
\mathcal{L}_{18}=\left[ \epsilon_{\mu \nu \rho \sigma} A^{\a\mu}
A^{\b\nu}  \left(\partial^\rho A^\alpha_{\a} \right) 
\left(\partial^\sigma A_{\b\alpha} \right)\right], \\
\mathcal{L}_{19}=\left[ \epsilon_{\mu \nu \rho \sigma} A^{\a\mu}
A^{\b\nu}  \left(\partial^\rho A^\alpha_{\a} \right) 
\left(\partial_\alpha A^\sigma_{\b} \right)\right] ,\\
\mathcal{L}_{20}=\left[ \epsilon_{\mu \nu \rho \sigma} A^{\a\mu}
A^{\b\nu}  \left(\partial^\alpha A^\rho_{\a} \right)  
\left(\partial_\alpha A^\sigma_{\b} \right)\right]  ,\\
\mathcal{L}_{21}=\left[ \epsilon_{\mu \nu \rho \sigma} A^{\a\mu}
A^\alpha_{\a}  \left(\partial^\nu A^{\rho}_{\b} \right) 
\left(\partial^\sigma A_\alpha ^{\b}\right)\right], \\
\mathcal{L}_{22}=\left[ \epsilon_{\mu \nu \rho \sigma} A^{\a\mu}
A^\alpha_{\b}  \left(\partial^\nu A^{\rho}_{\a} \right) 
\left(\partial^\sigma A_\alpha ^{\b}\right)\right], \\
\mathcal{L}_{23}=\left[ \epsilon_{\mu \nu \rho \sigma} A^{\mu}_{\a}
A^\alpha_{\b}  \left(\partial^\nu A^{\b\rho} \right) 
\left(\partial^\sigma A_\alpha ^{\a}\right)\right] ,\\
\mathcal{L}_{24}=\left[ \epsilon_{\mu \nu \rho \sigma} A^{\a\mu}
A^\alpha_{\a}  \left(\partial^\nu A^{\b\rho} \right) 
\left(\partial_\alpha A^\sigma_{\b} \right)\right] ,\\
\mathcal{L}_{25}=\left[ \epsilon_{\mu \nu \rho \sigma} A^{\a\mu}
A^\alpha_{\b}  \left(\partial^\nu A^{\rho}_{\a} \right) 
\left(\partial_\alpha A^{\b\sigma} \right)\right], \\
\mathcal{L}_{26}=\left[ \epsilon_{\mu \nu \rho \sigma} A^{\mu}_{\a}
A^\alpha_{\b}  \left(\partial^\nu A^{\b\rho} \right) 
\left(\partial_\alpha A^{\a\sigma} \right)\right] ,\\
\mathcal{L}_{27}=\left[ A_\mu^{\a} A_{\nu}^{\b}  \left(\partial^\mu
A^\alpha_{\b} \right) \left(\partial^\nu A_{\a \alpha} \right)
\right],\\
\mathcal{L}_{28}=\left[ A_\mu^{\a} A_{\nu}^{\b}  \left(\partial^\alpha
A^\mu_{\b} \right) \left(\partial_\alpha A^\nu_{\a} \right) \right].
\end{array} \right.
\end{equation}

Finally, with three derivatives, one finds
\begin{equation}
\left\{ \begin{array}{l}
\mathcal{L}_{1}=  \epsilon_{\a \b \c} \left[\left(\partial^\mu
A_\nu^{\a} \right) \left(\partial^\nu A_\rho^{\b} \right)
\left(\partial^\rho A_\mu^{\c} \right) \right], \\
\mathcal{L}_{2}=  \epsilon_{\a \b \c} \left[ \left(\partial^\mu
A_\nu^{\a} \right) \left(\partial^\nu A_\rho^{\b} \right)
\left(\partial_\mu A^{\c \rho} \right) \right], \\
\mathcal{L}_{3}=  \epsilon_{\a \b \c} \left[ \epsilon_{\mu \nu \rho
\sigma} \left(\partial^\mu A^{\a \alpha} \right) \left(\partial^\nu
A_\alpha^{\b} \right) \left(\partial^\rho A^{\c \sigma} \right) \right],
\\
\mathcal{L}_{4}=  \epsilon_{\a \b \c} \left[ \epsilon_{\mu \nu \rho
\sigma} \left(\partial^\mu A^{\a \alpha} \right) \left(\partial_\alpha
A^{\b \nu} \right) \left(\partial^\rho A^{\c \sigma} \right) \right], \\
\mathcal{L}_{5}=  \epsilon_{\a \b \c} \left[ \epsilon_{\mu \nu \rho
\sigma} \left(\partial^\alpha A^{\a \mu} \right)\left(\partial_\alpha
A^{\b \nu} \right)\left(\partial^\rho A^{\c \sigma} \right) \right],
\end{array} \right.
\end{equation}
completing our list of test Lagrangians.

\section{Construction of the healthy terms} \label{cht}
\subsection{Hessian Condition}
\label{PartHessianCondition}

Let us now apply the Hessian condition, as discussed in
Sec.~\ref{PartProcedure}. The first step is to calculate the Hessians
associated with the various test Lagrangians, defined by
Eq.~(\ref{EqDefHessian}). One sees that only those terms containing at
least two first-order derivatives of the vector field yield a nonvanishing value. In practice, one gets
\begin{equation}
\left\{ \begin{array}{l}
\mathcal{H}^{\mu\nu \al\be}_1= 2 g ^{0 \mu} g ^{0 \nu} g^{\al
\be},\\
\mathcal{H}^{\mu\nu \al\be}_2=-2 g^{\mu\nu} g^{\al \be},\\
\mathcal{H}^{\mu\nu \al\be}_3=2  g ^{0 \mu} g^{0 \nu} g^{\al \be},\\
\mathcal{H}^{\mu\nu \al\be}_4=0,
\end{array} \right.
\end{equation}
for the terms with two first-order derivatives and no additional fields, and
\begin{equation}
\left\{ \begin{array}{l}
\mathcal{H}^{\mu\nu \al\be}_1=2 g ^{0 \mu} g ^{0 \nu} g^{\al
\be}\left(A^{\b}\cdot A_{\b} \right),\\
\mathcal{H}^{\mu\nu \al\be}_2=2 g ^{0 \mu} g ^{0 \nu} \left(A^{\al}\cdot
A^{\be} \right),\\
\mathcal{H}^{\mu\nu \al\be}_3= -2g^{\mu\nu}g^{\al \be} \left(A^{\b}\cdot
A_{\b} \right),\\
\mathcal{H}^{\mu\nu \al\be}_4= -2g^{\mu\nu} \left(A^{\al}\cdot A^{\be}
\right),\\ \mathcal{H}^{\mu\nu \al\be}_5=2 g ^{0 \mu} g ^{0 \nu}g^{\al
\be} \left(A^{\b}\cdot A_{\b} \right),\\
\mathcal{H}^{\mu\nu \al\be}_6=2 g ^{0 \mu} g ^{0 \nu} \left(A^{\al}\cdot
A^{\be} \right),\\
\mathcal{H}^{\mu\nu \al\be}_7=0,\\
\mathcal{H}^{\mu\nu \al\be}_8=0,\\
\mathcal{H}^{\mu\nu \al\be}_9=A ^{0 \al} A ^{\mu \be} g^{0\nu}+A ^{0
\be} A ^{\nu \al} g^{0\mu},\\
\mathcal{H}^{\mu\nu \al\be}_{10}=A ^{0 \be} A ^{\mu \al} g^{0\nu}+A ^{0
\al} A ^{\nu \be} g^{0\mu},\\
\mathcal{H}^{\mu\nu \al\be}_{11}=A ^{0 \b} A ^{\mu}_{\b} g^{0\nu}g^{\al
\be}+A ^{0 \b} A ^{\nu}_{\b} g^{0\mu}g^{\al \be},\\
\mathcal{H}^{\mu\nu \al\be}_{12}=\epsilon^{0\mu}{}_{\rho\sigma} A^{\rho
\al} A^{\sigma \be} g^{0\nu}+\epsilon^{0\nu}{}_{\rho\sigma} A^{\rho \be}
A^{\sigma \al} g^{0\mu},\\
\mathcal{H}^{\mu\nu \al\be}_{13}=2 A^{0 \b} A^{0}_{\b} g^{\mu\nu} g^{\al
\be},\\
\mathcal{H}^{\mu\nu \al\be}_{14}=2 A^{0 \al} A^{0 \be} g^{\mu\nu} ,\\
\mathcal{H}^{\mu\nu \al\be}_{15}=A^{0 \b} A^{\nu}_{\b} g^{\mu 0} g^{\al
\be}+A^{0 \b} A^{\mu}_{\b} g^{\nu 0} g^{\al \be},\\
\mathcal{H}^{\mu\nu \al\be}_{16}=A^{0 \al} A^{\nu \be} g^{\mu 0}+A^{0
\be} A^{\mu \al} g^{\nu 0},\\
\mathcal{H}^{\mu\nu \al\be}_{17}=A^{0 \be} A^{\nu \al} g^{\mu 0}+A^{0
\al} A^{\mu \be} g^{\nu 0},\\
\mathcal{H}^{\mu\nu \al\be}_{18}=0,\\
\mathcal{H}^{\mu\nu \al\be}_{19}=\epsilon _{\rho \sigma}{}^{0 \nu}
A^{\rho \al}\ A^{\sigma \be} g^{\mu 0} + \epsilon _{\rho \sigma}{}^{0
\mu} A^{\rho \be}\ A^{\sigma \al} g^{\nu 0} ,\\
\mathcal{H}^{\mu\nu \al\be}_{20}= -2 \epsilon_{\rho \sigma}{}^{\mu\nu}
A^{\rho \al}A^{\sigma \be},\\
\mathcal{H}^{\mu\nu \al\be}_{21}=0,\\
\mathcal{H}^{\mu\nu \al\be}_{22}=0,\\
\mathcal{H}^{\mu\nu \al\be}_{23}=0,\\
\mathcal{H}^{\mu\nu \al\be}_{24}=0,\\
\mathcal{H}^{\mu\nu \al\be}_{25}=\epsilon_{\rho}{}^{0\mu\nu} A^{\rho
\al} A^{0\be} + \epsilon_{\rho}{}^{0\nu\mu} A^{\rho \be} A^{0\al} ,\\
\mathcal{H}^{\mu\nu \al\be}_{26}=\epsilon_{\rho}{}^{0\mu\nu} A^{\rho
\be} A^{0\al} + \epsilon_{\rho}{}^{0\nu\mu} A^{\rho \al} A^{0\be} ,\\
\mathcal{H}^{\mu\nu \al\be}_{27}=2 A^{0 \al} A^{0 \be} g^{\mu\nu} ,\\
\mathcal{H}^{\mu\nu \al\be}_{28}= -2 A^{\nu \al} A^{\mu\be},
\end{array} \right.
\end{equation}
for those with two first-order derivatives and two additional vector fields.

For the terms with three first-order derivatives, we have
\begin{equation}
\left\{ \begin{array}{l}
\mathcal{H}^{\mu\nu \al\be}_1=3 \epsilon^{\al \be}{}_{\c} (g^{0 \mu}
\partial^\nu A^{\c 0} - g^{0 \nu} \partial^\mu A^{\c 0}), \\
\mathcal{H}^{\mu\nu \al\be}_2=\epsilon^{\al \be}{}_{\c} (g^{0 \mu}
\partial^0 A^{\c \nu} - g^{0 \nu} \partial^0 A^{\c \mu}) + \epsilon^{\al
\be}{}_{\c} (\partial^\mu A^{\c \nu} - \partial^\nu A^{\c \mu}), \\
\mathcal{H}^{\mu\nu \al\be}_3=0,\\
\mathcal{H}^{\mu\nu \al\be}_4= \epsilon^{\al \be}{}_{\c} (\epsilon^{0
\nu \rho \sigma} g^{0 \mu} \partial_\rho A^{\c}_\sigma - \epsilon^{0 \mu
\rho \sigma} g^{0 \nu} \partial_\rho A^{\c}_\sigma) + 2 \epsilon^{\al
\be}{}_{\c} \epsilon^{\rho \mu 0 \nu} \partial_\rho A^{\c 0} , \\
\mathcal{H}^{\mu\nu \al\be}_5= -2 \epsilon^{\al \be}{}_{\c}
\epsilon^{\mu \nu \rho \sigma} \partial_\rho A^{\c}_\sigma + 4
\epsilon^{\al \be}{}_{\c} \epsilon^{\rho \mu 0 \nu} \partial^0
A^{\c}_\rho .
\end{array} \right.
\end{equation}

With these partial Hessians, we now construct a basis of terms
fulfilling the condition discussed above, i.e., such that
$\mathcal{H}^{0\mu\al\be} = 0$ for all values of $\mu$, $\al$ and $\be$; see
Sec.~\ref{PartProcedure}. To reach this goal, using notations already
introduced in Ref.~\cite{Allys:2015sht}, we produce a Lagrangian by
means of a linear combination of our test ones, namely,
\begin{equation}
\mathcal{L}_{\text{test}}=\sum_i x_i \mathcal{L}_i,
\end{equation}
for a yet-unknown set of constant parameters $x_i$. The Hessian is then
calculated for this Lagrangian, leading to algebraic equations for the
$x_i$ whose roots provide the required actions. It turns out to be
easier to separately compute the cases $\mu=0$ and $\mu=i$, as well as
$\al=\be$ and $\al\neq\be$.

Let us begin with the case $\al=\be$. Test Lagrangians with two
derivatives and no additional fields have only one Hessian component not
identically vanishing, namely,
\begin{equation}
\mathcal{H}^{00\al \al} =  4(x_1 + x_2 + x_3), 
\end{equation}
while for two additional vector fields, there are four independent Hessian
conditions, given by
\begin{align}
\mathcal{H}^{00\al\al} = & ~  4\left(x_1 + x_3 + x_5 \right) \left(
A^{\b} \cdot A_{\b}\right) + 2 \left( x_2 + x_4 + x_6 \right)
\left(A^{\al} \cdot A^{\al} \right) \nonumber\\
&- 2 \left( x_9 + x_{10} + x_{14} + x_{16} + x_{17}+ x_{27} + x_{28}
\right) \left( A^{0\al} A^{0\al} \right) \nonumber \\
& - 4\left( x_{11} + x_{13} + x_{15}  \right) \left( A^{0 \b} A^{0}_{\b}
\right),
\end{align}
\begin{align}
\mathcal{H}^{0i\al\al} =&  -\left( x_9 + x_{10} + x_{16} + x_{17}  +
2x_{28} \right) \left( A^{0\al} A^{i \al} \right) - 2 \left(
x_{11} + x_{15} \right) \left(A^{0 \b} A^{i}_{\b} \right) \nonumber\\
& -\left(x_{12} + x_{19} + 2 x_{20} \right) \left( \epsilon_{\rho
\sigma}{}^{0i}A^{\rho \al}A^{\sigma \al} \right).
\end{align}
On the other hand, the case $\al \not= \be$ implies 
\begin{equation}
\mathcal{H}^{00\al\be} = 2 \left(x_2 + x_4 + x_6 \right) \left( A^{\al}
\cdot A^{\be} \right) - 2 \left( x_9 + x_{10} +x_{14} + x_{16} + x_{17}
+ x_{27} + x_{28} \right) \left( A^{0\al} A^{0\be} \right),
\end{equation}
\begin{align}
\mathcal{H}^{0i \al\be} = -\left(x_9 + x_{17} + 2x_{28} \right) \left(
A^{0\be} A^{i\al} \right) - \left(x_{10} + x_{16} \right) \left(A^{0\al}
A^{i\be} \right) -\left(- x_{12} + x_{19} + 2 x_{20} \right)
\left(\epsilon^{0i}{}_{\rho\sigma} A^{\rho \al}A^{\sigma \be} \right).
\end{align}
Making these four terms vanish can be done, without loss of
generality (since all linear combinations of the resulting terms are all
also acceptable):
\begin{equation}
\begin{array}{l}
x_3 = -x_1 - x_5, \\
x_4 = - x_2 - x_6, \\
x_{12} = 0, \\
x_{13} = 0,\\
x_{14} = - x_{27} + x_{28},  \\
x_{15} = - x_{11} , \\
x_{16} = -x_{10} , \\
x_{17} = -x_9 - 2 x_{28},\\
x_{19} =  -2 x_{20}.
\end{array}
\end{equation}

With three derivatives, one finds that $\mathcal{H}^{00\al\al} $,
$\mathcal{H}^{00\al\be}$ ($\al\neq\be$) and $\mathcal{H}^{0i\al\al}$
identically vanish, whereas for $\al \neq \be$, we have
\begin{equation}
\mathcal{H}^{0i\al\be} = \epsilon_{\c}{}^{\al \be} \left[(-3x_1 - x_2)
\partial^i A^{0 \c} - (x_4 + 2x_5) \epsilon^{0 i \rho \sigma}
\partial_\rho A_\sigma^{\c}\right],
\end{equation}
thus leading to the conditions
\begin{equation}
\begin{array}{l}
x_2 = - 3x_1, \\
x_4 = - 2x_5.
\end{array}
\end{equation}

\subsection{Simplification of the Lagrangian}

For one gradient and two vector fields, we can define the current
\begin{equation}
J^\mu = \epsilon_{\a\b\c} \epsilon^{\mu\nu\rho\sigma} A_\nu^{\a}
A_\rho^{\b} A_\sigma^{\c},
\end{equation}
showing that $\mathcal{L}_2$ is a total derivative, namely,
\begin{equation}
\partial_\mu J^\mu = 3 \mathcal{L}_2.
\end{equation}

A similar technique applies for one derivative term and 4 additional
vector fields: in this case, one forms the following two currents,
\begin{equation}
\begin{array}{l}
J^{\mu}_1 = \epsilon^{\mu}{}_{\nu\rho\sigma} A^{\nu\a} A^{\rho\b} A
^{\sigma\c} A^{\alpha \d} A_{\alpha\d} \epsilon_{\a\b\c},\\
J^{\alpha}_{2} = \epsilon_{\mu\nu\rho\sigma}
A^{\mu\a}A^{\nu\b}A^{\rho\c} A^{\sigma\d} A^{\alpha}_{\d}
\epsilon_{\a\b\c},
\end{array}
\end{equation}
yielding
\begin{equation}
\begin{array}{l}
\partial_{\mu} J^{\mu}_1 = 3 \left( \mathcal{L}_{3}- 2 \mathcal{L}_{5} +
2 \mathcal{L}_{6} \right),\\
\partial_{\alpha} J^{\alpha}_2 = - \mathcal{L}_{8}.
\end{array}
\end{equation}
Finally, some terms involving two first-order derivatives can
be described by 
\begin{eqnarray}
J^{\mu_1} &=& \delta^{\mu_1\mu_2}_{\nu_1\nu_2} A^{\nu_1 \a}
\partial_{\mu_2} A^{\nu_2}_{\a}, \\
J^{\mu}_\epsilon &=& \epsilon^{\mu\nu\rho\sigma} A_\nu^{\a}
(\partial_\rho A_{\sigma \a}),
\end{eqnarray}
where we have used the definition $\delta^{\mu_1 \mu_2}_{\nu_1 \nu_2} \equiv
\delta^{\mu_1}_{\nu_1}\delta^{\mu_2}_{\nu_2}
-\delta^{\mu_1}_{\nu_2}\delta^{\mu_2}_{\nu_1}$
stemming from Eq.~\eqref{deltaMult}, leading to
\begin{eqnarray}
\partial_{\mu_1} J^{\mu_1} &=& \mathcal{L}_{1} - \mathcal{L}_{3}, \\
\partial_\mu J^\mu_\epsilon &=& \mathcal{L}_4.
\end{eqnarray}

Terms containing two derivatives and two fields are
slightly more involved. We first make use of the identity
\cite{Allys:2016jaq,Fleury:2014qfa}
\begin{equation}
\label{EqFP}
A^{\mu \alpha} \tilde{B}_{\nu \alpha} + B^{\mu \alpha} \tilde{A}_{\nu
\alpha} = \frac{1}{2} (B^{ \alpha \beta} \tilde{A}_{\alpha \beta}
)\delta^{\mu}_{\nu},
\end{equation}
valid for all antisymmetric tensors $A$ and $B$. This provides the relations
\begin{equation}
\left( G^{\mu\alpha\a}\tilde{G}_{\nu\alpha}^{\b} +
G^{\mu\alpha\b}\tilde{G}_{\nu\alpha}{}^{\a}\right) A_{\mu\a}
A^{\nu}_{\b} = \frac{1}{2} \left( G^{\alpha\beta \a}
\tilde{G}_{\alpha\beta}^{\b} \right) \left(A_{\a}\cdot A_{\b} \right)
\end{equation}
and
\begin{equation}
G^{\mu\alpha\a}\tilde{G}_{\nu\alpha\a} A_{\mu}^{\b}A^{\nu}_{\b} =
\frac{1}{4}\left( G^{\alpha\beta \a} \tilde{G}_{\alpha\beta \a}\right)
\left(A^{\b}\cdot A_{\b} \right),
\end{equation}
where $G^{\mu\alpha\a}$ is the Abelian form of the Faraday tensor
as defined below in Eq.~\eqref{EqFaradayAbelian}.
From these, one then derives the following two identities relating
the Lagrangians in Eq.~\eqref{EqLagBeforeHessian}:
\begin{equation}
\mathcal{L}_{25}+\mathcal{L}_{26}-\mathcal{L}_{22}-\mathcal{L}_{23}
=\mathcal{L}_{8}
\end{equation}
and
\begin{equation}
2\left(\mathcal{L}_{24}-\mathcal{L}_{21} \right) = \mathcal{L}_{7}.
\end{equation}

It is also possible to find total derivatives to reduce the number of
independent terms. First, one can use the fact that $\tilde{G}$ is
divergence-free, introducing the currents
\begin{equation}
\begin{array}{l}
J^{\mu}_{G,1}=\tilde{G}^{\mu\nu}_{\a}A_{\nu}^{\a} \left(A^{\b}\cdot
A_{\b} \right),\\
J^{\mu}_{G,2}=\tilde{G}^{\mu\nu}_{\a}A_{\nu\b} \left(A^{\a}\cdot A^{\b}
\right),
\end{array}
\end{equation}
providing
\begin{equation}
\begin{array}{l}
\partial_{\mu}J^{\mu}_{G,1}= \mathcal{L}_{7} - 2 \mathcal{L}_{22},\\
\partial_{\mu}J^{\mu}_{G,2}= \mathcal{L}_{8} - \mathcal{L}_{21} -
\mathcal{L}_{23}.
\end{array}
\end{equation}
One can subsequently use the antisymmetric forms written
from $\delta^{\mu_1 \mu_2}_{\nu_1 \nu_2}$: 
\begin{equation}
\begin{array}{l}
J^{\mu}_{\delta,1}=\delta^{\mu_1 \mu_2}_{\nu_1 \nu_2}
A_{\b}^{\lambda}A_{\lambda}^{\b} A^{\nu_1}_{\a} \partial_{\mu_2}
A^{\nu_2\a},\\
J^{\mu}_{\delta,2}=\delta^{\mu_1 \mu_2}_{\nu_1 \nu_2}
A_{\a}^{\lambda}A_{\lambda}^{\b} A^{\nu_1\a} \partial_{\mu_2}
A^{\nu_2}_{\b},\\
J^{\mu}_{\delta,3}=\delta^{\mu_1 \mu_2}_{\nu_1 \nu_2}
A^{\lambda\a}A^{\nu_1}_{\a} A^{\nu_2}_{\b}
\partial_{\mu_2}A_{\lambda}^{\b},
\end{array}
\end{equation}
resulting in
\begin{equation}
\begin{array}{l}
\partial_{\mu}J^{\mu}_{\delta,1} = \mathcal{L}_{1}-\mathcal{L}_{5}+2
\mathcal{L}_{10} - 2 \mathcal{L}_{16},\\
\partial_{\mu}J^{\mu}_{\delta,2} = \mathcal{L}_{2}-\mathcal{L}_{6}+ 
\mathcal{L}_{9}+ \mathcal{L}_{11} -  \mathcal{L}_{15} - 
\mathcal{L}_{17},\\
\partial_{\mu}J^{\mu}_{\delta,3} = \mathcal{L}_{14}+\mathcal{L}_{9}+
\mathcal{L}_{15}- \mathcal{L}_{27} -  \mathcal{L}_{17} - 
\mathcal{L}_{11}.
\end{array}
\end{equation}
Finally, we can write
\begin{equation}
J^{\mu}_{\epsilon,1}= \epsilon^{\mu}{}_{\nu\rho\sigma} A^{\nu \a}
A^{\rho\b}A^{\alpha}_{\a} \partial^{\sigma}A_{\alpha \b},
\end{equation}
implying
\begin{equation}
\partial_{\mu}J^{\mu}_{\epsilon,1}=\mathcal{L}_{18} + 
\mathcal{L}_{23} - \mathcal{L}_{21}.
\end{equation}

All the above conditions are linearly independent. They allow us to write
Lagrangians $\mathcal{L}_{9}$, $\mathcal{L}_{10}- \mathcal{L}_{16}$,
$\mathcal{L}_{11}- \mathcal{L}_{15}$, $\mathcal{L}_{18}$,
$\mathcal{L}_{21}$, $\mathcal{L}_{22}$, $\mathcal{L}_{24}$, and
$\mathcal{L}_{25}$ as functions of the other Lagrangians. Note, however,
that one can always add these to other terms of the final basis for
simplification purposes.

Lastly, the current
$J^{\mu} = \epsilon^{\mu}{}_{\nu\rho\sigma} \partial^{\nu} A^{\alpha \a}
\partial^{\rho} A_{\alpha}^{\b} A^{\sigma \c} \epsilon_{\a\b\c}$
permits us to simplify one of the terms containing three first-order
derivatives by making use of
$\partial_{\mu} J^{\mu} = \mathcal{L}_3$.

\subsection{A New Basis}

One can now rewrite our basis of Lagrangians satisfying the Hessian
condition, taking into account the extra relations stemming from the
total derivatives and the identity of Ref. \cite{Fleury:2014qfa}. We
group our terms to produce a new and more convenient basis, and for that
purpose, we use the Abelian form
of the Faraday tensor, namely,
\begin{equation}
\label{EqFaradayAbelian}
G_{\mu\nu}^{\a} = \partial_\mu A_\nu^{\a} - \partial_\nu A_\mu^{\a},
\end{equation}
as well as its Hodge dual $\tilde{G}_{\mu\nu}^{\a}=\frac12 \epsilon_{\mu
\nu\rho\sigma} G^{\rho\sigma\a}$, also defined in the usual way. Using
the Abelian form of the Faraday tensor to describe a non-Abelian vector field theory
may seem a bit unusual, but it considerably simplifies our
forthcoming considerations since this term naturally appears from the
first-order derivatives of the vector field and cancels in the scalar
sector. We later move on to a formulation using the actual non-Abelian 
Faraday tensor, as is given by Eq.~(\ref{EqFaradayNonAbelian}).
We also make use of the symmetric counterpart of the Abelian
Faraday tensor, namely, 
\begin{equation}
S_{\mu\nu}^{\a} = \partial_{\mu}A_{\nu}^{\a} +
\partial_{\nu}A_{\mu}^{\a}.
\label{S}
\end{equation}

For one first-order derivative of the vector field and two additional vector fields, we obtain
\begin{equation}
%\begin{array}{l}
\tilde{\mathcal{L}}_1= 2 \mathcal{L}_{1} = \epsilon_{\a \b
\c}\left[G^{\mu\nu \a} A^{\b}_\mu A^{\c}_\nu \right],
%\tilde{\mathcal{L}}_2= 2 \mathcal{L}_{2} = \epsilon_{\a\b\c}\left[
% \tilde{G}^{\mu\nu \a}  A^{\b}_\mu A^{\c}_\nu \right]  ,
%\end{array}
\end{equation}
and with four additional vector fields, we obtain
\begin{equation}
\left\{ \begin{array}{l}
\tilde{\mathcal{L}}_1=  2 \mathcal{L}_{1}
=\epsilon_{\a\b\c}\left[G^{\mu\nu \d} A^{\a}_\mu A^{\b}_\nu
\right]\left(A^{\c} \cdot A_{\d} \right),\\
\tilde{\mathcal{L}}_2= \mathcal{L}_{2} + \mathcal{L}_{3} = 
\epsilon_{\a\b\c}\left[S^{\mu\nu\a}A^{\b}_\mu A^{\d}_\nu
\right]\left(A^{\c} \cdot A_{\d} \right),\\
\tilde{\mathcal{L}}_3= \mathcal{L}_{3} - \mathcal{L}_{2} =
\epsilon_{\a\b\c}\left[ G^{\mu \nu \a} A^{\b}_{\mu} A^{\d}_{\nu}
\right]\left(A^{\c} \cdot A_{\d} \right) , \\
\tilde{\mathcal{L}}_4= \mathcal{L}_{4} =
\epsilon_{\a\b\c}\left[\tilde{G}^{\mu\nu\d} A^{\a}_\mu A^{\b}_\nu
\right]\left(A^{\c} \cdot A_{\d} \right),\\
\tilde{\mathcal{L}}_5= \mathcal{L}_{5} =
\epsilon_{\a\b\c}\left[\tilde{G}^{\mu\nu\a}  A^{\d}_{\mu} A^{\b}_{\nu}
\right]\left(A^{\c} \cdot A_{\d} \right),  \\
\tilde{\mathcal{L}}_6=\mathcal{L}_{6} - \mathcal{L}_{7} =
\epsilon_{\a\b\c}\left[\epsilon_{\mu \nu \rho \sigma} G^{\mu \alpha \d}
A_{\d}^\nu A^{\a\rho} A^{\b\sigma} A^{\c}_\alpha    \right].
\end{array} \right.
\end{equation}

Terms with two first-order derivatives and no additional fields can be written as
\begin{equation}
\mathcal{L}_1=2 \left( \mathcal{L}_{2} - \mathcal{L}_{3} \right) =
G^{\mu\nu}_{\a}G_{\mu\nu}^{\a},
\end{equation}
and with two additional fields, they are given by
\begin{equation}
\label{EqLagPostHessian}
\left\{
\begin{array}{l}
\tilde{\mathcal{L}}_1 =\mathcal{L}_{1} - \mathcal{L}_{5} = \delta^{\mu_1
\mu_2}_{\nu_1 \nu_2} A^{\lambda}_{\b} A^{\b}_{\lambda}
\left(\partial_{\mu_1} A^{\nu_1}_{\a} \right) \left(\partial_{\mu_2}
A^{\nu_2 \a} \right),\\
\tilde{\mathcal{L}}_2 =  2 \left(\mathcal{L}_{3} - \mathcal{L}_{5}
\right)=G^{\mu\nu}_{\a}G_{\mu\nu}^{\a} \left(A^{\b} \cdot A_{\b}
\right), \\
\tilde{\mathcal{L}}_3 = \mathcal{L}_{2} - \mathcal{L}_{6} =\delta^{\mu_1
\mu_2}_{\nu_1 \nu_2} A^{\lambda}_{\a} A_{\lambda\b}
\left(\partial_{\mu_1} A^{\nu_1\a} \right) \left(\partial_{\mu_2}
A^{\nu_2 \b} \right),\\
\tilde{\mathcal{L}}_4 = 2 \left(\mathcal{L}_{4} - \mathcal{L}_{6}
\right)=G^{\mu\nu}_{\a}G_{\mu\nu\b} \left(A^{\a} \cdot A^{\b} \right) \\
\tilde{\mathcal{L}}_5 = 2 \mathcal{L}_{7} =\tilde{G}_{\mu\nu}^{\a}
G^{\mu\nu}_{\a}\left(A^{\b} \cdot A_{\b} \right) ,\\
\tilde{\mathcal{L}}_6 =  2 \mathcal{L}_{8}  = \tilde{G}_{\mu\nu\a}
G^{\mu\nu}_{\b}\left(A^{\a} \cdot A^{\b} \right),\\
\tilde{\mathcal{L}}_{7} =\mathcal{L}_{18} + \mathcal{L}_{20} - 2
\mathcal{L}_{19} =   \left[ \epsilon_{\mu \nu \rho \sigma} A^{\a\mu}
A^{\b\nu}  G^{\rho \alpha}_{\a} G^\sigma_{\, \, \, \alpha \b}\right], \\
\tilde{\mathcal{L}}_{8} =  \mathcal{L}_{26} + \mathcal{L}_{23} =
\tilde{G}_{\mu\sigma}^{\b} A^{\mu}_{\a} A_{\alpha\b} S^{\alpha \sigma
\a} , \\
\tilde{\mathcal{L}}_{9} = \mathcal{L}_{26} - \mathcal{L}_{23} =
\tilde{G}_{\mu\sigma}^{\b}A^{\mu}_{\a} A_{\alpha\b}   G^{\alpha \sigma
\a} ,\\
\tilde{\mathcal{L}}_{10} = \mathcal{L}_{14} - \mathcal{L}_{27} =
\delta^{\mu_1 \mu_2}_{\nu_1\nu_2} A_{\mu_1}^{\a} A_{\mu_2}^{\b}
\left(\partial^{\mu_1}A^{\alpha}_{\a} \right) \left(\partial^{\mu_2}
A_{\alpha \b} \right),\\
\tilde{\mathcal{L}}_{11} = \mathcal{L}_{27} + \mathcal{L}_{28} - 2
\mathcal{L}_{17} = A^{\a}_\mu A^{\b}_\nu G^\mu_{\;\; \alpha \b}  G^{\nu
\alpha}_{\; \; \; \; \a} .
\end{array}
\right.
\end{equation}
As anticipated, we obtain 11
independent terms, which correspond to 28 terms to begin with, with 8
constraints and 9 Hessian conditions.

Finally, the three-gradient case yields
\begin{equation}
\left\{ \begin{array}{l}
\tilde{\mathcal{L}}_{1}= 2\left( \mathcal{L}_1 -3\mathcal{L}_2\right) = 
\epsilon_{\a\b\c} G^{\mu}{}_{\nu}{}^{\a} G^{\nu}{}_{\rho}{}^{\b}
G^{\rho}{}_{\mu}{}^{\c} ,\\
\tilde{\mathcal{L}}_{2} =  2 \mathcal{L}_4 -\mathcal{L}_3 -
\mathcal{L}_5 =   \epsilon_{\a\b\c} G^{\mu\alpha\a} G_{\alpha}{}^{\nu\b}
\tilde{G}_{\mu\nu}{}^{\c}.
\end{array} \right.
\end{equation}

\subsection{Scalar Contribution}
\label{PartScalarContribution}

Let us now consider the scalar part of the previously developed
Lagrangian, as explained in Sec.~\ref{PartProcedure}, making the
substitution $A_\mu^{\a} \rightarrow \partial_\mu\pi^{\a}$ and writing
only those terms that do not identically vanish, using the results of
the Appendix, where the useful Galileon
Lagrangians are provided (Sec.~\ref{PartAlternativeGalileonLag}), as
well as the linear combinations leading to second-order equations
(Sec.~\ref{PartResultLagGal}).

With one derivative and four vector fields, the only remaining term of
the scalar sector out of the original three is
\begin{equation}
\tilde{\mathcal{L}}_2 =  \epsilon_{\a\b\c} S^{\mu\nu\a}A^{\b}_\mu
A^{\d}_\nu \left(A^{\c} \cdot A_{\d} \right),
\label{limit}
\end{equation}
which does not yield second-order equations in the scalar limit. 

Lagrangians involving two derivatives of the vector fields provide
\begin{equation}
\left\{
\begin{array}{l}
\tilde{\mathcal{L}}_1 = \delta^{\mu_1 \mu_2}_{\nu_1 \nu_2}
A^{\lambda}_{\b} A^{\b}_{\lambda} \left(\partial_{\mu_1} A^{\nu_1}_{\a}
\right) \left(\partial_{\mu_2} A^{\nu_2 \a} \right),\\
\tilde{\mathcal{L}}_3 = \delta^{\mu_1 \mu_2}_{\nu_1 \nu_2}
A^{\lambda}_{\a} A_{\lambda\b} \left(\partial_{\mu_1} A^{\nu_1\a}
\right) \left(\partial_{\mu_2} A^{\nu_2 \b} \right),\\
\tilde{\mathcal{L}}_{10} = \delta^{\mu_1 \mu_2}_{\nu_1\nu_2}
A_{\mu_1}^{\a} A_{\mu_2}^{\b} \left(\partial^{\mu_1}A^{\alpha}_{\a}
\right) \left(\partial^{\mu_2} A_{\alpha \b} \right),
\end{array}
\right.
\end{equation}
leading to the corresponding scalar terms
\begin{equation}
\left\{
\begin{array}{l}
\left.\tilde{\mathcal{L}}_1\right|_{\pi}
=\mathcal{L}^{\text{Gal},3}_{4,\mathrm{I}},\\
\left.\tilde{\mathcal{L}}_3
\right|_{\pi}=\mathcal{L}^{\text{Gal},3}_{4,\mathrm{II}},\\
\left.\tilde{\mathcal{L}}_{10} \right|_{\pi} =
\mathcal{L}^{\text{Gal},2}_{4,\mathrm{II}} -
\mathcal{L}^{\text{Gal},2}_{4,\mathrm{III}}.
\end{array}
\right.
\end{equation}
One can derive two linear combinations having second-order equations,
namely,
\begin{equation}
 \left.\tilde{\mathcal{L}}_1\right|_{\pi} +2
 \left.\tilde{\mathcal{L}}_3\right|_{\pi} =
 \mathcal{L}^{\text{Gal},3}_{4,\mathrm{I}} +
 2\mathcal{L}^{\text{Gal},3}_{4,\mathrm{II}} \\ \end{equation}
 [see Eq.  \eqref{EqGal1}] and
 \begin{equation}
\left.\tilde{\mathcal{L}}_{10}\right|_{\pi} +
\left.\tilde{\mathcal{L}}_3\right|_{\pi}   =
\mathcal{L}^{\text{Gal},2}_{4,\mathrm{II}} - \frac12
\left(2\mathcal{L}^{\text{Gal},2}_{4,\mathrm{III}} +
\mathcal{L}^{\text{Gal},3}_{4,\mathrm{I}}  \right) + \frac12 \left(
\mathcal{L}^{\text{Gal},3}_{4,\mathrm{I}} +
2\mathcal{L}^{\text{Gal},3}_{4,\mathrm{II}}\right),
\label{L10pi}
\end{equation}
yielding second-order equations, as each of the three terms on the right-hand side
of Eq.  \eqref{L10pi} does so, as shown in the Appendix [see
Eqs.~\eqref{EqGal3}, \eqref{EqGal2} and \eqref{EqGal1}].

%\subsection{Note on the Faraday tensors}

\subsection{Final Flat Spacetime Model}
\label{PartSwitchFaradayYM}

Let us regroup the results of the above sections to produce the final
theory in flat spacetime with the Minkowskian metric. We first gather
most of the new terms induced by the nonlinear contributions into an
arbitrary function $f(A_{\mu}^{\a}, G_{\mu\nu}^\a,
\tilde{G}_{\mu\nu}^\a)$. Indeed, this is possible because they not only
appear in the systematic procedure we have exposed, but they also
satisfy all our conditions; this is equivalent to the general proof
discussed in Ref.~\cite{Allys:2016jaq}, where the typical term is built
out of Levi-Civita tensors, necessarily inducing terms proportional to
$\epsilon^{00\dots}$ in the Hessian, and hence vanishing contributions.

Up to now, we have used the Abelian form
of the Faraday tensor to express the
relevant Lagrangians, although there can be situations in which working
with the non-Abelian counterpart in Eq. (\ref{EqFaradayNonAbelian}) can be more
convenient, in particular, in view of the fact that this is the relevant
tensor that appears naturally when one extends the theory to its gauged
version. This is quite simple since the arbitrary function
$f(A_{\mu}^{\a}, G_{\mu\nu}^\a, \tilde{G}_{\mu\nu}^\a)$ can be
equivalently written as a new function $\tilde{f}(A_{\mu}^{\a},
F_{\mu\nu}^{\a}, \tilde{F}_{\mu\nu}^\a)$ using
Eq.~(\ref{EqFaradayNonAbelian}). It is worth noting that such a change
of variable implies no other terms than those already included in the
original function.

Gathering the above considerations into a compact form, we obtain
a first generic term, reminiscent of the Abelian case, namely,
\begin{equation}
\mathcal{L}_2 = f(A_{\mu}^{\a}, G_{\mu\nu}^{\a}, \tilde{G}_{\mu\nu} ^{\a}) =
\tilde{f}(A_{\mu}^{\a}, F_{\mu\nu}^{\a}, \tilde{F}_{\mu\nu} ^{\a}).
\end{equation}

In addition to this term, all the remaining previously derived terms involving
contractions with up to six Lorentz indices are
\begin{equation}
\left\{
\begin{array}{l}
\hat{\mathcal{L}}_1 = \delta^{\mu_1 \mu_2}_{\nu_1 \nu_2}
A^{\lambda}_{\b} A^{\b}_{\lambda} \left(\partial_{\mu_1} A^{\nu_1}_{\a}
\right) \left(\partial_{\mu_2} A^{\nu_2 \a} \right) + 2  \delta^{\mu_1
\mu_2}_{\nu_1 \nu_2} A^{\lambda}_{\a} A_{\lambda\b}
\left(\partial_{\mu_1} A^{\nu_1\a} \right) \left(\partial_{\mu_2}
A^{\nu_2 \b} \right) , \\
\hat{\mathcal{L}}_2 =  \delta^{\mu_1 \mu_2}_{\nu_1 \nu_2}
A^{\lambda}_{\a} A_{\lambda\b} \left(\partial_{\mu_1} A^{\nu_1\a}
\right) \left(\partial_{\mu_2} A^{\nu_2 \b} \right) + \delta^{\mu_1
\mu_2}_{\nu_1\nu_2} A_{\mu_1}^{\a} A_{\mu_2}^{\b}
\left(\partial^{\mu_1}A^{\alpha}_{\a} \right) \left(\partial^{\mu_2}
A_{\alpha \b} \right),\\
\hat{\mathcal{L}}_3 = \tilde{G}_{\mu\sigma}^{\b}A^\mu_{\a} A_{\alpha\b} 
 S^{\alpha\sigma \a},\\
\end{array}
\right.
\end{equation}
the first two actually being equivalent in the pure scalar sector since they
lead to the same equations of motion, i.e., those stemming from the
Galileon Lagrangian containing four scalar fields in the ${\bm 3}$
representation of SU(2). Note that there is no term containing only one
gradient.

With this general basis, which we expand upon in the final
discussion section, we can now turn to the covariantization required to
apply this category of theories to cosmologically relevant situations.

\section{Covariantization}  \label{covproc}
\subsection{Procedure}

Below we follow a procedure similar to that proposed for the Galileon
case~\cite{Deffayet:2009mn,Deffayet:2009wt,deRham:2011by}, the
generalized Proca model
\cite{Jimenez:2013qsa,Gleyzes:2014dya,Heisenberg:2014rta,Allys:2015sht},
and the
multi-Galileon situation~\cite{Deffayet:2010zh,Padilla:2012dx,Sivanesan:2013tba}.
The principle is simple: one first transforms all partial derivatives into
covariant ones and then checks that only those terms leading to at most second-order
equations of motion are kept.

The pure vector part now contains $A$ and $\nabla A$ terms, which
translate into $A$, $\partial A$, $g$ and $\partial g$ terms. None of
these terms could lead to any derivative of order higher than two in the
equations of motion. On the other hand, the Faraday tensor terms do not
yield metric derivatives since partial derivatives can be replaced by
covariant ones by virtue of the antisymmetry of these terms. We also
leave these terms aside.

As for the scalar part, derivatives of order three or more could appear
for the curvature. To fix this potential problem, we write the equations of
motion in terms of covariant derivatives and commute them in order to
generate the curvature tensor, which contains only second-order
derivatives of the metric: the problem is with the derivatives of
the curvature terms. As these particular contributions stem from terms
implying at least fourth-order derivatives of the scalar field, it is
easy to identify them and to write down the required counterterms.

In practice, this does not show that the resulting equations of motion of
the metric do not involve higher-order derivatives of the scalar field.
We merely apply the results of Ref.~\cite{Jimenez:2013qsa}, where it was
shown that if the equations of motion for the scalar field are safe,
then so are those for the metric. This result translates directly to
our case.

For many of the terms discussed below, it turns out to be
easier to write the Lagrangian as a function of the vector
field rather than of its scalar part, even though we are ultimately
interested in the latter. Indeed, the scalar Euler-Lagrange
equation
\begin{equation}
0=\frac{\partial\mathcal{L}}{\partial \pi_{\al}} - \nabla_\nu
\frac{\partial \mathcal{L}}{\partial (\nabla_\nu \pi_{\al})} +
\nabla_\nu \nabla_\mu \frac{\partial \mathcal{L}}{\partial (\nabla_\mu
\nabla_\nu \pi_{\al})} 
\end{equation}
can be written as
\begin{equation}
0=- \nabla_\nu \frac{\partial \mathcal{L}}{\partial (\nabla_\nu
\pi_{\al})} + \nabla_\nu \nabla_\mu \frac{\partial \mathcal{L}}{\partial
(\nabla_\mu \nabla_\nu \pi_{\al})} = - \nabla_\nu \left( \frac{\partial
\mathcal{L}}{\partial A_{\nu\al}} -  \nabla_\mu
\frac{\partial\mathcal{L}}{\partial (\nabla_\mu A_{\nu\al})} \right)
\end{equation}
since the action is assumed to be local in $A_\mu$ and therefore cannot
contain terms involving nonderivative functions of the scalar field
$\pi$.

In the following sections, we write those terms containing only the
curvature and its derivative, or only its derivative, by the respective
notation $\left.\mathcal{F}\right|_{R}$ or
$\left.\mathcal{F}\right|_{\nabla R}$, where $\mathcal{F}$ is the term
whose restriction is being considered. We concentrate on terms
which are nonvanishing in the scalar sector only.

\subsection{Terms in $\mathcal{L}^{\text{Gal}}$}

The Lagrangians we consider give, in the scalar sector, 
\begin{equation}
\left\{
\begin{array}{l}
\left.\hat{\mathcal{L}}_1\right|_{\pi}  = 
\mathcal{L}^{\text{Gal},3}_{4,\mathrm{I}} + 2
\mathcal{L}^{\text{Gal},3}_{4,\mathrm{II}} , \\
\left.\hat{\mathcal{L}}_2\right|_{\pi}  =
\mathcal{L}^{\text{Gal},3}_{4,\mathrm{II}} +
\mathcal{L}^{\text{Gal},2}_{4,\mathrm{II}}  -
\mathcal{L}^{\text{Gal},2}_{4,\mathrm{III}},
\end{array}
\right.
\label{EqLagFinGal}
\end{equation}
where we use the Galileon Lagrangians of the Appendix. In the following, working in the vector sector,
we substitute $\partial_\mu\pi^\a\to A_\mu^\a$.
Equation~\eqref{EqLagFinGal} implies that only three independent
counterterms are needed, i.e., those associated with
$\mathcal{L}^{\text{Gal},3}_{4,\mathrm{I}}$,
$\mathcal{L}^{\text{Gal},3}_{4,\mathrm{II}}$ and
$(\mathcal{L}^{\text{Gal},2}_{4,\mathrm{II}}  -
\mathcal{L}^{\text{Gal},2}_{4,\mathrm{III}})$. We now proceed to find
these counterterms.

First, we have
\begin{equation}
\left.\left\{ \nabla_{\nu} \nabla_{\mu} \left[\frac{\partial
\mathcal{L}^{\text{Gal},3}_{4,\mathrm{I}}}{\partial \left( \nabla_{\mu}
A_{\nu\al} \right)}\right]\right\}\right\vert_{R} = - 2 A^{\lambda}_{\b}
A^{\b}_{\lambda}  R_{\mu\nu} \nabla^{\nu}A^{\mu\al} - 2 A_{\b}^{\lambda}
A^{\b}_{\lambda} A^{\mu\al} \nabla^{\nu}R_{\mu\nu}.
\end{equation}
Introducing
\begin{equation}
\mathcal{L}^{\text{Gal},3}_{4,\mathrm{I},\mathrm{CT}} = \frac{1}{4}
A^{\lambda}_{\b} A ^{\b}_{\lambda} A^{\mu}_{\a} A^{\a}_{\mu} R,
\end{equation}
we find that
\begin{equation}
\left.\left\{ \nabla_{\nu} \left[\frac{\partial
\mathcal{L}^{\text{Gal},3}_{4,\mathrm{I},\mathrm{CT}}}{\partial \left(
A_{\nu\al} \right)} \right] \right\}\right|_{\nabla R}= 
A_{\b}^{\lambda}A^{\b}_{\lambda} A^{\mu\al} \nabla^{\nu}\left(g_{\mu\nu}
R \right),
\end{equation}
which finally implies the equation of motion (EOM)
\begin{equation}
EOM_\pi\left.\left(\mathcal{L}^{\text{Gal},3}_{4,\mathrm{I}} +
\mathcal{L}^{\text{Gal},3}_{4,\mathrm{I},CT}  \right)\right|_{\nabla R}
=- 2A_{\b}^{\lambda}A^{\b}_{\lambda} A^{\mu\alpha}
\nabla^{\nu}\left(R_{\mu\nu} - \frac{1}{2}g_{\mu\nu} R \right) = 0,
\end{equation}
vanishing by virtue of the properties of the Einstein tensor.

Similarly, for $\mathcal{L}^{\text{Gal},3}_{4,\mathrm{II}}$, we have
\begin{equation}
\left.\left\{\nabla_{\nu} \nabla_{\mu} \left[\frac{\partial
\mathcal{L}^{\text{Gal},3}_{4,\mathrm{II}}}{\partial \left( \nabla_{\mu}
A_{\nu\al} \right)} \right] \right\}\right|_{R} = - 2 A^{\lambda}_{\b}
A^{\al}_{\lambda}  R_{\mu\nu}\nabla^{\nu}A^{\mu\b} - 2 A_{\b}^{\lambda}
A^{\al}_{\lambda} A^{\mu\b} \nabla^{\nu}R_{\mu\nu}.
\end{equation}
Introducing
\begin{equation}
\mathcal{L}^{\text{Gal},3}_{4,\mathrm{II},\mathrm{CT}} = \frac{1}{4}
A^{\lambda}_{\b} A _{\lambda\a} A^{\mu\b} A^{\a}_{\mu} R,
\end{equation}
which verifies
\begin{equation}
\left.\left\{ \nabla_{\nu} \left[\frac{\partial
\mathcal{L}^{\text{Gal},3}_{4,\mathrm{II},\mathrm{CT}}}{\partial \left( 
A_{\nu\al} \right)} \right] \right\}\right|_{\nabla R}  = 
A_{\b}^{\lambda}A^{\al}_{\lambda} A^{\mu\b} \nabla^{\nu}\left(g_{\mu\nu}
R \right),
\end{equation}
we obtain
\begin{equation}
EOM_\pi\left.\left(\mathcal{L}^{\text{Gal},3}_{4,\mathrm{II}} +
\mathcal{L}^{\text{Gal},3}_{4,\mathrm{II},CT}  \right)\right|_{\nabla
R}=- 2A_{\b}^{\lambda}A^{\al}_{\lambda} A^{\mu\b}
\nabla^{\nu}\left(R_{\mu\nu} - \frac{1}{2}g_{\mu\nu} R \right) = 0.
\end{equation}

Finally, using the previous notation
\begin{equation}
\tilde{\mathcal{L}}_{10} =\mathcal{L}^{\text{Gal},2}_{4,\mathrm{II}} -
\mathcal{L}^{\text{Gal},2}_{4,\mathrm{III}},
\end{equation}
we have
\begin{equation}
\left.\left\{\nabla_{\nu} \nabla_{\mu} \left[\frac{\partial
\tilde{\mathcal{L}}_{10} }{\partial \left( \nabla_{\mu} A_{\nu\al}
\right)} \right] \right\}\right|_{R} =-2 A^{\mu\al}A^{\lambda\b}
R^{\nu}{}_{\rho\lambda\mu} \nabla_\nu A^{\rho}_{\b}-2
A^{\mu\al}A^{\lambda\b}A^{\rho}_{\b}\nabla_\nu R^{\nu}{}_{\rho\lambda\mu}.
\end{equation}
We introduce the counterterm
\begin{equation}
\mathcal{L}_{10,\mathrm{CT}} = -\frac12 A^{\mu\a} A^{\nu \b}
A^{\rho}_{\a} A^{\sigma}_{\b} R_{\mu\nu\rho \sigma},
\end{equation}
giving
\begin{equation}
\nabla_{\rho}\left. \left( \frac{\partial
\mathcal{L}_{10,\mathrm{CT}}}{\partial \left(A_{\rho \al} \right)}
\right)\right|_{\nabla R} = -2 A^{\mu\a} A_{\lambda \a} A^{\rho \al}
\nabla^{\nu} R^{\lambda}{}_{\rho \mu \nu} = 2 A^{\mu\al}
A^{\lambda\b}A^{\rho}_{\b} \nabla_\nu R^{\nu}{}_{\rho\lambda\mu},
\end{equation}
which, as expected, results in
\begin{equation}
EOM_\pi\left.\left(\tilde{\mathcal{L}}_{10}  +
\mathcal{L}_{10,\mathrm{CT}} \right)\right|_{\nabla R}= 0.
\end{equation}

Then, to obtain the covariantized form of the action, it is sufficient
to add the counterterms obtained in this part to the action given
previously in flat spacetime. The result is summarized in
Sec.~\ref{FinalModel}.

\subsection{Coupling with Curvature}

Once the derivatives have been covariantized, one must also include
possible direct coupling terms between the vector field and the
curvature tensors, which we do below in a way entirely similar to that
of Ref.~\cite{Allys:2015sht}. First, we demand contractions with tensors
whose divergences vanish on all indices (to ensure that integration by parts provides no
higher-order contributions in the equations of motion)
\cite{deRham:2011by,Jimenez:2013qsa}: this means the Einstein tensor as
well as
\begin{equation}
L_{\mu\nu\rho\sigma}=2R_{\mu\nu\rho\sigma} + 2(R_{\mu\sigma}g_{\rho\nu}
+ R_{\rho\nu}g_{\mu\sigma} - R_{\mu\rho}g_{\nu\sigma} -
R_{\nu\sigma}g_{\mu\rho}) + R(g_{\mu\rho}g_{\nu\sigma} -
g_{\mu\sigma}g_{\rho\nu}),
\end{equation}
whose symmetries are those of the Riemann tensor, to which it is dual in
the sense that it can be written as
\begin{equation}
L^{\alpha\beta\gamma\delta}= - \frac12
\epsilon^{\alpha\beta\mu\nu}\epsilon^{\gamma\delta
\rho\sigma}R_{\mu\nu\rho\sigma}.
\end{equation}

Even limiting ourselves to the same number of fields as in the flat
spacetime situation, many terms are a priori possible. To begin with,
all contractions involving a single vector field are impossible. With two
such fields, the reasoning is exactly equivalent to the Abelian case,
which means the Lagrangians
\begin{equation}
\mathcal{L}^\mathrm{curv}_{1} = G_{\mu\nu}A^{\mu\a}A^{\nu}_{\a}
\end{equation}
and
\begin{equation}
\mathcal{L}^\mathrm{curv}_{2} = L_{\mu\nu\rho\sigma} G^{\mu\nu\a}
G^{\rho\sigma}_{\a}
\end{equation}
are acceptable.

Terms in which at least one of the Abelian-like Faraday tensors is replaced by its Hodge
dual can always be rewritten as a contraction between the Riemann tensor
and two Abelian-like Faraday tensors, which cannot give second-order equations of
motion~\cite{Jimenez:2013qsa}. One could envisage a contraction with
a term like $G^{\mu\rho\a} G^{\nu\sigma}_{\a}$, but which is proportional
to $\mathcal{L}^\mathrm{curv}_{2}$: to show this, one needs to use the
following identity,
\begin{equation}
%\epsilon^{abcd}\epsilon^{efgh}-\epsilon^{aefg}\epsilon^{bcdh} +
%\epsilon^{acdh} \epsilon^{befg}+\epsilon^{abdh}\epsilon^{ecfg} -
%\epsilon^{abch}\epsilon^{edfg}=0,
\epsilon^{\alpha\beta\gamma\delta} \epsilon^{\rho\sigma\mu\nu} -
\epsilon^{\alpha\rho\sigma\mu} \epsilon^{\beta\gamma\delta\nu} +
\epsilon^{\alpha\gamma\delta\nu} \epsilon^{\beta\rho\sigma\mu} +
\epsilon^{\alpha\beta\delta\nu} \epsilon^{\rho\gamma\sigma\mu} -
\epsilon^{\alpha\beta\gamma\nu} \epsilon^{\rho\delta\sigma\mu}=0,
\end{equation}
and the first Bianchi identity.

With three fields, one can obtain a new nonvanishing term, in
contrast to the Abelian case. This is mostly due to the fact
that it is possible to have an antisymmetry in the exchange of two
underived vector fields. We get
\begin{equation}
\mathcal{L}^\mathrm{curv}_{3} = L_{\mu\nu\rho\sigma} \epsilon_{\a\b\c}
G^{\mu \nu \a} A^{\rho\b} A^{\sigma\c},
\end{equation}
which is shown to be proportional to $L_{\mu\nu\rho\sigma}
\epsilon_{\a\b\c} G^{\mu\rho\a} A^{\nu\b} A^{\sigma\c}$, by making use
of the previous identity on the Levi-Civita tensor.

Four fields provide, again in contrast to the Abelian situation,
the extra contribution
\begin{equation}
\mathcal{L}^\mathrm{curv}_{4} = L_{\mu\nu\rho\sigma}  A^{\mu\a} A^{\nu
\b} A^{\rho}_{\a}A^{\sigma}_{\b}.
\end{equation}
It is worth noticing at this point that it is possible to go from the
expression of $\mathcal{L}^\mathrm{curv}_{2}$ and
$\mathcal{L}^\mathrm{curv}_{3}$ using $G_{\mu\nu}^\a$ (the Abelian form of the
Faraday tensor) to that using $F_{\mu\nu}^\a$ (the non-Abelian one),
both of which are equal in an Abelian theory: it is sufficient for
this purpose to include the terms $\mathcal{L}^\mathrm{curv}_{3}$ and
$\mathcal{L}^\mathrm{curv}_{4}$ only (they are generated by the
transformation from $G_{\mu\nu}^\a$ to $F_{\mu\nu}^\a$).

\section{Final model, discussion}  \label{FinalModel}

Let us summarize the results obtained for the generalized SU(2) Proca
theory. First, we showed that any function of the vector field, Faraday
tensor, and its Hodge dual (either in their Abelian or non-Abelian
formulation) was possible, i.e.,
\begin{equation}
\mathcal{L}_2 = f(A_{\mu}^{\a}, G_{\mu\nu}^\a, \tilde{G}_{\mu\nu} ^\a) =
\tilde{f}(A_{\mu}^{\a}, F_{\mu\nu}^\a, \tilde{F}_{\mu\nu} ^\a).
\end{equation}
Such a  general $\mathcal{L}_2$ term involving only gauge-invariant
quantities for the derivatives is also present in the Abelian case; we
will not discuss it any further since it appears similarly (and for the
same reasons) in both the Abelian and non-Abelian theories.

Before presenting the other terms contained in the non-Abelian action,
let us pursue the summary of what was found for its Abelian
counterpart, as worked out in
Refs.~\cite{Heisenberg:2014rta,Allys:2015sht,Jimenez:2016isa,Allys:2016jaq}; 
as usual, we denote $\mathcal{L}_{n+2}$ the Lagrangians containing $n\geq
1$ first-order derivatives of the vector field.
First, the relation between the more general scalar and vector theories, i.e.,
the Galileon and generalized Proca models, provide, in this case, a deeper
understanding through the use of the Stückelberg trick to go from one sector to
another (i.e., switching between $\partial_\mu \pi$ and $A_\mu$). In the scalar
Galileon theory, only one term exists in the Lagrangians $\mathcal{L}_3$ to
$\mathcal{L}_5$, each of which generates a contribution to the vector
sector by the Stückelberg trick, i.e., those with a prefactor $f_i(X)$ in the
conclusion of Ref.~\cite{Allys:2016jaq}. An additional freedom
stems from the fact that a given scalar Lagrangian can give different
vector Lagrangians when permuting the second-order derivatives before
introducing the vector field: although $\partial_\mu\partial_\nu \pi =
\partial_\nu\partial_\mu \pi$, this symmetry is absent in the pure
vector case since $\partial_\mu A_\nu \not= \partial_\nu A_\mu$. This
property led to one additional contribution to the vector sector of each
$\mathcal{L}_4$ to $\mathcal{L}_6$. These contributions appear with the 
prefactor $g_i(X)$ in Ref.~\cite{Allys:2016jaq}; they vanish in the
pure scalar sector.

Coming back to the non-Abelian situation, and in addition to
$\mathcal{L}_2$, we derived those relevant Lagrangians implying up to 6
contracted Lorentz indices and being nontrivial in flat spacetime.
Contrary to the Abelian case, we found no such Lagrangian for $n=1$. For
$n=2$, there are three possible terms; i.e., $\mathcal{L}_4$ contains
\begin{equation}
\left\{
\begin{array}{l}
\mathcal{L}_4^1 = \delta^{\mu_1 \mu_2}_{\nu_1 \nu_2} A^{\lambda}_{\b}
A^{\b}_{\lambda} \left(\nabla_{\mu_1} A^{\nu_1}_{\a} \right)
\left(\nabla_{\mu_2} A^{\nu_2 \a} \right) + \frac14 A^{\lambda}_{\b} A
^{\b}_{\lambda} A^{\mu}_{\a} A^{\a}_{\mu} R \\
~~~~~~~ + 2 \delta^{\mu_1 \mu_2}_{\nu_1 \nu_2} A^{\lambda}_{\a} A_{\lambda\b} \left(\nabla_{\mu_1} A^{\nu_1\a} \right) \left(\nabla_{\mu_2} A^{\nu_2 \b} \right) + \frac12 A^{\lambda}_{\b} A _{\lambda\a} A^{\mu\b} A^{\a}_{\mu} R , \\
\mathcal{L}_4^2 = \delta^{\mu_1 \mu_2}_{\nu_1 \nu_2} A^{\lambda}_{\a} A_{\lambda\b} \left(\nabla_{\mu_1} A^{\nu_1\a} \right) \left(\nabla_{\mu_2} A^{\nu_2 \b} \right) + \frac{1}{4} A^{\lambda}_{\b} A _{\lambda\a} A^{\mu\b} A^{\a}_{\mu} R \\
~~~~~~~ + \delta^{\mu_1 \mu_2}_{\nu_1\nu_2} A_{\mu_1}^{\a} A_{\mu_2}^{\b} \left(\nabla^{\nu_1}A^{\alpha}_{\a} \right) \left(\nabla^{\nu_2} A_{\alpha \b} \right) -\frac12 A^{\mu\a} A^{\nu \b} A^{\rho}_{\a} A^{\sigma}_{\b} R_{\mu\nu\rho \sigma},\\
\mathcal{L}_4^3 = \tilde{G}_{\mu\sigma}^{\b}A^\mu_{\a} A_{\alpha\b}  S^{\alpha\sigma \a},\\
\end{array}
\right.
\end{equation}
the first two terms giving, once developed, the following forms:
\begin{equation}
\left\{
\begin{array}{l}
\mathcal{L}_4^1 = (A_{\b} \cdot A^{\b}) \left[ \left(\nabla\cdot A_\a \right)\left(\nabla\cdot A^\a \right) - (\nabla_\mu A^\nu_{\a})(\nabla^\mu A_\nu^{\a}) +\frac14
A_{\a} \cdot A^{\a} R \right] \\ 
\hspace{1cm} + 2 (A_{\a} \cdot A_{\b}) \left[\left(\nabla\cdot A^\a \right)\left(\nabla\cdot A^\b \right) - (\nabla_\mu A^{\nu \a})(\nabla^\mu A_\nu^{\b}) +\frac12 A^{\a}
\cdot A^{\b} R\right], \\ 
\mathcal{L}_4^2 = (A_{\a} \cdot A_{\b}) \left[\left(\nabla\cdot A^\a \right)\left(\nabla\cdot A^\b \right) - (\nabla_\mu A^{\nu \a})(\nabla^\mu A_\nu^{\b}) +\frac14 A^{\a}
\cdot A^{\b} R\right] \\
\hspace{1cm} + (A^{\mu \a} A^{\nu\b}) \left[\left(\nabla_\mu A^\alpha_\a \right)\left(\nabla_\nu A_{\alpha\b} \right) - \left(\nabla_\nu A^\alpha_\a \right)\left(\nabla_\mu A_{\alpha\b} \right) 
-\frac12 A^\rho_{\b} A^{\sigma \b} 
R_{\mu\nu\rho\sigma}\right],
\end{array}
\right.
\end{equation}
which are more easily compared with the equivalent results for the Abelian case.
Finally, we also found four extra possibilities for the Lagrangians,
implying a coupling with the curvature
\begin{equation}
\begin{array}{l}
\mathcal{L}^\mathrm{curv}_{1} = G_{\mu\nu}A^{\mu\a}A^{\nu}_{\a},\\
\mathcal{L}^\mathrm{curv}_{2} = L_{\mu\nu\rho\sigma} F^{\mu\nu}_{\a}
F_{\mu\nu}^{\a},\\
\mathcal{L}^\mathrm{curv}_{3} = L_{\mu\nu\rho\sigma} \epsilon_{\a\b\c}
F^{\mu \nu \a} A^{\rho\b} A^{\sigma\c},\\
\mathcal{L}^\mathrm{curv}_{4} = L_{\mu\nu\rho\sigma} A^{\mu\a} A^{\nu
\b} A^{\rho}_{\a}A^{\sigma}_{\b},
\end{array}
\end{equation}
thereby completing the full action at that order.

Let us first consider the actions whose equations of motion involve only
second-order derivatives for the scalar (not first-order ones), which is
equivalent to having only two vector fields together with the relevant 
gradients in the action. The
multi-Galileon SU(2) model in the adjoint representation has been
considered in \cite{Padilla:2010ir}, where it was shown that building a
Lagrangian is only possible at the order of $\mathcal{L}_4$ (not to
mention the order $\mathcal{L}_2$ already discussed above). The
equivalent formulations of this Lagrangian are detailed in
Appendix~\ref{AppendixGalileon}. Following the previous considerations,
no Lagrangian in the vector sector should appear at the order of
$\mathcal{L}_3$ since there is no such associated Lagrangian for the
multi-Galileon at that order; we explicitly confirmed this expectation.
In addition, two Lagrangians should appear at the order of
$\mathcal{L}_4$, one associated with the multi-Galileon dynamics and one
associated with the commutation of second-order derivatives of the scalar
field. In fact, three Lagrangians have been found, two of them giving
the multi-Galileon dynamics in the scalar sector. We then interpret
these two previous terms as contributions which are equivalent in the
scalar case but not in the vector case. The fact that there are two nonvanishing
Lagrangians in the scalar sector is also due to
a commutation of the second-order derivatives of the scalar fields but
in a current term, which implies that it is not possible to describe
this commutation with a Lagrangian vanishing in the pure scalar
sector. This additional term is specific to the non-Abelian case:
the term in $\delta^{\mu_1 \mu_2}_{\nu_1\nu_2} A_{\mu_1}^{\a}
A_{\mu_2}^{\b} \left(\nabla^{\nu_1}A^{\alpha}_{\a} \right)
\left(\nabla^{\nu_2} A_{\alpha \b} \right)$ vanishes in the Abelian
case, while $\mathcal{L}_4^1$ and $\mathcal{L}_4^2$ both reduce to
$\mathcal{L}_4^{\text{Abelian}} = \delta^{\mu_1 \mu_2}_{\nu_1 \nu_2}
A^{\lambda}A_{\lambda} \left(\nabla_{\mu_1} A^{\nu_1}\right)
\left(\nabla_{\mu_2} A^{\nu_2} \right)$.

To go further, let us first consider terms implying more derivatives,
i.e., having $n\geq 3$. At the order of $\mathcal{L}_5$, and since there
is no possible dynamics for the SU(2) adjoint multi-Galileon, we expect that
no term having a nonvanishing pure scalar contribution is possible.
This suggests that the only possible term is
\begin{equation}
\mathcal{L}_5 = \epsilon_{\a\b\c} \left(A^\a \cdot A^\d\right)
\tilde{G}^{\alpha\mu}_\d\tilde{G}^{\beta}{}_\mu^\b S_{\alpha\beta}^\c,
\end{equation}
with the other SU(2) index contractions giving a vanishing result. At the
order of $\mathcal{L}_6$, the only possibility seems to be the
independent possible contractions of SU(2) indices on
$\mathcal{L}_6^{\text{Abelian}}=\left(A\cdot A
\right)\tilde{G}^{\alpha\beta}\tilde{G}^{\mu\nu}
S_{\alpha\mu}S_{\beta\nu}$, since there is no possibility of having a term that
does not vanish in the pure scalar sector. However, one
should verify that there is no other term vanishing in the pure scalar
sector, not included in $\mathcal{L}_2$, and whose dynamics is not
described by the previous ones. This kind of terms would be
specific to a non-Abelian theory, as is the second term of $\mathcal{L}_4^2$, 
and they would vanish for a vector field in a trivial group representation.

Concerning the Lagrangians with more than two vector fields together
with the relevant gradients, one has to pay attention to the fact that
fully factorizing an $f\left(A_\mu^\a\right)$ as in the Abelian case
is not guaranteed to lead to a valid procedure, although factorizing
such an arbitrary function in front of any valid contribution also leads
to another valid contribution. In addition, one could think that if there
is no valid Lagrangian with only a few nongradient vector fields at
a given derivative order, it is fairly probable that there is also
no such valid Lagrangian at all at this order. For instance, we showed
explicitly that terms at the order of $\mathcal{L}_3$ are not possible
with up to $4$ vector fields, and this questions the possibility of having
such a term even with a higher number of vector fields. An interesting
point is that if a Lagrangian is allowed which does not vanish in the
pure scalar sector, it corresponds to a possible term in the
multi-Galileon action, which shows that both theories are closely
related.

To conclude, this discussion showed that even if the full action of the
model has not been obtained yet, discussing the low order terms permits us
to identify and understand the whole Lagrangian structure. The above
discussion is not specific to the SU(2) case and therefore can be
extended to other group representations. For a theory with a vector
field transforming under any representation of any group, a systematic
study of all possible terms in the action should be performed in
parallel with the corresponding multi-Galileon theory.

\section*{Acknowledgments}  This work was supported by COLCIENCIAS - ECOS NORD grant number RC 0899-2012 with the help of ICETEX, and by
COLCIENCIAS grant numbers 110656933958 RC 0384-2013 and 123365843539 RC
FP44842-081-2014. P.P. would like to thank the Labex Institut Lagrange
de Paris (reference ANR-10-LABX-63) part of the Idex SUPER, within which part of
this work has been completed.

\appendix
\section{SU(2) Galileon Lagrangian Equivalent Formulations}
\label{AppendixGalileon}
\subsection{Introduction}

The purpose of this appendix is to write explicitly all the Lagrangians
describing the multi-Galileon dynamics in the 3-dimensional
representation of SU(2), focusing on the Lagrangians containing only four
Galileon fields, i.e., those which are useful in this article. A
Lagrangian describing this dynamics is given in Ref. \cite{Padilla:2012dx},
namely,
\begin{equation}
\label{EqLagIniPSZ}
\mathcal{L}^{\pi}_{m} = \alpha^{i_1 \cdots i_m} \delta^{\mu_2 \cdots
\mu_m}_{\left[ \nu_2\cdots \nu_m\right]} \pi_{i_1}
\partial_{\mu_2}\partial^{\nu_2}\pi_{i_2}\cdots
\partial_{\mu_m}\partial^{\nu_m}\pi_{i_m},
\end{equation}
with $m$ running from 1 to 5, and with the notation 
\begin{equation}
\frac{1}{(D-n)!}\epsilon^{i_1\cdots i_n \sigma_1 \cdots \sigma_{D-n}}
\epsilon_{j_1\cdots j_n \sigma_1 \cdots \sigma_{D-n}} =  n! \delta^{j_1
\cdots j_n}_{[i_1 \cdots i_n]} = \delta^{j_1 \cdots j_n}_{i_1 \cdots
i_n} = \delta^{j_1}_{i_1} \cdots \delta^{j_n}_{i_n} \pm \cdots,
\label{deltaMult}
\end{equation}
for $n$ running from 1 to 4 (in a four-dimensional spacetime). Other
equivalent formulations are possible, which is the purpose of
this appendix.

This investigation is necessary for two reasons. First, the
formulation given in Eq. \eqref{EqLagIniPSZ} cannot be obtained from a
vector Lagrangian using the switch $A_{\mu}^{\a} \rightarrow
\partial_{\mu} \pi^{\a}$ since a scalar field without derivatives is
present. Second, if different Lagrangians are equivalent in the scalar
sector, they could give Lagrangians that are not equivalent in the vector sector.
We thus expect that different Lagrangians valid in the vector sector
become different but equivalent formulations of the multi-Galileon
dynamics when considering the pure scalar part of the action.

For this purpose, we use the results of Ref.~\cite{Deffayet:2011gz},
which describe equivalent formulations of the Galileon theory in the Abelian
case, introducing a Lagrangian similar to that in Eq.
\eqref{EqLagIniPSZ}, together with the following Lagrangians:
\begin{equation}
\label{EqGalAbelien1}
\mathcal{L}^{\text{Gal},1}_{m}= \delta^{\mu_1 \cdots \mu_{m-1}}_{\left[
\nu_1\cdots \nu_{m-1}\right]} \partial_{\mu_{1}}\pi
\partial^{\nu_{1}}\pi \partial_{\mu_2}\partial^{\nu_2} \pi \cdots
\partial_{\mu_{m-1}}\partial^{\nu_{m-1}} \pi,
\end{equation}
\begin{equation}
\mathcal{L}^{\text{Gal},2}_{m}= \delta^{\mu_1 \cdots \mu_{m-2}}_{\left[
\nu_1\cdots \nu_{m-2}\right]} \partial_{\mu_{1}}\pi
\partial_{\lambda}\pi \partial^{\nu_1}\partial^{\lambda} \pi \cdots
\partial_{\mu_{m-2}}\partial^{\nu_{m-2}} \pi,
\end{equation}
\begin{equation}
\label{EqGalAbelien3}
\mathcal{L}^{\text{Gal},3}_{m}= \delta^{\mu_1 \cdots \mu_{m-2}}_{\left[
\nu_1\cdots \nu_{m-2}\right]} \partial_{\lambda}\pi
\partial^{\lambda}\pi \partial_{\mu_1}\partial^{\nu_1} \pi \cdots
\partial_{\mu_{m-2}}\partial^{\nu_{m-2}} \pi,
\end{equation}
for $m\geq2$, the case $m=1$ giving $\mathcal{L}=\pi$. These Lagrangians all give
second-order equations of motion.

\subsection{Lagrangians}
\label{PartAlternativeGalileonLag}

We first write all possible Lagrangians appearing when we add the
group indices to the previous Lagrangians, restricting ourselves to the
case $m=4$. They are more numerous than in the multi-Galileon case since we
have an additional freedom when choosing the group index contractions.

The only possible Lagrangian associated with the formulation of
Ref.~\cite{Padilla:2012dx} is
\begin{equation}
\mathcal{L}^{\text{PSZ}}_{4} = \delta^{\mu_1 \cdots \mu_3}_{\nu_1 \cdots
\nu_3} \pi_{\a} \partial_{\mu_1} \partial^{\nu_1} \pi^{\a}
\partial_{\mu_2} \partial^{\nu_2} \pi_{\b} \partial_{\mu_3}
\partial^{\nu_3} \pi^{\b} .
\end{equation}
The Lagrangians appearing in Ref.~\cite{Deffayet:2011gz}, given in
Eqs.~\eqref{EqGalAbelien1} to~\eqref{EqGalAbelien3}, can be endowed with SU(2)
indices in several ways, namely, two possibilities for
$\mathcal{L}^{\mathrm{Gal},1}_{4}$:
\begin{equation}
\mathcal{L}^{\text{Gal},1}_{4,\mathrm{I}} = \delta^{\mu_1 \cdots
\mu_3}_{\nu_1 \cdots \nu_3} \partial _{\mu_1}\pi_{\a} \partial^{\nu_1}
\pi^{\a} \partial_{\mu_2} \partial^{\nu_2} \pi_{\b} \partial_{\mu_3}
\partial^{\nu_3} \pi^{\b}
\end{equation}
and
\begin{equation}
\mathcal{L}^{\text{Gal},1}_{4,\mathrm{II}} = \delta^{\mu_1 \cdots
\mu_3}_{\nu_1 \cdots \nu_3} \partial _{\mu_1}\pi_{\a} \partial^{\nu_1}
\pi_{\b} \partial_{\mu_2} \partial^{\nu_2} \pi^{\a} \partial_{\mu_3}
\partial^{\nu_3} \pi^{\b};
\end{equation}
three possibilities for $\mathcal{L}^{\text{Gal},2}_{4}$:
\begin{equation}
\mathcal{L}^{\text{Gal},2}_{4,\mathrm{I}} = \delta^{\mu_1 \mu_2}_{\nu_1
\nu_2}  \partial_{\mu_1} \pi_{\a} \partial_{\lambda} \pi^{\a}
\partial^{\lambda}\partial^{\nu_1} \pi_{\b} \partial_{\mu_2}
\partial^{\nu_2} \pi^{\b},
\end{equation}
\begin{equation}
\mathcal{L}^{\text{Gal},2}_{4,\mathrm{II}} = \delta^{\mu_1 \mu_2}_{\nu_1
\nu_2}  \partial_{\mu_1} \pi_{\a} \partial_{\lambda} \pi_{\b}
\partial^{\lambda}\partial^{\nu_1} \pi^{\a} \partial_{\mu_2}
\partial^{\nu_2} \pi^{\b},
\end{equation}
and
\begin{equation}
\mathcal{L}^{\text{Gal},2}_{4,\mathrm{III}} = \delta^{\mu_1
\mu_2}_{\nu_1 \nu_2}  \partial_{\mu_1} \pi_{\a} \partial_{\lambda}
\pi_{\b} \partial^{\lambda}\partial^{\nu_1} \pi^{\b} \partial_{\mu_2}
\partial^{\nu_2} \pi^{\a};
\end{equation}
and finally two possibilities for $\mathcal{L}^{\text{Gal},3}_{4}$:
\begin{equation}
\mathcal{L}^{\text{Gal},3}_{4,\mathrm{I}} = \partial_{\lambda} \pi_{\a}
\partial^{\lambda} \pi^{\a} \delta ^{\mu_1 \mu_2} _{\nu_1 \nu_2}
\partial_{\mu_1} \partial^{\nu_1} \pi_{\b} \partial_{\mu_2} \partial
^{\nu_2} \pi^{\b}
\end{equation}
and
\begin{equation}
\mathcal{L}^{\text{Gal},3}_{4,\mathrm{II}} = \partial_{\lambda} \pi_{\a}
\partial^{\lambda} \pi_{\b} \delta ^{\mu_1 \mu_2} _{\nu_1 \nu_2}
\partial_{\mu_1} \partial^{\nu_1} \pi^{\a} \partial_{\mu_2} \partial
^{\nu_2} \pi^{\b}.
\end{equation}

Looking for the Lagrangians implying second-order equations of motion,
one can quickly verify that $\mathcal{L}^{\text{PSZ}}_{4}$,
$\mathcal{L}^{\text{Gal},1}_{4,\mathrm{I}}$ and
$\mathcal{L}^{\text{Gal},1}_{4,\mathrm{II}}$ have this property due to
the symmetry properties of $\delta^{\mu_1 \cdots \mu_3}_{\nu_1 \cdots
\nu_3}$. However, the other Lagrangians do not give a priori second-order 
equations of motion\footnote{The automatic cancellation between
third-order derivatives discussed in Ref.~\cite{Deffayet:2011gz} is not
valid anymore since this cancellation can be spoiled by the group
indices.}. We then investigate, in the following, the relations among
the different Lagrangians.

\subsection{Relations among the Lagrangians}
\subsubsection{Between PSZ and Gal,1}

We first relate $\mathcal{L}^{\text{PSZ}}_{4}$ and the Lagrangians
$\mathcal{L}^{\text{Gal},1}_{4}$ by means of conserved currents. Indeed,
\begin{equation}
J^{\mu_1}_{0,\mathrm{I}}=J^{\text{PSZ-Gal},\mu_1} _{4,\mathrm{I}} =
\delta^{\mu_1 \cdots \mu_3}_{\nu_1 \cdots \nu_3} \pi_{\a}
\partial^{\nu_1} \pi^{\a} \partial_{\mu_2} \partial^{\nu_2} \pi_{\b}
\partial_{\mu_3} \partial^{\nu_3} \pi^{\b} 
\end{equation}
gives
\begin{equation}
\partial_{\mu_1} J^{\mu_1}_{0,\mathrm{I}} =\partial_{\mu_1}
J^{\text{PSZ-Gal},\mu_1} _{4,\mathrm{I}} = \mathcal{L}^{\text{PSZ}}_{4} 
+ \mathcal{L}^{\text{Gal},1}_{4,\mathrm{I}},
\end{equation}
and
\begin{equation}
J^{\mu_1}_{0,\mathrm{II}}=J^{\text{PSZ-Gal},\mu_1} _{4,\mathrm{II}} =
\delta^{\mu_1 \cdots \mu_3}_{\nu_1 \cdots \nu_3} \pi_{\a} 
\partial^{\nu_1} \pi_{\b} \partial_{\mu_2} \partial^{\nu_2} \pi^{\a}
\partial_{\mu_3} \partial^{\nu_3} \pi^{\b}
\end{equation}
gives
\begin{equation}
\partial_{\mu_1} J^{\mu_1}_{0,\mathrm{II}} = \partial_{\mu_1}
J^{\text{PSZ-Gal},\mu_1} _{4,\mathrm{II}} = \mathcal{L}^{\text{PSZ}}_{4}  +
\mathcal{L}^{\text{Gal},1}_{4,\mathrm{II}} .
\end{equation}

It is also possible to make a direct correspondence between
$\mathcal{L}^{\text{Gal},1}_{4,\mathrm{I}}$ and $
\mathcal{L}^{\text{Gal},1}_{4,\mathrm{II}}$ with the current
\begin{equation}
J^{\mu_2}_{0,\mathrm{I\to II}}=J_{4,\mathrm{I\to
II}}^{\text{Gal},1,\mu_2} = \delta^{\mu_1 \cdots \mu_3}_{\nu_1 \cdots
\nu_3} \partial_{\mu_1} \pi_{\a} \partial^{\nu_1} \pi^{\a}
\partial^{\nu_2} \pi_{\b} \partial_{\mu_3} \partial ^{\nu_3} \pi^{\b},
\end{equation} 
yielding
\begin{equation}
\partial_{\mu_2} J^{\mu_2}_{0,\mathrm{I\to II}} = \partial_{\mu_2}
J_{4,\mathrm{I\to II}}^{\text{Gal},1,\mu_2}  =
\mathcal{L}^{\text{Gal},1}_{4,\mathrm{I}} -
\mathcal{L}^{\text{Gal},1}_{4,\mathrm{II}}.
\end{equation}

\subsubsection{Between Gal,2 and Gal,3}

Introducing
\begin{equation}
J^{\mu_1}_{1} = J^{\text{Gal},2-3,\mu_1}_{4,\mathrm{I}} = \partial_{\lambda}
\pi_{\a} \partial^{\lambda} \pi ^{\a} \delta ^{\mu_1 \mu_2} _{\nu_1
\nu_2} \partial^{\nu_1} \pi_{\b} \partial_{\mu_2} \partial^{\nu_2}
\pi^{\b},
\end{equation}
we get
\begin{equation}
\label{EqGal3}
\partial_{\mu_1} J^{\mu_1}_{1}  = \partial_{\mu_1}
J^{\text{Gal},2-3,\mu_1}_{4,\mathrm{I}} = 2
\mathcal{L}^{\text{Gal},2}_{4,\mathrm{III}} +
\mathcal{L}^{\text{Gal},3}_{4,\mathrm{I}}.
\end{equation}

In a similar way, from
\begin{equation}
J^{\mu_1}_{2} = J^{\text{Gal},2-3,\mu_1}_{4,\mathrm{II}} =
\partial_{\lambda} \pi_{\a} \partial^{\lambda} \pi_{\b} \delta ^{\mu_1
\mu_2} _{\nu_1 \nu_2} \partial^{\nu_1} \pi^{\a} \partial_{\mu_2}
\partial^{\nu_2} \pi^{\b},
\end{equation}
we obtain
\begin{equation}
\partial_{\mu_1} J^{\mu_1}_{2}  =\partial_{\mu_1}
J^{\text{Gal},2-3,\mu_1}_{4,\mathrm{II}} = 
\mathcal{L}^{\text{Gal},2}_{4,\mathrm{I}} +
\mathcal{L}^{\text{Gal},2}_{4,\mathrm{II}} +
\mathcal{L}^{\text{Gal},3}_{4,\mathrm{II}}.
\end{equation}

\subsubsection{Between Gal,1, Gal,2 and Gal,3 through Kronecker properties}

We use the following identity given in Ref.~\cite{Deffayet:2011gz}:
\begin{equation}
\label{EqDevAntiSym}
\delta^{\mu_1 \cdots \mu_n} _{\nu_1 \cdots \nu_n} = \delta^{\mu_1}
_{\nu_1} \delta^{\mu_2 \cdots \mu_n} _{\nu_2 \cdots \nu_n} +
\sum_{i=2}^{n} \left(-1 \right)^{i-1} \delta^{\mu_1}_{\nu_i} \delta
^{\mu_2 \cdots \mu_n} _{\nu_1 \nu_2 \cdots \nu_{i-1} \nu_{i+1} \cdots
\nu_n},
\end{equation}
which gives, for $n=3$,
\begin{equation}
\label{EqAlgebraicAntisym}
\delta^{\mu_1 \cdots \mu_3} _{\nu_1 \cdots \nu_3} =  \delta^{\mu_1}
_{\nu_1}  \delta^{\mu_2 \mu_3} _{\nu_2 \nu_3} - \delta^{\mu_1} _{\nu_2}
\delta^{\mu_2 \mu_3}_{\nu_1 \nu_3} + \delta ^{\mu_1} _{\nu_3} \delta
^{\mu_2 \mu_3} _{\nu_1 \nu_2}.
\end{equation}
It is then possible to obtain two additional relations among the
different Lagrangians. Indeed, applying this identity to
$\mathcal{L}^{\text{Gal},1}_{4,\mathrm{I}}$ and
$\mathcal{L}^{\text{Gal},1}_{4,\mathrm{II}}$, we get
\begin{equation}
\mathcal{L}^{\text{Gal},1}_{4,\mathrm{I}} = -2
\mathcal{L}^{\text{Gal},2}_{4,\mathrm{I}} +
\mathcal{L}^{\text{Gal},3}_{4,\mathrm{I}}
\end{equation}
and
\begin{equation}
\mathcal{L}^{\text{Gal},1}_{4,\mathrm{II}} =
-\mathcal{L}^{\text{Gal},2}_{4,\mathrm{II}}  -
\mathcal{L}^{\text{Gal},2}_{4,\mathrm{III}} +
\mathcal{L}^{\text{Gal},3}_{4,\mathrm{II}}.
\end{equation}

\subsection{Lagrangians with Second-Order Equations of Motion}
\label{PartResultLagGal}

Using the results of the previous subsections, we can summarize the Lagrangians
that give second-order equations of motion:
\begin{equation}
\mathcal{L}^{\text{PSZ}}_{4},
\end{equation}
\begin{equation}
\mathcal{L}^{\text{Gal},1}_{4,\mathrm{I}}  = -2
\mathcal{L}^{\text{Gal},2}_{4,\mathrm{I}} +
\mathcal{L}^{\text{Gal},3}_{4,\mathrm{I}}= -
\mathcal{L}^{\text{PSZ}}_{4} - \partial_{\mu} J^{\mu}_{0,\mathrm{I}},
\end{equation}
\begin{equation}
\mathcal{L}^{\text{Gal},1}_{4,\mathrm{II}} = 
-\mathcal{L}^{\text{Gal},2}_{4,\mathrm{II}}  -
\mathcal{L}^{\text{Gal},2}_{4,\mathrm{III}} +
\mathcal{L}^{\text{Gal},3}_{4,\mathrm{II}} = -
\mathcal{L}^{\text{PSZ}}_{4} - \partial_{\mu} J^{\mu}_{0,\mathrm{II}},
\end{equation}
\begin{equation}
\label{EqGal2}
\mathcal{L}^{\text{Gal},2}_{4,\mathrm{II}} =
\frac{1}{4}\mathcal{L}^{\text{Gal},1}_{4,\mathrm{I}}
-\frac{1}{2}\mathcal{L}^{\text{Gal},1}_{4,\mathrm{II}} -\frac{1}{4}\partial_\mu
J^{\mu}_1 +\frac{1}{2}\partial_\mu J^{\mu}_2,
\end{equation}
\begin{equation}
\mathcal{L}^{\text{Gal},2}_{4,\mathrm{I}} +
\mathcal{L}^{\text{Gal},2}_{4,\mathrm{III}}  =
-\frac{1}{2}\mathcal{L}^{\text{Gal},1}_{4,\mathrm{I}}
+\frac{1}{2}\partial_\mu J^{\mu}_1,
\end{equation}
and
\begin{equation}
\label{EqGal1}
\mathcal{L}^{\text{Gal},3}_{4,\mathrm{I}} +2
\mathcal{L}^{\text{Gal},3}_{4,\mathrm{II}} =
\frac{1}{2}\mathcal{L}^{\text{Gal},1}_{4,\mathrm{I}}
+\mathcal{L}^{\text{Gal},1}_{4,\mathrm{II}} +\frac{1}{2}\partial_\mu
J^{\mu}_1 +\partial_\mu J^{\mu}_2.
\end{equation}

\bibliographystyle{apsrev4-1}
\bibliography{bibli} 

%merlin.mbs apsrev4-1.bst 2010-07-25 4.21a (PWD, AO, DPC) hacked
%Control: key (0)
%Control: author (72) initials jnrlst
%Control: editor formatted (1) identically to author
%Control: production of article title (-1) disabled
%Control: page (0) single
%Control: year (1) truncated
%Control: production of eprint (0) enabled
\begin{thebibliography}{127}%
\makeatletter
\providecommand \@ifxundefined [1]{%
 \@ifx{#1\undefined}
}%
\providecommand \@ifnum [1]{%
 \ifnum #1\expandafter \@firstoftwo
 \else \expandafter \@secondoftwo
 \fi
}%
\providecommand \@ifx [1]{%
 \ifx #1\expandafter \@firstoftwo
 \else \expandafter \@secondoftwo
 \fi
}%
\providecommand \natexlab [1]{#1}%
\providecommand \enquote  [1]{``#1''}%
\providecommand \bibnamefont  [1]{#1}%
\providecommand \bibfnamefont [1]{#1}%
\providecommand \citenamefont [1]{#1}%
\providecommand \href@noop [0]{\@secondoftwo}%
\providecommand \href [0]{\begingroup \@sanitize@url \@href}%
\providecommand \@href[1]{\@@startlink{#1}\@@href}%
\providecommand \@@href[1]{\endgroup#1\@@endlink}%
\providecommand \@sanitize@url [0]{\catcode `\\12\catcode `\$12\catcode
  `\&12\catcode `\#12\catcode `\^12\catcode `\_12\catcode `\%12\relax}%
\providecommand \@@startlink[1]{}%
\providecommand \@@endlink[0]{}%
\providecommand \url  [0]{\begingroup\@sanitize@url \@url }%
\providecommand \@url [1]{\endgroup\@href {#1}{\urlprefix }}%
\providecommand \urlprefix  [0]{URL }%
\providecommand \Eprint [0]{\href }%
\providecommand \doibase [0]{http://dx.doi.org/}%
\providecommand \selectlanguage [0]{\@gobble}%
\providecommand \bibinfo  [0]{\@secondoftwo}%
\providecommand \bibfield  [0]{\@secondoftwo}%
\providecommand \translation [1]{[#1]}%
\providecommand \BibitemOpen [0]{}%
\providecommand \bibitemStop [0]{}%
\providecommand \bibitemNoStop [0]{.\EOS\space}%
\providecommand \EOS [0]{\spacefactor3000\relax}%
\providecommand \BibitemShut  [1]{\csname bibitem#1\endcsname}%
\let\auto@bib@innerbib\@empty
%</preamble>
\bibitem [{\citenamefont {Bezrukov}\ \emph {et~al.}(2012)\citenamefont
  {Bezrukov}, \citenamefont {Kalmykov}, \citenamefont {Kniehl},\ and\
  \citenamefont {Shaposhnikov}}]{Bezrukov:2012sa}%
  \BibitemOpen
  \bibfield  {author} {\bibinfo {author} {\bibfnamefont {F.}~\bibnamefont
  {Bezrukov}}, \bibinfo {author} {\bibfnamefont {M.~Y.}\ \bibnamefont
  {Kalmykov}}, \bibinfo {author} {\bibfnamefont {B.~A.}\ \bibnamefont
  {Kniehl}}, \ and\ \bibinfo {author} {\bibfnamefont {M.}~\bibnamefont
  {Shaposhnikov}},\ }\bibfield  {booktitle} {\emph {\bibinfo {booktitle}
  {{Helmholtz Alliance Linear Collider Forum: Proceedings of the Workshops
  Hamburg, Munich, Hamburg 2010-2012, Germany}}},\ }\href {\doibase
  10.1007/JHEP10(2012)140} {\bibfield  {journal} {\bibinfo  {journal} {JHEP}\
  }\textbf {\bibinfo {volume} {1210}},\ \bibinfo {pages} {140} (\bibinfo {year}
  {2012})},\ \Eprint {http://arxiv.org/abs/1205.2893} {arXiv:1205.2893
  [hep-ph]} \BibitemShut {NoStop}%
%%CITATION = ARXIV:1205.2893;%%
\bibitem [{\citenamefont {Bezrukov}\ and\ \citenamefont
  {Shaposhnikov}(2008)}]{Bezrukov:2007ep}%
  \BibitemOpen
  \bibfield  {author} {\bibinfo {author} {\bibfnamefont {F.~L.}\ \bibnamefont
  {Bezrukov}}\ and\ \bibinfo {author} {\bibfnamefont {M.}~\bibnamefont
  {Shaposhnikov}},\ }\href {\doibase 10.1016/j.physletb.2007.11.072} {\bibfield
   {journal} {\bibinfo  {journal} {Phys. Lett.}\ }\textbf {\bibinfo {volume}
  {B659}},\ \bibinfo {pages} {703} (\bibinfo {year} {2008})},\ \Eprint
  {http://arxiv.org/abs/0710.3755} {arXiv:0710.3755 [hep-th]} \BibitemShut
  {NoStop}%
%%CITATION = ARXIV:0710.3755;%%
\bibitem [{\citenamefont {Mazumdar}\ and\ \citenamefont
  {Rocher}(2011)}]{Mazumdar:2010sa}%
  \BibitemOpen
  \bibfield  {author} {\bibinfo {author} {\bibfnamefont {A.}~\bibnamefont
  {Mazumdar}}\ and\ \bibinfo {author} {\bibfnamefont {J.}~\bibnamefont
  {Rocher}},\ }\href {\doibase 10.1016/j.physrep.2010.08.001} {\bibfield
  {journal} {\bibinfo  {journal} {Phys. Rept.}\ }\textbf {\bibinfo {volume}
  {497}},\ \bibinfo {pages} {85} (\bibinfo {year} {2011})},\ \Eprint
  {http://arxiv.org/abs/1001.0993} {arXiv:1001.0993 [hep-ph]} \BibitemShut
  {NoStop}%
%%CITATION = ARXIV:1001.0993;%%
\bibitem [{\citenamefont {Lyth}\ and\ \citenamefont
  {Riotto}(1999)}]{Lyth:1998xn}%
  \BibitemOpen
  \bibfield  {author} {\bibinfo {author} {\bibfnamefont {D.~H.}\ \bibnamefont
  {Lyth}}\ and\ \bibinfo {author} {\bibfnamefont {A.}~\bibnamefont {Riotto}},\
  }\href {\doibase 10.1016/S0370-1573(98)00128-8} {\bibfield  {journal}
  {\bibinfo  {journal} {Phys. Rept.}\ }\textbf {\bibinfo {volume} {314}},\
  \bibinfo {pages} {1} (\bibinfo {year} {1999})},\ \Eprint
  {http://arxiv.org/abs/hep-ph/9807278} {arXiv:hep-ph/9807278 [hep-ph]}
  \BibitemShut {NoStop}%
%%CITATION = HEP-PH/9807278;%%
\bibitem [{\citenamefont {Hertzberg}\ and\ \citenamefont
  {Wilczek}(2014)}]{Hertzberg:2014sza}%
  \BibitemOpen
  \bibfield  {author} {\bibinfo {author} {\bibfnamefont {M.~P.}\ \bibnamefont
  {Hertzberg}}\ and\ \bibinfo {author} {\bibfnamefont {F.}~\bibnamefont
  {Wilczek}},\ }\href@noop {} {\  (\bibinfo {year} {2014})},\ \Eprint
  {http://arxiv.org/abs/1407.6010} {arXiv:1407.6010 [hep-ph]} \BibitemShut
  {NoStop}%
%%CITATION = ARXIV:1407.6010;%%
\bibitem [{\citenamefont {Ferrara}\ \emph {et~al.}(2013)\citenamefont
  {Ferrara}, \citenamefont {Kallosh}, \citenamefont {Linde},\ and\
  \citenamefont {Porrati}}]{Ferrara:2013rsa}%
  \BibitemOpen
  \bibfield  {author} {\bibinfo {author} {\bibfnamefont {S.}~\bibnamefont
  {Ferrara}}, \bibinfo {author} {\bibfnamefont {R.}~\bibnamefont {Kallosh}},
  \bibinfo {author} {\bibfnamefont {A.}~\bibnamefont {Linde}}, \ and\ \bibinfo
  {author} {\bibfnamefont {M.}~\bibnamefont {Porrati}},\ }\href {\doibase
  10.1103/PhysRevD.88.085038} {\bibfield  {journal} {\bibinfo  {journal} {Phys.
  Rev.}\ }\textbf {\bibinfo {volume} {D88}},\ \bibinfo {pages} {085038}
  (\bibinfo {year} {2013})},\ \Eprint {http://arxiv.org/abs/1307.7696}
  {arXiv:1307.7696 [hep-th]} \BibitemShut {NoStop}%
%%CITATION = ARXIV:1307.7696;%%
\bibitem [{\citenamefont {Olive}(1990)}]{Olive:1989nu}%
  \BibitemOpen
  \bibfield  {author} {\bibinfo {author} {\bibfnamefont {K.~A.}\ \bibnamefont
  {Olive}},\ }\href {\doibase 10.1016/0370-1573(90)90144-Q} {\bibfield
  {journal} {\bibinfo  {journal} {Phys. Rept.}\ }\textbf {\bibinfo {volume}
  {190}},\ \bibinfo {pages} {307} (\bibinfo {year} {1990})}\BibitemShut
  {NoStop}%
%%CITATION = PRPLC,190,307;%%
\bibitem [{\citenamefont {Baumann}\ and\ \citenamefont
  {McAllister}(2015)}]{Baumann:2014nda}%
  \BibitemOpen
  \bibfield  {author} {\bibinfo {author} {\bibfnamefont {D.}~\bibnamefont
  {Baumann}}\ and\ \bibinfo {author} {\bibfnamefont {L.}~\bibnamefont
  {McAllister}},\ }\href
  {https://inspirehep.net/record/1289899/files/arXiv:1404.2601.pdf} {\emph
  {\bibinfo {title} {{Inflation and String Theory}}}}\ (\bibinfo  {publisher}
  {Cambridge University Press},\ \bibinfo {year} {2015})\ \Eprint
  {http://arxiv.org/abs/1404.2601} {arXiv:1404.2601 [hep-th]} \BibitemShut
  {NoStop}%
%%CITATION = ARXIV:1404.2601;%%
\bibitem [{\citenamefont {Ashtekar}\ and\ \citenamefont
  {Sloan}(2010)}]{Ashtekar:2009mm}%
  \BibitemOpen
  \bibfield  {author} {\bibinfo {author} {\bibfnamefont {A.}~\bibnamefont
  {Ashtekar}}\ and\ \bibinfo {author} {\bibfnamefont {D.}~\bibnamefont
  {Sloan}},\ }\href {\doibase 10.1016/j.physletb.2010.09.058} {\bibfield
  {journal} {\bibinfo  {journal} {Phys. Lett.}\ }\textbf {\bibinfo {volume}
  {B694}},\ \bibinfo {pages} {108} (\bibinfo {year} {2010})},\ \Eprint
  {http://arxiv.org/abs/0912.4093} {arXiv:0912.4093 [gr-qc]} \BibitemShut
  {NoStop}%
%%CITATION = ARXIV:0912.4093;%%
\bibitem [{\citenamefont {Barrau}(2010)}]{Barrau:2010nd}%
  \BibitemOpen
  \bibfield  {author} {\bibinfo {author} {\bibfnamefont {A.}~\bibnamefont
  {Barrau}},\ }\bibfield  {booktitle} {\emph {\bibinfo {booktitle}
  {{Proceedings, 35th International Conference on High energy physics (ICHEP
  2010): Paris, France, July 22-28, 2010}}},\ }\href@noop {} {\bibfield
  {journal} {\bibinfo  {journal} {PoS}\ }\textbf {\bibinfo {volume}
  {ICHEP2010}},\ \bibinfo {pages} {461} (\bibinfo {year} {2010})},\ \Eprint
  {http://arxiv.org/abs/1011.5516} {arXiv:1011.5516 [gr-qc]} \BibitemShut
  {NoStop}%
%%CITATION = ARXIV:1011.5516;%%
\bibitem [{\citenamefont {Ostrogradski}(1850)}]{ostro}%
  \BibitemOpen
  \bibfield  {author} {\bibinfo {author} {\bibfnamefont {M.}~\bibnamefont
  {Ostrogradski}},\ }\href@noop {} {\bibfield  {journal} {\bibinfo  {journal}
  {Mem. Ac. St. Petersbourg}\ }\textbf {\bibinfo {volume} {VI}},\ \bibinfo
  {pages} {385} (\bibinfo {year} {1850})}\BibitemShut {NoStop}%
\bibitem [{\citenamefont {Horndeski}(1974)}]{Horndeski:1974wa}%
  \BibitemOpen
  \bibfield  {author} {\bibinfo {author} {\bibfnamefont {G.~W.}\ \bibnamefont
  {Horndeski}},\ }\href {\doibase 10.1007/BF01807638} {\bibfield  {journal}
  {\bibinfo  {journal} {Int. J. Theor. Phys.}\ }\textbf {\bibinfo {volume}
  {10}},\ \bibinfo {pages} {363} (\bibinfo {year} {1974})}\BibitemShut
  {NoStop}%
%%CITATION = IJTPB,10,363;%%
\bibitem [{\citenamefont {Woodard}(2007)}]{Woodard:2006nt}%
  \BibitemOpen
  \bibfield  {author} {\bibinfo {author} {\bibfnamefont {R.~P.}\ \bibnamefont
  {Woodard}},\ }\bibfield  {booktitle} {\emph {\bibinfo {booktitle} {{The
  invisible universe: Dark matter and dark energy. Proceedings, 3rd Aegean
  School, Karfas, Greece, September 26-October 1, 2005}}},\ }\href {\doibase
  10.1007/978-3-540-71013-4_14} {\bibfield  {journal} {\bibinfo  {journal}
  {Lect. Notes Phys.}\ }\textbf {\bibinfo {volume} {720}},\ \bibinfo {pages}
  {403} (\bibinfo {year} {2007})},\ \Eprint
  {http://arxiv.org/abs/astro-ph/0601672} {arXiv:astro-ph/0601672 [astro-ph]}
  \BibitemShut {NoStop}%
%%CITATION = ASTRO-PH/0601672;%%
\bibitem [{\citenamefont {Woodard}(2015)}]{Woodard:2015zca}%
  \BibitemOpen
  \bibfield  {author} {\bibinfo {author} {\bibfnamefont {R.~P.}\ \bibnamefont
  {Woodard}},\ }\href {\doibase 10.4249/scholarpedia.32243} {\bibfield
  {journal} {\bibinfo  {journal} {Scholarpedia}\ }\textbf {\bibinfo {volume}
  {10}},\ \bibinfo {pages} {32243} (\bibinfo {year} {2015})},\ \Eprint
  {http://arxiv.org/abs/1506.02210} {arXiv:1506.02210 [hep-th]} \BibitemShut
  {NoStop}%
%%CITATION = ARXIV:1506.02210;%%
\bibitem [{\citenamefont {Nicolis}\ \emph {et~al.}(2009)\citenamefont
  {Nicolis}, \citenamefont {Rattazzi},\ and\ \citenamefont
  {Trincherini}}]{Nicolis:2008in}%
  \BibitemOpen
  \bibfield  {author} {\bibinfo {author} {\bibfnamefont {A.}~\bibnamefont
  {Nicolis}}, \bibinfo {author} {\bibfnamefont {R.}~\bibnamefont {Rattazzi}}, \
  and\ \bibinfo {author} {\bibfnamefont {E.}~\bibnamefont {Trincherini}},\
  }\href {\doibase 10.1103/PhysRevD.79.064036} {\bibfield  {journal} {\bibinfo
  {journal} {Phys.Rev.}\ }\textbf {\bibinfo {volume} {D79}},\ \bibinfo {pages}
  {064036} (\bibinfo {year} {2009})},\ \Eprint {http://arxiv.org/abs/0811.2197}
  {arXiv:0811.2197 [hep-th]} \BibitemShut {NoStop}%
%%CITATION = ARXIV:0811.2197;%%
\bibitem [{\citenamefont {Deffayet}\ \emph {et~al.}(2011)\citenamefont
  {Deffayet}, \citenamefont {Gao}, \citenamefont {Steer},\ and\ \citenamefont
  {Zahariade}}]{Deffayet:2011gz}%
  \BibitemOpen
  \bibfield  {author} {\bibinfo {author} {\bibfnamefont {C.}~\bibnamefont
  {Deffayet}}, \bibinfo {author} {\bibfnamefont {X.}~\bibnamefont {Gao}},
  \bibinfo {author} {\bibfnamefont {D.}~\bibnamefont {Steer}}, \ and\ \bibinfo
  {author} {\bibfnamefont {G.}~\bibnamefont {Zahariade}},\ }\href {\doibase
  10.1103/PhysRevD.84.064039} {\bibfield  {journal} {\bibinfo  {journal}
  {Phys.Rev.}\ }\textbf {\bibinfo {volume} {D84}},\ \bibinfo {pages} {064039}
  (\bibinfo {year} {2011})},\ \Eprint {http://arxiv.org/abs/1103.3260}
  {arXiv:1103.3260 [hep-th]} \BibitemShut {NoStop}%
%%CITATION = ARXIV:1103.3260;%%
\bibitem [{\citenamefont {Deffayet}\ and\ \citenamefont
  {Steer}(2013)}]{Deffayet:2013lga}%
  \BibitemOpen
  \bibfield  {author} {\bibinfo {author} {\bibfnamefont {C.}~\bibnamefont
  {Deffayet}}\ and\ \bibinfo {author} {\bibfnamefont {D.~A.}\ \bibnamefont
  {Steer}},\ }\href {\doibase 10.1088/0264-9381/30/21/214006} {\bibfield
  {journal} {\bibinfo  {journal} {Class. Quant. Grav.}\ }\textbf {\bibinfo
  {volume} {30}},\ \bibinfo {pages} {214006} (\bibinfo {year} {2013})},\
  \Eprint {http://arxiv.org/abs/1307.2450} {arXiv:1307.2450 [hep-th]}
  \BibitemShut {NoStop}%
%%CITATION = ARXIV:1307.2450;%%
\bibitem [{\citenamefont {Deffayet}\ \emph
  {et~al.}(2009{\natexlab{a}})\citenamefont {Deffayet}, \citenamefont
  {Esposito-Farese},\ and\ \citenamefont {Vikman}}]{Deffayet:2009wt}%
  \BibitemOpen
  \bibfield  {author} {\bibinfo {author} {\bibfnamefont {C.}~\bibnamefont
  {Deffayet}}, \bibinfo {author} {\bibfnamefont {G.}~\bibnamefont
  {Esposito-Farese}}, \ and\ \bibinfo {author} {\bibfnamefont {A.}~\bibnamefont
  {Vikman}},\ }\href {\doibase 10.1103/PhysRevD.79.084003} {\bibfield
  {journal} {\bibinfo  {journal} {Phys.Rev.}\ }\textbf {\bibinfo {volume}
  {D79}},\ \bibinfo {pages} {084003} (\bibinfo {year} {2009}{\natexlab{a}})},\
  \Eprint {http://arxiv.org/abs/0901.1314} {arXiv:0901.1314 [hep-th]}
  \BibitemShut {NoStop}%
%%CITATION = ARXIV:0901.1314;%%
\bibitem [{\citenamefont {Deffayet}\ \emph
  {et~al.}(2009{\natexlab{b}})\citenamefont {Deffayet}, \citenamefont {Deser},\
  and\ \citenamefont {Esposito-Farese}}]{Deffayet:2009mn}%
  \BibitemOpen
  \bibfield  {author} {\bibinfo {author} {\bibfnamefont {C.}~\bibnamefont
  {Deffayet}}, \bibinfo {author} {\bibfnamefont {S.}~\bibnamefont {Deser}}, \
  and\ \bibinfo {author} {\bibfnamefont {G.}~\bibnamefont {Esposito-Farese}},\
  }\href {\doibase 10.1103/PhysRevD.80.064015} {\bibfield  {journal} {\bibinfo
  {journal} {Phys.Rev.}\ }\textbf {\bibinfo {volume} {D80}},\ \bibinfo {pages}
  {064015} (\bibinfo {year} {2009}{\natexlab{b}})},\ \Eprint
  {http://arxiv.org/abs/0906.1967} {arXiv:0906.1967 [gr-qc]} \BibitemShut
  {NoStop}%
%%CITATION = ARXIV:0906.1967;%%
\bibitem [{\citenamefont {Kobayashi}\ \emph {et~al.}(2011)\citenamefont
  {Kobayashi}, \citenamefont {Yamaguchi},\ and\ \citenamefont
  {Yokoyama}}]{Kobayashi:2011nu}%
  \BibitemOpen
  \bibfield  {author} {\bibinfo {author} {\bibfnamefont {T.}~\bibnamefont
  {Kobayashi}}, \bibinfo {author} {\bibfnamefont {M.}~\bibnamefont
  {Yamaguchi}}, \ and\ \bibinfo {author} {\bibfnamefont {J.}~\bibnamefont
  {Yokoyama}},\ }\href {\doibase 10.1143/PTP.126.511} {\bibfield  {journal}
  {\bibinfo  {journal} {Prog. Theor. Phys.}\ }\textbf {\bibinfo {volume}
  {126}},\ \bibinfo {pages} {511} (\bibinfo {year} {2011})},\ \Eprint
  {http://arxiv.org/abs/1105.5723} {arXiv:1105.5723 [hep-th]} \BibitemShut
  {NoStop}%
%%CITATION = ARXIV:1105.5723;%%
\bibitem [{\citenamefont {Kobayashi}\ \emph {et~al.}(2013)\citenamefont
  {Kobayashi}, \citenamefont {Tanahashi},\ and\ \citenamefont
  {Yamaguchi}}]{Kobayashi:2013ina}%
  \BibitemOpen
  \bibfield  {author} {\bibinfo {author} {\bibfnamefont {T.}~\bibnamefont
  {Kobayashi}}, \bibinfo {author} {\bibfnamefont {N.}~\bibnamefont
  {Tanahashi}}, \ and\ \bibinfo {author} {\bibfnamefont {M.}~\bibnamefont
  {Yamaguchi}},\ }\href {\doibase 10.1103/PhysRevD.88.083504} {\bibfield
  {journal} {\bibinfo  {journal} {Phys. Rev.}\ }\textbf {\bibinfo {volume}
  {D88}},\ \bibinfo {pages} {083504} (\bibinfo {year} {2013})},\ \Eprint
  {http://arxiv.org/abs/1308.4798} {arXiv:1308.4798 [hep-th]} \BibitemShut
  {NoStop}%
%%CITATION = ARXIV:1308.4798;%%
\bibitem [{\citenamefont {Creminelli}\ \emph {et~al.}(2010)\citenamefont
  {Creminelli}, \citenamefont {Nicolis},\ and\ \citenamefont
  {Trincherini}}]{Creminelli:2010ba}%
  \BibitemOpen
  \bibfield  {author} {\bibinfo {author} {\bibfnamefont {P.}~\bibnamefont
  {Creminelli}}, \bibinfo {author} {\bibfnamefont {A.}~\bibnamefont {Nicolis}},
  \ and\ \bibinfo {author} {\bibfnamefont {E.}~\bibnamefont {Trincherini}},\
  }\href {\doibase 10.1088/1475-7516/2010/11/021} {\bibfield  {journal}
  {\bibinfo  {journal} {JCAP}\ }\textbf {\bibinfo {volume} {1011}},\ \bibinfo
  {pages} {021} (\bibinfo {year} {2010})},\ \Eprint
  {http://arxiv.org/abs/1007.0027} {arXiv:1007.0027 [hep-th]} \BibitemShut
  {NoStop}%
%%CITATION = ARXIV:1007.0027;%%
\bibitem [{\citenamefont {Kobayashi}\ \emph {et~al.}(2010)\citenamefont
  {Kobayashi}, \citenamefont {Yamaguchi},\ and\ \citenamefont
  {Yokoyama}}]{Kobayashi:2010cm}%
  \BibitemOpen
  \bibfield  {author} {\bibinfo {author} {\bibfnamefont {T.}~\bibnamefont
  {Kobayashi}}, \bibinfo {author} {\bibfnamefont {M.}~\bibnamefont
  {Yamaguchi}}, \ and\ \bibinfo {author} {\bibfnamefont {J.}~\bibnamefont
  {Yokoyama}},\ }\href {\doibase 10.1103/PhysRevLett.105.231302} {\bibfield
  {journal} {\bibinfo  {journal} {Phys. Rev. Lett.}\ }\textbf {\bibinfo
  {volume} {105}},\ \bibinfo {pages} {231302} (\bibinfo {year} {2010})},\
  \Eprint {http://arxiv.org/abs/1008.0603} {arXiv:1008.0603 [hep-th]}
  \BibitemShut {NoStop}%
%%CITATION = ARXIV:1008.0603;%%
\bibitem [{\citenamefont {Mizuno}\ and\ \citenamefont
  {Koyama}(2010)}]{Mizuno:2010ag}%
  \BibitemOpen
  \bibfield  {author} {\bibinfo {author} {\bibfnamefont {S.}~\bibnamefont
  {Mizuno}}\ and\ \bibinfo {author} {\bibfnamefont {K.}~\bibnamefont
  {Koyama}},\ }\href {\doibase 10.1103/PhysRevD.82.103518} {\bibfield
  {journal} {\bibinfo  {journal} {Phys. Rev.}\ }\textbf {\bibinfo {volume}
  {D82}},\ \bibinfo {pages} {103518} (\bibinfo {year} {2010})},\ \Eprint
  {http://arxiv.org/abs/1009.0677} {arXiv:1009.0677 [hep-th]} \BibitemShut
  {NoStop}%
%%CITATION = ARXIV:1009.0677;%%
\bibitem [{\citenamefont {Burrage}\ \emph {et~al.}(2011)\citenamefont
  {Burrage}, \citenamefont {de~Rham}, \citenamefont {Seery},\ and\
  \citenamefont {Tolley}}]{Burrage:2010cu}%
  \BibitemOpen
  \bibfield  {author} {\bibinfo {author} {\bibfnamefont {C.}~\bibnamefont
  {Burrage}}, \bibinfo {author} {\bibfnamefont {C.}~\bibnamefont {de~Rham}},
  \bibinfo {author} {\bibfnamefont {D.}~\bibnamefont {Seery}}, \ and\ \bibinfo
  {author} {\bibfnamefont {A.~J.}\ \bibnamefont {Tolley}},\ }\href {\doibase
  10.1088/1475-7516/2011/01/014} {\bibfield  {journal} {\bibinfo  {journal}
  {JCAP}\ }\textbf {\bibinfo {volume} {1101}},\ \bibinfo {pages} {014}
  (\bibinfo {year} {2011})},\ \Eprint {http://arxiv.org/abs/1009.2497}
  {arXiv:1009.2497 [hep-th]} \BibitemShut {NoStop}%
%%CITATION = ARXIV:1009.2497;%%
\bibitem [{\citenamefont {Creminelli}\ \emph {et~al.}(2011)\citenamefont
  {Creminelli}, \citenamefont {D'Amico}, \citenamefont {Musso}, \citenamefont
  {Norena},\ and\ \citenamefont {Trincherini}}]{Creminelli:2010qf}%
  \BibitemOpen
  \bibfield  {author} {\bibinfo {author} {\bibfnamefont {P.}~\bibnamefont
  {Creminelli}}, \bibinfo {author} {\bibfnamefont {G.}~\bibnamefont {D'Amico}},
  \bibinfo {author} {\bibfnamefont {M.}~\bibnamefont {Musso}}, \bibinfo
  {author} {\bibfnamefont {J.}~\bibnamefont {Norena}}, \ and\ \bibinfo {author}
  {\bibfnamefont {E.}~\bibnamefont {Trincherini}},\ }\href {\doibase
  10.1088/1475-7516/2011/02/006} {\bibfield  {journal} {\bibinfo  {journal}
  {JCAP}\ }\textbf {\bibinfo {volume} {1102}},\ \bibinfo {pages} {006}
  (\bibinfo {year} {2011})},\ \Eprint {http://arxiv.org/abs/1011.3004}
  {arXiv:1011.3004 [hep-th]} \BibitemShut {NoStop}%
%%CITATION = ARXIV:1011.3004;%%
\bibitem [{\citenamefont {Kamada}\ \emph {et~al.}(2011)\citenamefont {Kamada},
  \citenamefont {Kobayashi}, \citenamefont {Yamaguchi},\ and\ \citenamefont
  {Yokoyama}}]{Kamada:2010qe}%
  \BibitemOpen
  \bibfield  {author} {\bibinfo {author} {\bibfnamefont {K.}~\bibnamefont
  {Kamada}}, \bibinfo {author} {\bibfnamefont {T.}~\bibnamefont {Kobayashi}},
  \bibinfo {author} {\bibfnamefont {M.}~\bibnamefont {Yamaguchi}}, \ and\
  \bibinfo {author} {\bibfnamefont {J.}~\bibnamefont {Yokoyama}},\ }\href
  {\doibase 10.1103/PhysRevD.83.083515} {\bibfield  {journal} {\bibinfo
  {journal} {Phys. Rev.}\ }\textbf {\bibinfo {volume} {D83}},\ \bibinfo {pages}
  {083515} (\bibinfo {year} {2011})},\ \Eprint {http://arxiv.org/abs/1012.4238}
  {arXiv:1012.4238 [astro-ph.CO]} \BibitemShut {NoStop}%
%%CITATION = ARXIV:1012.4238;%%
\bibitem [{\citenamefont {Libanov}\ \emph {et~al.}(2016)\citenamefont
  {Libanov}, \citenamefont {Mironov},\ and\ \citenamefont
  {Rubakov}}]{Libanov:2016kfc}%
  \BibitemOpen
  \bibfield  {author} {\bibinfo {author} {\bibfnamefont {M.}~\bibnamefont
  {Libanov}}, \bibinfo {author} {\bibfnamefont {S.}~\bibnamefont {Mironov}}, \
  and\ \bibinfo {author} {\bibfnamefont {V.}~\bibnamefont {Rubakov}},\ }\href
  {\doibase 10.1088/1475-7516/2016/08/037} {\bibfield  {journal} {\bibinfo
  {journal} {JCAP}\ }\textbf {\bibinfo {volume} {1608}},\ \bibinfo {pages}
  {037} (\bibinfo {year} {2016})},\ \Eprint {http://arxiv.org/abs/1605.05992}
  {arXiv:1605.05992 [hep-th]} \BibitemShut {NoStop}%
%%CITATION = ARXIV:1605.05992;%%
\bibitem [{\citenamefont {Banerjee}\ and\ \citenamefont
  {Saridakis}(2016)}]{Banerjee:2016hom}%
  \BibitemOpen
  \bibfield  {author} {\bibinfo {author} {\bibfnamefont {S.}~\bibnamefont
  {Banerjee}}\ and\ \bibinfo {author} {\bibfnamefont {E.~N.}\ \bibnamefont
  {Saridakis}},\ }\href@noop {} {\  (\bibinfo {year} {2016})},\ \Eprint
  {http://arxiv.org/abs/1604.06932} {arXiv:1604.06932 [gr-qc]} \BibitemShut
  {NoStop}%
%%CITATION = ARXIV:1604.06932;%%
\bibitem [{\citenamefont {Hirano}\ \emph {et~al.}(2016)\citenamefont {Hirano},
  \citenamefont {Kobayashi},\ and\ \citenamefont {Yokoyama}}]{Hirano:2016gmv}%
  \BibitemOpen
  \bibfield  {author} {\bibinfo {author} {\bibfnamefont {S.}~\bibnamefont
  {Hirano}}, \bibinfo {author} {\bibfnamefont {T.}~\bibnamefont {Kobayashi}}, \
  and\ \bibinfo {author} {\bibfnamefont {S.}~\bibnamefont {Yokoyama}},\
  }\href@noop {} {\  (\bibinfo {year} {2016})},\ \Eprint
  {http://arxiv.org/abs/1604.00141} {arXiv:1604.00141 [astro-ph.CO]}
  \BibitemShut {NoStop}%
%%CITATION = ARXIV:1604.00141;%%
\bibitem [{\citenamefont {Brandenberger}\ and\ \citenamefont
  {Peter}(2016)}]{Brandenberger:2016vhg}%
  \BibitemOpen
  \bibfield  {author} {\bibinfo {author} {\bibfnamefont {R.}~\bibnamefont
  {Brandenberger}}\ and\ \bibinfo {author} {\bibfnamefont {P.}~\bibnamefont
  {Peter}},\ }\href@noop {} {\  (\bibinfo {year} {2016})},\ \Eprint
  {http://arxiv.org/abs/1603.05834} {arXiv:1603.05834 [hep-th]} \BibitemShut
  {NoStop}%
%%CITATION = ARXIV:1603.05834;%%
\bibitem [{\citenamefont {Nishi}\ and\ \citenamefont
  {Kobayashi}(2016)}]{Nishi:2016wty}%
  \BibitemOpen
  \bibfield  {author} {\bibinfo {author} {\bibfnamefont {S.}~\bibnamefont
  {Nishi}}\ and\ \bibinfo {author} {\bibfnamefont {T.}~\bibnamefont
  {Kobayashi}},\ }\href {\doibase 10.1088/1475-7516/2016/04/018} {\bibfield
  {journal} {\bibinfo  {journal} {JCAP}\ }\textbf {\bibinfo {volume} {1604}},\
  \bibinfo {pages} {018} (\bibinfo {year} {2016})},\ \Eprint
  {http://arxiv.org/abs/1601.06561} {arXiv:1601.06561 [hep-th]} \BibitemShut
  {NoStop}%
%%CITATION = ARXIV:1601.06561;%%
\bibitem [{\citenamefont {Chow}\ and\ \citenamefont
  {Khoury}(2009)}]{Chow:2009fm}%
  \BibitemOpen
  \bibfield  {author} {\bibinfo {author} {\bibfnamefont {N.}~\bibnamefont
  {Chow}}\ and\ \bibinfo {author} {\bibfnamefont {J.}~\bibnamefont {Khoury}},\
  }\href {\doibase 10.1103/PhysRevD.80.024037} {\bibfield  {journal} {\bibinfo
  {journal} {Phys. Rev.}\ }\textbf {\bibinfo {volume} {D80}},\ \bibinfo {pages}
  {024037} (\bibinfo {year} {2009})},\ \Eprint {http://arxiv.org/abs/0905.1325}
  {arXiv:0905.1325 [hep-th]} \BibitemShut {NoStop}%
%%CITATION = ARXIV:0905.1325;%%
\bibitem [{\citenamefont {Silva}\ and\ \citenamefont
  {Koyama}(2009)}]{Silva:2009km}%
  \BibitemOpen
  \bibfield  {author} {\bibinfo {author} {\bibfnamefont {F.~P.}\ \bibnamefont
  {Silva}}\ and\ \bibinfo {author} {\bibfnamefont {K.}~\bibnamefont {Koyama}},\
  }\href {\doibase 10.1103/PhysRevD.80.121301} {\bibfield  {journal} {\bibinfo
  {journal} {Phys. Rev.}\ }\textbf {\bibinfo {volume} {D80}},\ \bibinfo {pages}
  {121301} (\bibinfo {year} {2009})},\ \Eprint {http://arxiv.org/abs/0909.4538}
  {arXiv:0909.4538 [astro-ph.CO]} \BibitemShut {NoStop}%
%%CITATION = ARXIV:0909.4538;%%
\bibitem [{\citenamefont {Kobayashi}(2010)}]{Kobayashi:2010wa}%
  \BibitemOpen
  \bibfield  {author} {\bibinfo {author} {\bibfnamefont {T.}~\bibnamefont
  {Kobayashi}},\ }\href {\doibase 10.1103/PhysRevD.81.103533} {\bibfield
  {journal} {\bibinfo  {journal} {Phys. Rev.}\ }\textbf {\bibinfo {volume}
  {D81}},\ \bibinfo {pages} {103533} (\bibinfo {year} {2010})},\ \Eprint
  {http://arxiv.org/abs/1003.3281} {arXiv:1003.3281 [astro-ph.CO]} \BibitemShut
  {NoStop}%
%%CITATION = ARXIV:1003.3281;%%
\bibitem [{\citenamefont {Gannouji}\ and\ \citenamefont
  {Sami}(2010)}]{Gannouji:2010au}%
  \BibitemOpen
  \bibfield  {author} {\bibinfo {author} {\bibfnamefont {R.}~\bibnamefont
  {Gannouji}}\ and\ \bibinfo {author} {\bibfnamefont {M.}~\bibnamefont
  {Sami}},\ }\href {\doibase 10.1103/PhysRevD.82.024011} {\bibfield  {journal}
  {\bibinfo  {journal} {Phys. Rev.}\ }\textbf {\bibinfo {volume} {D82}},\
  \bibinfo {pages} {024011} (\bibinfo {year} {2010})},\ \Eprint
  {http://arxiv.org/abs/1004.2808} {arXiv:1004.2808 [gr-qc]} \BibitemShut
  {NoStop}%
%%CITATION = ARXIV:1004.2808;%%
\bibitem [{\citenamefont {Tsujikawa}(2010)}]{Tsujikawa:2010zza}%
  \BibitemOpen
  \bibfield  {author} {\bibinfo {author} {\bibfnamefont {S.}~\bibnamefont
  {Tsujikawa}},\ }\href {\doibase 10.1007/978-3-642-10598-2_3} {\bibfield
  {journal} {\bibinfo  {journal} {Lect. Notes Phys.}\ }\textbf {\bibinfo
  {volume} {800}},\ \bibinfo {pages} {99} (\bibinfo {year} {2010})},\ \Eprint
  {http://arxiv.org/abs/1101.0191} {arXiv:1101.0191 [gr-qc]} \BibitemShut
  {NoStop}%
%%CITATION = ARXIV:1101.0191;%%
\bibitem [{\citenamefont {De~Felice}\ and\ \citenamefont
  {Tsujikawa}(2010)}]{DeFelice:2010pv}%
  \BibitemOpen
  \bibfield  {author} {\bibinfo {author} {\bibfnamefont {A.}~\bibnamefont
  {De~Felice}}\ and\ \bibinfo {author} {\bibfnamefont {S.}~\bibnamefont
  {Tsujikawa}},\ }\href {\doibase 10.1103/PhysRevLett.105.111301} {\bibfield
  {journal} {\bibinfo  {journal} {Phys. Rev. Lett.}\ }\textbf {\bibinfo
  {volume} {105}},\ \bibinfo {pages} {111301} (\bibinfo {year} {2010})},\
  \Eprint {http://arxiv.org/abs/1007.2700} {arXiv:1007.2700 [astro-ph.CO]}
  \BibitemShut {NoStop}%
%%CITATION = ARXIV:1007.2700;%%
\bibitem [{\citenamefont {Ali}\ \emph {et~al.}(2010)\citenamefont {Ali},
  \citenamefont {Gannouji},\ and\ \citenamefont {Sami}}]{Ali:2010gr}%
  \BibitemOpen
  \bibfield  {author} {\bibinfo {author} {\bibfnamefont {A.}~\bibnamefont
  {Ali}}, \bibinfo {author} {\bibfnamefont {R.}~\bibnamefont {Gannouji}}, \
  and\ \bibinfo {author} {\bibfnamefont {M.}~\bibnamefont {Sami}},\ }\href
  {\doibase 10.1103/PhysRevD.82.103015} {\bibfield  {journal} {\bibinfo
  {journal} {Phys. Rev.}\ }\textbf {\bibinfo {volume} {D82}},\ \bibinfo {pages}
  {103015} (\bibinfo {year} {2010})},\ \Eprint {http://arxiv.org/abs/1008.1588}
  {arXiv:1008.1588 [astro-ph.CO]} \BibitemShut {NoStop}%
%%CITATION = ARXIV:1008.1588;%%
\bibitem [{\citenamefont {Padilla}\ \emph
  {et~al.}(2011{\natexlab{a}})\citenamefont {Padilla}, \citenamefont {Saffin},\
  and\ \citenamefont {Zhou}}]{Padilla:2010tj}%
  \BibitemOpen
  \bibfield  {author} {\bibinfo {author} {\bibfnamefont {A.}~\bibnamefont
  {Padilla}}, \bibinfo {author} {\bibfnamefont {P.~M.}\ \bibnamefont {Saffin}},
  \ and\ \bibinfo {author} {\bibfnamefont {S.-Y.}\ \bibnamefont {Zhou}},\
  }\href {\doibase 10.1007/JHEP01(2011)099} {\bibfield  {journal} {\bibinfo
  {journal} {JHEP}\ }\textbf {\bibinfo {volume} {1101}},\ \bibinfo {pages}
  {099} (\bibinfo {year} {2011}{\natexlab{a}})},\ \Eprint
  {http://arxiv.org/abs/1008.3312} {arXiv:1008.3312 [hep-th]} \BibitemShut
  {NoStop}%
%%CITATION = ARXIV:1008.3312;%%
\bibitem [{\citenamefont {De~Felice}\ and\ \citenamefont
  {Tsujikawa}(2011)}]{DeFelice:2010nf}%
  \BibitemOpen
  \bibfield  {author} {\bibinfo {author} {\bibfnamefont {A.}~\bibnamefont
  {De~Felice}}\ and\ \bibinfo {author} {\bibfnamefont {S.}~\bibnamefont
  {Tsujikawa}},\ }\href {\doibase 10.1103/PhysRevD.84.124029} {\bibfield
  {journal} {\bibinfo  {journal} {Phys. Rev.}\ }\textbf {\bibinfo {volume}
  {D84}},\ \bibinfo {pages} {124029} (\bibinfo {year} {2011})},\ \Eprint
  {http://arxiv.org/abs/1008.4236} {arXiv:1008.4236 [hep-th]} \BibitemShut
  {NoStop}%
%%CITATION = ARXIV:1008.4236;%%
\bibitem [{\citenamefont {Mota}\ \emph {et~al.}(2010)\citenamefont {Mota},
  \citenamefont {Sandstad},\ and\ \citenamefont {Zlosnik}}]{Mota:2010bs}%
  \BibitemOpen
  \bibfield  {author} {\bibinfo {author} {\bibfnamefont {D.~F.}\ \bibnamefont
  {Mota}}, \bibinfo {author} {\bibfnamefont {M.}~\bibnamefont {Sandstad}}, \
  and\ \bibinfo {author} {\bibfnamefont {T.}~\bibnamefont {Zlosnik}},\ }\href
  {\doibase 10.1007/JHEP12(2010)051} {\bibfield  {journal} {\bibinfo  {journal}
  {JHEP}\ }\textbf {\bibinfo {volume} {1012}},\ \bibinfo {pages} {051}
  (\bibinfo {year} {2010})},\ \Eprint {http://arxiv.org/abs/1009.6151}
  {arXiv:1009.6151 [astro-ph.CO]} \BibitemShut {NoStop}%
%%CITATION = ARXIV:1009.6151;%%
\bibitem [{\citenamefont {Nesseris}\ \emph {et~al.}(2010)\citenamefont
  {Nesseris}, \citenamefont {De~Felice},\ and\ \citenamefont
  {Tsujikawa}}]{Nesseris:2010pc}%
  \BibitemOpen
  \bibfield  {author} {\bibinfo {author} {\bibfnamefont {S.}~\bibnamefont
  {Nesseris}}, \bibinfo {author} {\bibfnamefont {A.}~\bibnamefont {De~Felice}},
  \ and\ \bibinfo {author} {\bibfnamefont {S.}~\bibnamefont {Tsujikawa}},\
  }\href {\doibase 10.1103/PhysRevD.82.124054} {\bibfield  {journal} {\bibinfo
  {journal} {Phys. Rev.}\ }\textbf {\bibinfo {volume} {D82}},\ \bibinfo {pages}
  {124054} (\bibinfo {year} {2010})},\ \Eprint {http://arxiv.org/abs/1010.0407}
  {arXiv:1010.0407 [astro-ph.CO]} \BibitemShut {NoStop}%
%%CITATION = ARXIV:1010.0407;%%
\bibitem [{\citenamefont {Gabadadze}\ and\ \citenamefont
  {Yu}(2016)}]{Gabadadze:2016llq}%
  \BibitemOpen
  \bibfield  {author} {\bibinfo {author} {\bibfnamefont {G.}~\bibnamefont
  {Gabadadze}}\ and\ \bibinfo {author} {\bibfnamefont {S.}~\bibnamefont {Yu}},\
  }\href@noop {} {\  (\bibinfo {year} {2016})},\ \Eprint
  {http://arxiv.org/abs/1608.01060} {arXiv:1608.01060 [hep-th]} \BibitemShut
  {NoStop}%
%%CITATION = ARXIV:1608.01060;%%
\bibitem [{\citenamefont {Neveu}\ \emph {et~al.}(2016)\citenamefont {Neveu},
  \citenamefont {Ruhlmann-Kleider}, \citenamefont {Astier}, \citenamefont
  {Besancon}, \citenamefont {Guy}, \citenamefont {M$\ddot{o}$ller},\ and\
  \citenamefont {Babichev}}]{Neveu:2016gxp}%
  \BibitemOpen
  \bibfield  {author} {\bibinfo {author} {\bibfnamefont {J.}~\bibnamefont
  {Neveu}}, \bibinfo {author} {\bibfnamefont {V.}~\bibnamefont
  {Ruhlmann-Kleider}}, \bibinfo {author} {\bibfnamefont {P.}~\bibnamefont
  {Astier}}, \bibinfo {author} {\bibfnamefont {M.}~\bibnamefont {Besancon}},
  \bibinfo {author} {\bibfnamefont {J.}~\bibnamefont {Guy}}, \bibinfo {author}
  {\bibfnamefont {A.}~\bibnamefont {M$\ddot{o}$ller}}, \ and\ \bibinfo {author}
  {\bibfnamefont {E.}~\bibnamefont {Babichev}},\ }\href@noop {} {\  (\bibinfo
  {year} {2016})},\ \Eprint {http://arxiv.org/abs/1605.02627} {arXiv:1605.02627
  [gr-qc]} \BibitemShut {NoStop}%
%%CITATION = ARXIV:1605.02627;%%
\bibitem [{\citenamefont {Salvatelli}\ \emph {et~al.}(2016)\citenamefont
  {Salvatelli}, \citenamefont {Piazza},\ and\ \citenamefont
  {Marinoni}}]{Salvatelli:2016mgy}%
  \BibitemOpen
  \bibfield  {author} {\bibinfo {author} {\bibfnamefont {V.}~\bibnamefont
  {Salvatelli}}, \bibinfo {author} {\bibfnamefont {F.}~\bibnamefont {Piazza}},
  \ and\ \bibinfo {author} {\bibfnamefont {C.}~\bibnamefont {Marinoni}},\
  }\href {\doibase 10.1088/1475-7516/2016/09/027} {\bibfield  {journal}
  {\bibinfo  {journal} {JCAP}\ }\textbf {\bibinfo {volume} {1609}},\ \bibinfo
  {pages} {027} (\bibinfo {year} {2016})},\ \Eprint
  {http://arxiv.org/abs/1602.08283} {arXiv:1602.08283 [astro-ph.CO]}
  \BibitemShut {NoStop}%
%%CITATION = ARXIV:1602.08283;%%
\bibitem [{\citenamefont {Shahalam}\ \emph {et~al.}(2016)\citenamefont
  {Shahalam}, \citenamefont {Pacif},\ and\ \citenamefont
  {Myrzakulov}}]{Shahalam:2016kkg}%
  \BibitemOpen
  \bibfield  {author} {\bibinfo {author} {\bibfnamefont {M.}~\bibnamefont
  {Shahalam}}, \bibinfo {author} {\bibfnamefont {S.~K.~J.}\ \bibnamefont
  {Pacif}}, \ and\ \bibinfo {author} {\bibfnamefont {R.}~\bibnamefont
  {Myrzakulov}},\ }\href {\doibase 10.1140/epjc/s10052-016-4254-y} {\bibfield
  {journal} {\bibinfo  {journal} {Eur. Phys. J.}\ }\textbf {\bibinfo {volume}
  {C76}},\ \bibinfo {pages} {410} (\bibinfo {year} {2016})},\ \Eprint
  {http://arxiv.org/abs/1602.03176} {arXiv:1602.03176 [gr-qc]} \BibitemShut
  {NoStop}%
%%CITATION = ARXIV:1602.03176;%%
\bibitem [{\citenamefont {Minamitsuji}(2016)}]{Minamitsuji:2016qyc}%
  \BibitemOpen
  \bibfield  {author} {\bibinfo {author} {\bibfnamefont {M.}~\bibnamefont
  {Minamitsuji}},\ }\href {\doibase 10.1007/s10714-016-2025-6} {\bibfield
  {journal} {\bibinfo  {journal} {Gen. Rel. Grav.}\ }\textbf {\bibinfo {volume}
  {48}},\ \bibinfo {pages} {26} (\bibinfo {year} {2016})}\BibitemShut {NoStop}%
%%CITATION = GRGVA,48,26;%%
\bibitem [{\citenamefont {Saridakis}\ and\ \citenamefont
  {Tsoukalas}(2016)}]{Saridakis:2016ahq}%
  \BibitemOpen
  \bibfield  {author} {\bibinfo {author} {\bibfnamefont {E.~N.}\ \bibnamefont
  {Saridakis}}\ and\ \bibinfo {author} {\bibfnamefont {M.}~\bibnamefont
  {Tsoukalas}},\ }\href {\doibase 10.1103/PhysRevD.93.124032} {\bibfield
  {journal} {\bibinfo  {journal} {Phys. Rev.}\ }\textbf {\bibinfo {volume}
  {D93}},\ \bibinfo {pages} {124032} (\bibinfo {year} {2016})},\ \Eprint
  {http://arxiv.org/abs/1601.06734} {arXiv:1601.06734 [gr-qc]} \BibitemShut
  {NoStop}%
%%CITATION = ARXIV:1601.06734;%%
\bibitem [{\citenamefont {Biswas}\ and\ \citenamefont
  {Debnath}(2016)}]{Biswas:2016bwq}%
  \BibitemOpen
  \bibfield  {author} {\bibinfo {author} {\bibfnamefont {M.}~\bibnamefont
  {Biswas}}\ and\ \bibinfo {author} {\bibfnamefont {U.}~\bibnamefont
  {Debnath}},\ }\href {\doibase 10.1088/0253-6102/65/1/121} {\bibfield
  {journal} {\bibinfo  {journal} {Commun. Theor. Phys.}\ }\textbf {\bibinfo
  {volume} {65}},\ \bibinfo {pages} {121} (\bibinfo {year} {2016})}\BibitemShut
  {NoStop}%
%%CITATION = CTPMD,65,121;%%
\bibitem [{\citenamefont {Gleyzes}\ \emph {et~al.}(2015)\citenamefont
  {Gleyzes}, \citenamefont {Langlois}, \citenamefont {Piazza},\ and\
  \citenamefont {Vernizzi}}]{Gleyzes:2014dya}%
  \BibitemOpen
  \bibfield  {author} {\bibinfo {author} {\bibfnamefont {J.}~\bibnamefont
  {Gleyzes}}, \bibinfo {author} {\bibfnamefont {D.}~\bibnamefont {Langlois}},
  \bibinfo {author} {\bibfnamefont {F.}~\bibnamefont {Piazza}}, \ and\ \bibinfo
  {author} {\bibfnamefont {F.}~\bibnamefont {Vernizzi}},\ }\href {\doibase
  10.1103/PhysRevLett.114.211101} {\bibfield  {journal} {\bibinfo  {journal}
  {Phys.Rev.Lett.}\ }\textbf {\bibinfo {volume} {114}},\ \bibinfo {pages}
  {211101} (\bibinfo {year} {2015})},\ \Eprint {http://arxiv.org/abs/1404.6495}
  {arXiv:1404.6495 [hep-th]} \BibitemShut {NoStop}%
%%CITATION = ARXIV:1404.6495;%%
\bibitem [{\citenamefont {Langlois}\ and\ \citenamefont
  {Noui}(2016{\natexlab{a}})}]{Langlois:2015cwa}%
  \BibitemOpen
  \bibfield  {author} {\bibinfo {author} {\bibfnamefont {D.}~\bibnamefont
  {Langlois}}\ and\ \bibinfo {author} {\bibfnamefont {K.}~\bibnamefont
  {Noui}},\ }\href {\doibase 10.1088/1475-7516/2016/02/034} {\bibfield
  {journal} {\bibinfo  {journal} {JCAP}\ }\textbf {\bibinfo {volume} {1602}},\
  \bibinfo {pages} {034} (\bibinfo {year} {2016}{\natexlab{a}})},\ \Eprint
  {http://arxiv.org/abs/1510.06930} {arXiv:1510.06930 [gr-qc]} \BibitemShut
  {NoStop}%
%%CITATION = ARXIV:1510.06930;%%
\bibitem [{\citenamefont {Langlois}\ and\ \citenamefont
  {Noui}(2016{\natexlab{b}})}]{Langlois:2015skt}%
  \BibitemOpen
  \bibfield  {author} {\bibinfo {author} {\bibfnamefont {D.}~\bibnamefont
  {Langlois}}\ and\ \bibinfo {author} {\bibfnamefont {K.}~\bibnamefont
  {Noui}},\ }\href {\doibase 10.1088/1475-7516/2016/07/016} {\bibfield
  {journal} {\bibinfo  {journal} {JCAP}\ }\textbf {\bibinfo {volume} {1607}},\
  \bibinfo {pages} {016} (\bibinfo {year} {2016}{\natexlab{b}})},\ \Eprint
  {http://arxiv.org/abs/1512.06820} {arXiv:1512.06820 [gr-qc]} \BibitemShut
  {NoStop}%
%%CITATION = ARXIV:1512.06820;%%
\bibitem [{\citenamefont {Motohashi}\ \emph {et~al.}(2016)\citenamefont
  {Motohashi}, \citenamefont {Noui}, \citenamefont {Suyama}, \citenamefont
  {Yamaguchi},\ and\ \citenamefont {Langlois}}]{Motohashi:2016ftl}%
  \BibitemOpen
  \bibfield  {author} {\bibinfo {author} {\bibfnamefont {H.}~\bibnamefont
  {Motohashi}}, \bibinfo {author} {\bibfnamefont {K.}~\bibnamefont {Noui}},
  \bibinfo {author} {\bibfnamefont {T.}~\bibnamefont {Suyama}}, \bibinfo
  {author} {\bibfnamefont {M.}~\bibnamefont {Yamaguchi}}, \ and\ \bibinfo
  {author} {\bibfnamefont {D.}~\bibnamefont {Langlois}},\ }\href {\doibase
  10.1088/1475-7516/2016/07/033} {\bibfield  {journal} {\bibinfo  {journal}
  {JCAP}\ }\textbf {\bibinfo {volume} {1607}},\ \bibinfo {pages} {033}
  (\bibinfo {year} {2016})},\ \Eprint {http://arxiv.org/abs/1603.09355}
  {arXiv:1603.09355 [hep-th]} \BibitemShut {NoStop}%
%%CITATION = ARXIV:1603.09355;%%
\bibitem [{\citenamefont {Crisostomi}\ \emph {et~al.}(2016)\citenamefont
  {Crisostomi}, \citenamefont {Koyama},\ and\ \citenamefont
  {Tasinato}}]{Crisostomi:2016czh}%
  \BibitemOpen
  \bibfield  {author} {\bibinfo {author} {\bibfnamefont {M.}~\bibnamefont
  {Crisostomi}}, \bibinfo {author} {\bibfnamefont {K.}~\bibnamefont {Koyama}},
  \ and\ \bibinfo {author} {\bibfnamefont {G.}~\bibnamefont {Tasinato}},\
  }\href {\doibase 10.1088/1475-7516/2016/04/044} {\bibfield  {journal}
  {\bibinfo  {journal} {JCAP}\ }\textbf {\bibinfo {volume} {1604}},\ \bibinfo
  {pages} {044} (\bibinfo {year} {2016})},\ \Eprint
  {http://arxiv.org/abs/1602.03119} {arXiv:1602.03119 [hep-th]} \BibitemShut
  {NoStop}%
%%CITATION = ARXIV:1602.03119;%%
\bibitem [{\citenamefont {Harko}\ \emph {et~al.}(2016)\citenamefont {Harko},
  \citenamefont {Lobo}, \citenamefont {Saridakis},\ and\ \citenamefont
  {Tsoukalas}}]{Harko:2016xip}%
  \BibitemOpen
  \bibfield  {author} {\bibinfo {author} {\bibfnamefont {T.}~\bibnamefont
  {Harko}}, \bibinfo {author} {\bibfnamefont {F.~S.~N.}\ \bibnamefont {Lobo}},
  \bibinfo {author} {\bibfnamefont {E.~N.}\ \bibnamefont {Saridakis}}, \ and\
  \bibinfo {author} {\bibfnamefont {M.}~\bibnamefont {Tsoukalas}},\ }\href@noop
  {} {\  (\bibinfo {year} {2016})},\ \Eprint {http://arxiv.org/abs/1609.01503}
  {arXiv:1609.01503 [gr-qc]} \BibitemShut {NoStop}%
%%CITATION = ARXIV:1609.01503;%%
\bibitem [{\citenamefont {Babichev}\ \emph {et~al.}(2016)\citenamefont
  {Babichev}, \citenamefont {Charmousis},\ and\ \citenamefont
  {Leh\'ebel}}]{Babichev:2016rlq}%
  \BibitemOpen
  \bibfield  {author} {\bibinfo {author} {\bibfnamefont {E.}~\bibnamefont
  {Babichev}}, \bibinfo {author} {\bibfnamefont {C.}~\bibnamefont
  {Charmousis}}, \ and\ \bibinfo {author} {\bibfnamefont {A.}~\bibnamefont
  {Leh\'ebel}},\ }\href {\doibase 10.1088/0264-9381/33/15/154002} {\bibfield
  {journal} {\bibinfo  {journal} {Class. Quant. Grav.}\ }\textbf {\bibinfo
  {volume} {33}},\ \bibinfo {pages} {154002} (\bibinfo {year} {2016})},\
  \Eprint {http://arxiv.org/abs/1604.06402} {arXiv:1604.06402 [gr-qc]}
  \BibitemShut {NoStop}%
%%CITATION = ARXIV:1604.06402;%%
\bibitem [{\citenamefont {Sakstein}\ \emph {et~al.}(2016)\citenamefont
  {Sakstein}, \citenamefont {Wilcox}, \citenamefont {Bacon}, \citenamefont
  {Koyama},\ and\ \citenamefont {Nichol}}]{Sakstein:2016ggl}%
  \BibitemOpen
  \bibfield  {author} {\bibinfo {author} {\bibfnamefont {J.}~\bibnamefont
  {Sakstein}}, \bibinfo {author} {\bibfnamefont {H.}~\bibnamefont {Wilcox}},
  \bibinfo {author} {\bibfnamefont {D.}~\bibnamefont {Bacon}}, \bibinfo
  {author} {\bibfnamefont {K.}~\bibnamefont {Koyama}}, \ and\ \bibinfo {author}
  {\bibfnamefont {R.~C.}\ \bibnamefont {Nichol}},\ }\href {\doibase
  10.1088/1475-7516/2016/07/019} {\bibfield  {journal} {\bibinfo  {journal}
  {JCAP}\ }\textbf {\bibinfo {volume} {1607}},\ \bibinfo {pages} {019}
  (\bibinfo {year} {2016})},\ \Eprint {http://arxiv.org/abs/1603.06368}
  {arXiv:1603.06368 [astro-ph.CO]} \BibitemShut {NoStop}%
%%CITATION = ARXIV:1603.06368;%%
\bibitem [{\citenamefont {Kobayashi}(2016)}]{Kobayashi:2016xpl}%
  \BibitemOpen
  \bibfield  {author} {\bibinfo {author} {\bibfnamefont {T.}~\bibnamefont
  {Kobayashi}},\ }\href {\doibase 10.1103/PhysRevD.94.043511} {\bibfield
  {journal} {\bibinfo  {journal} {Phys. Rev.}\ }\textbf {\bibinfo {volume}
  {D94}},\ \bibinfo {pages} {043511} (\bibinfo {year} {2016})},\ \Eprint
  {http://arxiv.org/abs/1606.05831} {arXiv:1606.05831 [hep-th]} \BibitemShut
  {NoStop}%
%%CITATION = ARXIV:1606.05831;%%
\bibitem [{\citenamefont {Lagos}\ \emph {et~al.}(2016)\citenamefont {Lagos},
  \citenamefont {Baker}, \citenamefont {Ferreira},\ and\ \citenamefont
  {Noller}}]{Lagos:2016wyv}%
  \BibitemOpen
  \bibfield  {author} {\bibinfo {author} {\bibfnamefont {M.}~\bibnamefont
  {Lagos}}, \bibinfo {author} {\bibfnamefont {T.}~\bibnamefont {Baker}},
  \bibinfo {author} {\bibfnamefont {P.~G.}\ \bibnamefont {Ferreira}}, \ and\
  \bibinfo {author} {\bibfnamefont {J.}~\bibnamefont {Noller}},\ }\href
  {\doibase 10.1088/1475-7516/2016/08/007} {\bibfield  {journal} {\bibinfo
  {journal} {JCAP}\ }\textbf {\bibinfo {volume} {1608}},\ \bibinfo {pages}
  {007} (\bibinfo {year} {2016})},\ \Eprint {http://arxiv.org/abs/1604.01396}
  {arXiv:1604.01396 [gr-qc]} \BibitemShut {NoStop}%
%%CITATION = ARXIV:1604.01396;%%
\bibitem [{\citenamefont {Frusciante}\ \emph {et~al.}(2016)\citenamefont
  {Frusciante}, \citenamefont {Papadomanolakis},\ and\ \citenamefont
  {Silvestri}}]{Frusciante:2016xoj}%
  \BibitemOpen
  \bibfield  {author} {\bibinfo {author} {\bibfnamefont {N.}~\bibnamefont
  {Frusciante}}, \bibinfo {author} {\bibfnamefont {G.}~\bibnamefont
  {Papadomanolakis}}, \ and\ \bibinfo {author} {\bibfnamefont {A.}~\bibnamefont
  {Silvestri}},\ }\href {\doibase 10.1088/1475-7516/2016/07/018} {\bibfield
  {journal} {\bibinfo  {journal} {JCAP}\ }\textbf {\bibinfo {volume} {1607}},\
  \bibinfo {pages} {018} (\bibinfo {year} {2016})},\ \Eprint
  {http://arxiv.org/abs/1601.04064} {arXiv:1601.04064 [gr-qc]} \BibitemShut
  {NoStop}%
%%CITATION = ARXIV:1601.04064;%%
\bibitem [{\citenamefont {Qiu}(2016)}]{Qiu:2015aha}%
  \BibitemOpen
  \bibfield  {author} {\bibinfo {author} {\bibfnamefont {T.}~\bibnamefont
  {Qiu}},\ }\href {\doibase 10.1103/PhysRevD.93.123515} {\bibfield  {journal}
  {\bibinfo  {journal} {Phys. Rev.}\ }\textbf {\bibinfo {volume} {D93}},\
  \bibinfo {pages} {123515} (\bibinfo {year} {2016})},\ \Eprint
  {http://arxiv.org/abs/1512.02887} {arXiv:1512.02887 [hep-th]} \BibitemShut
  {NoStop}%
%%CITATION = ARXIV:1512.02887;%%
\bibitem [{\citenamefont {Akita}\ and\ \citenamefont
  {Kobayashi}(2016)}]{Akita:2015mho}%
  \BibitemOpen
  \bibfield  {author} {\bibinfo {author} {\bibfnamefont {Y.}~\bibnamefont
  {Akita}}\ and\ \bibinfo {author} {\bibfnamefont {T.}~\bibnamefont
  {Kobayashi}},\ }\href {\doibase 10.1103/PhysRevD.93.043519} {\bibfield
  {journal} {\bibinfo  {journal} {Phys. Rev.}\ }\textbf {\bibinfo {volume}
  {D93}},\ \bibinfo {pages} {043519} (\bibinfo {year} {2016})},\ \Eprint
  {http://arxiv.org/abs/1512.01380} {arXiv:1512.01380 [hep-th]} \BibitemShut
  {NoStop}%
%%CITATION = ARXIV:1512.01380;%%
\bibitem [{\citenamefont {Horndeski}(1976)}]{Horndeski:1976gi}%
  \BibitemOpen
  \bibfield  {author} {\bibinfo {author} {\bibfnamefont {G.~W.}\ \bibnamefont
  {Horndeski}},\ }\href {\doibase 10.1063/1.522837} {\bibfield  {journal}
  {\bibinfo  {journal} {J. Math. Phys.}\ }\textbf {\bibinfo {volume} {17}},\
  \bibinfo {pages} {1980} (\bibinfo {year} {1976})}\BibitemShut {NoStop}%
%%CITATION = JMAPA,17,1980;%%
\bibitem [{\citenamefont {Deffayet}\ \emph {et~al.}(2014)\citenamefont
  {Deffayet}, \citenamefont {Gümrükçüoğlu}, \citenamefont {Mukohyama},\
  and\ \citenamefont {Wang}}]{Deffayet:2013tca}%
  \BibitemOpen
  \bibfield  {author} {\bibinfo {author} {\bibfnamefont {C.}~\bibnamefont
  {Deffayet}}, \bibinfo {author} {\bibfnamefont {A.~E.}\ \bibnamefont
  {Gümrükçüoğlu}}, \bibinfo {author} {\bibfnamefont {S.}~\bibnamefont
  {Mukohyama}}, \ and\ \bibinfo {author} {\bibfnamefont {Y.}~\bibnamefont
  {Wang}},\ }\href {\doibase 10.1007/JHEP04(2014)082} {\bibfield  {journal}
  {\bibinfo  {journal} {JHEP}\ }\textbf {\bibinfo {volume} {1404}},\ \bibinfo
  {pages} {082} (\bibinfo {year} {2014})},\ \Eprint
  {http://arxiv.org/abs/1312.6690} {arXiv:1312.6690 [hep-th]} \BibitemShut
  {NoStop}%
%%CITATION = ARXIV:1312.6690;%%
\bibitem [{\citenamefont {Heisenberg}(2014)}]{Heisenberg:2014rta}%
  \BibitemOpen
  \bibfield  {author} {\bibinfo {author} {\bibfnamefont {L.}~\bibnamefont
  {Heisenberg}},\ }\href {\doibase 10.1088/1475-7516/2014/05/015} {\bibfield
  {journal} {\bibinfo  {journal} {JCAP}\ }\textbf {\bibinfo {volume} {1405}},\
  \bibinfo {pages} {015} (\bibinfo {year} {2014})},\ \Eprint
  {http://arxiv.org/abs/1402.7026} {arXiv:1402.7026 [hep-th]} \BibitemShut
  {NoStop}%
%%CITATION = ARXIV:1402.7026;%%
\bibitem [{\citenamefont {Tasinato}(2014{\natexlab{a}})}]{Tasinato:2014eka}%
  \BibitemOpen
  \bibfield  {author} {\bibinfo {author} {\bibfnamefont {G.}~\bibnamefont
  {Tasinato}},\ }\href {\doibase 10.1007/JHEP04(2014)067} {\bibfield  {journal}
  {\bibinfo  {journal} {JHEP}\ }\textbf {\bibinfo {volume} {1404}},\ \bibinfo
  {pages} {067} (\bibinfo {year} {2014}{\natexlab{a}})},\ \Eprint
  {http://arxiv.org/abs/1402.6450} {arXiv:1402.6450 [hep-th]} \BibitemShut
  {NoStop}%
%%CITATION = ARXIV:1402.6450;%%
\bibitem [{\citenamefont {Allys}\ \emph
  {et~al.}(2016{\natexlab{a}})\citenamefont {Allys}, \citenamefont {Peter},\
  and\ \citenamefont {Rodr\'{\i}guez}}]{Allys:2015sht}%
  \BibitemOpen
  \bibfield  {author} {\bibinfo {author} {\bibfnamefont {E.}~\bibnamefont
  {Allys}}, \bibinfo {author} {\bibfnamefont {P.}~\bibnamefont {Peter}}, \ and\
  \bibinfo {author} {\bibfnamefont {Y.}~\bibnamefont {Rodr\'{\i}guez}},\ }\href
  {\doibase 10.1088/1475-7516/2016/02/004} {\bibfield  {journal} {\bibinfo
  {journal} {JCAP}\ }\textbf {\bibinfo {volume} {1602}},\ \bibinfo {pages}
  {004} (\bibinfo {year} {2016}{\natexlab{a}})},\ \Eprint
  {http://arxiv.org/abs/1511.03101} {arXiv:1511.03101 [hep-th]} \BibitemShut
  {NoStop}%
%%CITATION = ARXIV:1511.03101;%%
\bibitem [{\citenamefont {Beltr\'an~Jimenez}\ and\ \citenamefont
  {Heisenberg}(2016)}]{Jimenez:2016isa}%
  \BibitemOpen
  \bibfield  {author} {\bibinfo {author} {\bibfnamefont {J.}~\bibnamefont
  {Beltr\'an~Jimenez}}\ and\ \bibinfo {author} {\bibfnamefont {L.}~\bibnamefont
  {Heisenberg}},\ }\href {\doibase 10.1016/j.physletb.2016.04.017} {\bibfield
  {journal} {\bibinfo  {journal} {Phys. Lett.}\ }\textbf {\bibinfo {volume}
  {B757}},\ \bibinfo {pages} {405} (\bibinfo {year} {2016})},\ \Eprint
  {http://arxiv.org/abs/1602.03410} {arXiv:1602.03410 [hep-th]} \BibitemShut
  {NoStop}%
%%CITATION = ARXIV:1602.03410;%%
\bibitem [{\citenamefont {Allys}\ \emph
  {et~al.}(2016{\natexlab{b}})\citenamefont {Allys}, \citenamefont
  {Beltr\'an~Almeida}, \citenamefont {Peter},\ and\ \citenamefont
  {Rodr\'{\i}guez}}]{Allys:2016jaq}%
  \BibitemOpen
  \bibfield  {author} {\bibinfo {author} {\bibfnamefont {E.}~\bibnamefont
  {Allys}}, \bibinfo {author} {\bibfnamefont {J.~P.}\ \bibnamefont
  {Beltr\'an~Almeida}}, \bibinfo {author} {\bibfnamefont {P.}~\bibnamefont
  {Peter}}, \ and\ \bibinfo {author} {\bibfnamefont {Y.}~\bibnamefont
  {Rodr\'{\i}guez}},\ }\href {\doibase 10.1088/1475-7516/2016/09/026}
  {\bibfield  {journal} {\bibinfo  {journal} {JCAP}\ }\textbf {\bibinfo
  {volume} {1609}},\ \bibinfo {pages} {026} (\bibinfo {year}
  {2016}{\natexlab{b}})},\ \Eprint {http://arxiv.org/abs/1605.08355}
  {arXiv:1605.08355 [hep-th]} \BibitemShut {NoStop}%
%%CITATION = ARXIV:1605.08355;%%
\bibitem [{\citenamefont {Heisenberg}\ \emph
  {et~al.}(2016{\natexlab{a}})\citenamefont {Heisenberg}, \citenamefont
  {Kase},\ and\ \citenamefont {Tsujikawa}}]{Heisenberg:2016eld}%
  \BibitemOpen
  \bibfield  {author} {\bibinfo {author} {\bibfnamefont {L.}~\bibnamefont
  {Heisenberg}}, \bibinfo {author} {\bibfnamefont {R.}~\bibnamefont {Kase}}, \
  and\ \bibinfo {author} {\bibfnamefont {S.}~\bibnamefont {Tsujikawa}},\ }\href
  {\doibase 10.1016/j.physletb.2016.07.052} {\bibfield  {journal} {\bibinfo
  {journal} {Phys. Lett.}\ }\textbf {\bibinfo {volume} {B760}},\ \bibinfo
  {pages} {617} (\bibinfo {year} {2016}{\natexlab{a}})},\ \Eprint
  {http://arxiv.org/abs/1605.05565} {arXiv:1605.05565 [hep-th]} \BibitemShut
  {NoStop}%
%%CITATION = ARXIV:1605.05565;%%
\bibitem [{\citenamefont {Kimura}\ \emph {et~al.}(2016)\citenamefont {Kimura},
  \citenamefont {Naruko},\ and\ \citenamefont {Yoshida}}]{Kimura:2016rzw}%
  \BibitemOpen
  \bibfield  {author} {\bibinfo {author} {\bibfnamefont {R.}~\bibnamefont
  {Kimura}}, \bibinfo {author} {\bibfnamefont {A.}~\bibnamefont {Naruko}}, \
  and\ \bibinfo {author} {\bibfnamefont {D.}~\bibnamefont {Yoshida}},\
  }\href@noop {} {\  (\bibinfo {year} {2016})},\ \Eprint
  {http://arxiv.org/abs/1608.07066} {arXiv:1608.07066 [gr-qc]} \BibitemShut
  {NoStop}%
%%CITATION = ARXIV:1608.07066;%%
\bibitem [{\citenamefont {Tasinato}(2014{\natexlab{b}})}]{Tasinato:2014mia}%
  \BibitemOpen
  \bibfield  {author} {\bibinfo {author} {\bibfnamefont {G.}~\bibnamefont
  {Tasinato}},\ }\href {\doibase 10.1088/0264-9381/31/22/225004} {\bibfield
  {journal} {\bibinfo  {journal} {Class. Quant. Grav.}\ }\textbf {\bibinfo
  {volume} {31}},\ \bibinfo {pages} {225004} (\bibinfo {year}
  {2014}{\natexlab{b}})},\ \Eprint {http://arxiv.org/abs/1404.4883}
  {arXiv:1404.4883 [hep-th]} \BibitemShut {NoStop}%
%%CITATION = ARXIV:1404.4883;%%
\bibitem [{\citenamefont {Hull}\ \emph {et~al.}(2015)\citenamefont {Hull},
  \citenamefont {Koyama},\ and\ \citenamefont {Tasinato}}]{Hull:2014bga}%
  \BibitemOpen
  \bibfield  {author} {\bibinfo {author} {\bibfnamefont {M.}~\bibnamefont
  {Hull}}, \bibinfo {author} {\bibfnamefont {K.}~\bibnamefont {Koyama}}, \ and\
  \bibinfo {author} {\bibfnamefont {G.}~\bibnamefont {Tasinato}},\ }\href
  {\doibase 10.1007/JHEP03(2015)154} {\bibfield  {journal} {\bibinfo  {journal}
  {JHEP}\ }\textbf {\bibinfo {volume} {1503}},\ \bibinfo {pages} {154}
  (\bibinfo {year} {2015})},\ \Eprint {http://arxiv.org/abs/1408.6871}
  {arXiv:1408.6871 [hep-th]} \BibitemShut {NoStop}%
%%CITATION = ARXIV:1408.6871;%%
\bibitem [{\citenamefont {De~Felice}\ \emph
  {et~al.}(2016{\natexlab{a}})\citenamefont {De~Felice}, \citenamefont
  {Heisenberg}, \citenamefont {Kase}, \citenamefont {Tsujikawa}, \citenamefont
  {Zhang},\ and\ \citenamefont {Zhao}}]{DeFelice:2016cri}%
  \BibitemOpen
  \bibfield  {author} {\bibinfo {author} {\bibfnamefont {A.}~\bibnamefont
  {De~Felice}}, \bibinfo {author} {\bibfnamefont {L.}~\bibnamefont
  {Heisenberg}}, \bibinfo {author} {\bibfnamefont {R.}~\bibnamefont {Kase}},
  \bibinfo {author} {\bibfnamefont {S.}~\bibnamefont {Tsujikawa}}, \bibinfo
  {author} {\bibfnamefont {Y.-l.}\ \bibnamefont {Zhang}}, \ and\ \bibinfo
  {author} {\bibfnamefont {G.-B.}\ \bibnamefont {Zhao}},\ }\href {\doibase
  10.1103/PhysRevD.93.104016} {\bibfield  {journal} {\bibinfo  {journal} {Phys.
  Rev.}\ }\textbf {\bibinfo {volume} {D93}},\ \bibinfo {pages} {104016}
  (\bibinfo {year} {2016}{\natexlab{a}})},\ \Eprint
  {http://arxiv.org/abs/1602.00371} {arXiv:1602.00371 [gr-qc]} \BibitemShut
  {NoStop}%
%%CITATION = ARXIV:1602.00371;%%
\bibitem [{\citenamefont {De~Felice}\ \emph
  {et~al.}(2016{\natexlab{b}})\citenamefont {De~Felice}, \citenamefont
  {Heisenberg}, \citenamefont {Kase}, \citenamefont {Mukohyama}, \citenamefont
  {Tsujikawa},\ and\ \citenamefont {Zhang}}]{DeFelice:2016yws}%
  \BibitemOpen
  \bibfield  {author} {\bibinfo {author} {\bibfnamefont {A.}~\bibnamefont
  {De~Felice}}, \bibinfo {author} {\bibfnamefont {L.}~\bibnamefont
  {Heisenberg}}, \bibinfo {author} {\bibfnamefont {R.}~\bibnamefont {Kase}},
  \bibinfo {author} {\bibfnamefont {S.}~\bibnamefont {Mukohyama}}, \bibinfo
  {author} {\bibfnamefont {S.}~\bibnamefont {Tsujikawa}}, \ and\ \bibinfo
  {author} {\bibfnamefont {Y.-l.}\ \bibnamefont {Zhang}},\ }\href {\doibase
  10.1088/1475-7516/2016/06/048} {\bibfield  {journal} {\bibinfo  {journal}
  {JCAP}\ }\textbf {\bibinfo {volume} {1606}},\ \bibinfo {pages} {048}
  (\bibinfo {year} {2016}{\natexlab{b}})},\ \Eprint
  {http://arxiv.org/abs/1603.05806} {arXiv:1603.05806 [gr-qc]} \BibitemShut
  {NoStop}%
%%CITATION = ARXIV:1603.05806;%%
\bibitem [{\citenamefont {De~Felice}\ \emph
  {et~al.}(2016{\natexlab{c}})\citenamefont {De~Felice}, \citenamefont
  {Heisenberg}, \citenamefont {Kase}, \citenamefont {Mukohyama}, \citenamefont
  {Tsujikawa},\ and\ \citenamefont {Zhang}}]{DeFelice:2016uil}%
  \BibitemOpen
  \bibfield  {author} {\bibinfo {author} {\bibfnamefont {A.}~\bibnamefont
  {De~Felice}}, \bibinfo {author} {\bibfnamefont {L.}~\bibnamefont
  {Heisenberg}}, \bibinfo {author} {\bibfnamefont {R.}~\bibnamefont {Kase}},
  \bibinfo {author} {\bibfnamefont {S.}~\bibnamefont {Mukohyama}}, \bibinfo
  {author} {\bibfnamefont {S.}~\bibnamefont {Tsujikawa}}, \ and\ \bibinfo
  {author} {\bibfnamefont {Y.-l.}\ \bibnamefont {Zhang}},\ }\href {\doibase
  10.1103/PhysRevD.94.044024} {\bibfield  {journal} {\bibinfo  {journal} {Phys.
  Rev.}\ }\textbf {\bibinfo {volume} {D94}},\ \bibinfo {pages} {044024}
  (\bibinfo {year} {2016}{\natexlab{c}})},\ \Eprint
  {http://arxiv.org/abs/1605.05066} {arXiv:1605.05066 [gr-qc]} \BibitemShut
  {NoStop}%
%%CITATION = ARXIV:1605.05066;%%
\bibitem [{\citenamefont {Heisenberg}\ \emph
  {et~al.}(2016{\natexlab{b}})\citenamefont {Heisenberg}, \citenamefont
  {Kase},\ and\ \citenamefont {Tsujikawa}}]{Heisenberg:2016wtr}%
  \BibitemOpen
  \bibfield  {author} {\bibinfo {author} {\bibfnamefont {L.}~\bibnamefont
  {Heisenberg}}, \bibinfo {author} {\bibfnamefont {R.}~\bibnamefont {Kase}}, \
  and\ \bibinfo {author} {\bibfnamefont {S.}~\bibnamefont {Tsujikawa}},\
  }\href@noop {} {\  (\bibinfo {year} {2016}{\natexlab{b}})},\ \Eprint
  {http://arxiv.org/abs/1607.03175} {arXiv:1607.03175 [gr-qc]} \BibitemShut
  {NoStop}%
%%CITATION = ARXIV:1607.03175;%%
\bibitem [{\citenamefont {Watanabe}\ \emph {et~al.}(2009)\citenamefont
  {Watanabe}, \citenamefont {Kanno},\ and\ \citenamefont
  {Soda}}]{Watanabe:2009ct}%
  \BibitemOpen
  \bibfield  {author} {\bibinfo {author} {\bibfnamefont {M.-a.}\ \bibnamefont
  {Watanabe}}, \bibinfo {author} {\bibfnamefont {S.}~\bibnamefont {Kanno}}, \
  and\ \bibinfo {author} {\bibfnamefont {J.}~\bibnamefont {Soda}},\ }\href
  {\doibase 10.1103/PhysRevLett.102.191302} {\bibfield  {journal} {\bibinfo
  {journal} {Phys. Rev. Lett.}\ }\textbf {\bibinfo {volume} {102}},\ \bibinfo
  {pages} {191302} (\bibinfo {year} {2009})},\ \Eprint
  {http://arxiv.org/abs/0902.2833} {arXiv:0902.2833 [hep-th]} \BibitemShut
  {NoStop}%
%%CITATION = ARXIV:0902.2833;%%
\bibitem [{\citenamefont {Dimopoulos}(2006)}]{Dimopoulos:2006ms}%
  \BibitemOpen
  \bibfield  {author} {\bibinfo {author} {\bibfnamefont {K.}~\bibnamefont
  {Dimopoulos}},\ }\href {\doibase 10.1103/PhysRevD.74.083502} {\bibfield
  {journal} {\bibinfo  {journal} {Phys. Rev.}\ }\textbf {\bibinfo {volume}
  {D74}},\ \bibinfo {pages} {083502} (\bibinfo {year} {2006})},\ \Eprint
  {http://arxiv.org/abs/hep-ph/0607229} {arXiv:hep-ph/0607229 [hep-ph]}
  \BibitemShut {NoStop}%
%%CITATION = HEP-PH/0607229;%%
\bibitem [{\citenamefont {Dimopoulos}\ \emph {et~al.}(2010)\citenamefont
  {Dimopoulos}, \citenamefont {Karciauskas},\ and\ \citenamefont
  {Wagstaff}}]{Dimopoulos:2009am}%
  \BibitemOpen
  \bibfield  {author} {\bibinfo {author} {\bibfnamefont {K.}~\bibnamefont
  {Dimopoulos}}, \bibinfo {author} {\bibfnamefont {M.}~\bibnamefont
  {Karciauskas}}, \ and\ \bibinfo {author} {\bibfnamefont {J.~M.}\ \bibnamefont
  {Wagstaff}},\ }\href {\doibase 10.1103/PhysRevD.81.023522} {\bibfield
  {journal} {\bibinfo  {journal} {Phys. Rev.}\ }\textbf {\bibinfo {volume}
  {D81}},\ \bibinfo {pages} {023522} (\bibinfo {year} {2010})},\ \Eprint
  {http://arxiv.org/abs/0907.1838} {arXiv:0907.1838 [hep-ph]} \BibitemShut
  {NoStop}%
%%CITATION = ARXIV:0907.1838;%%
\bibitem [{\citenamefont {Golovnev}\ \emph {et~al.}(2008)\citenamefont
  {Golovnev}, \citenamefont {Mukhanov},\ and\ \citenamefont
  {Vanchurin}}]{Golovnev:2008cf}%
  \BibitemOpen
  \bibfield  {author} {\bibinfo {author} {\bibfnamefont {A.}~\bibnamefont
  {Golovnev}}, \bibinfo {author} {\bibfnamefont {V.}~\bibnamefont {Mukhanov}},
  \ and\ \bibinfo {author} {\bibfnamefont {V.}~\bibnamefont {Vanchurin}},\
  }\href {\doibase 10.1088/1475-7516/2008/06/009} {\bibfield  {journal}
  {\bibinfo  {journal} {JCAP}\ }\textbf {\bibinfo {volume} {0806}},\ \bibinfo
  {pages} {009} (\bibinfo {year} {2008})},\ \Eprint
  {http://arxiv.org/abs/0802.2068} {arXiv:0802.2068 [astro-ph]} \BibitemShut
  {NoStop}%
%%CITATION = ARXIV:0802.2068;%%
\bibitem [{\citenamefont {Armendariz-Picon}(2004)}]{ArmendarizPicon:2004pm}%
  \BibitemOpen
  \bibfield  {author} {\bibinfo {author} {\bibfnamefont {C.}~\bibnamefont
  {Armendariz-Picon}},\ }\href {\doibase 10.1088/1475-7516/2004/07/007}
  {\bibfield  {journal} {\bibinfo  {journal} {JCAP}\ }\textbf {\bibinfo
  {volume} {0407}},\ \bibinfo {pages} {007} (\bibinfo {year} {2004})},\ \Eprint
  {http://arxiv.org/abs/astro-ph/0405267} {arXiv:astro-ph/0405267 [astro-ph]}
  \BibitemShut {NoStop}%
%%CITATION = ASTRO-PH/0405267;%%
\bibitem [{\citenamefont {Rodriguez}\ \emph {et~al.}(2015)\citenamefont
  {Rodriguez}, \citenamefont {Gomez},\ and\ \citenamefont
  {Nieto}}]{Rodriguez:2015xra}%
  \BibitemOpen
  \bibfield  {author} {\bibinfo {author} {\bibfnamefont {Y.}~\bibnamefont
  {Rodriguez}}, \bibinfo {author} {\bibfnamefont {L.~G.}\ \bibnamefont
  {Gomez}}, \ and\ \bibinfo {author} {\bibfnamefont {C.~M.}\ \bibnamefont
  {Nieto}},\ }in\ \href
  {https://inspirehep.net/record/1346381/files/arXiv:1502.07292.pdf} {\emph
  {\bibinfo {booktitle} {{Proceedings, 2nd Argentinian-Brazilian Meeting on
  Gravitation, Relativistic Astrophysics and Cosmology (GRACo II): Buenos
  Aires, Argentina, April 22-25, 2014}}}}\ (\bibinfo {year} {2015})\ pp.\
  \bibinfo {pages} {139--144},\ \Eprint {http://arxiv.org/abs/1502.07292}
  {arXiv:1502.07292 [astro-ph.CO]} \BibitemShut {NoStop}%
%%CITATION = ARXIV:1502.07292;%%
\bibitem [{\citenamefont {Maleknejad}\ and\ \citenamefont
  {Sheikh-Jabbari}(2013)}]{Maleknejad:2011jw}%
  \BibitemOpen
  \bibfield  {author} {\bibinfo {author} {\bibfnamefont {A.}~\bibnamefont
  {Maleknejad}}\ and\ \bibinfo {author} {\bibfnamefont {M.~M.}\ \bibnamefont
  {Sheikh-Jabbari}},\ }\href {\doibase 10.1016/j.physletb.2013.05.001}
  {\bibfield  {journal} {\bibinfo  {journal} {Phys. Lett.}\ }\textbf {\bibinfo
  {volume} {B723}},\ \bibinfo {pages} {224} (\bibinfo {year} {2013})},\ \Eprint
  {http://arxiv.org/abs/1102.1513} {arXiv:1102.1513 [hep-ph]} \BibitemShut
  {NoStop}%
%%CITATION = ARXIV:1102.1513;%%
\bibitem [{\citenamefont {Maleknejad}\ and\ \citenamefont
  {Sheikh-Jabbari}(2011)}]{Maleknejad:2011sq}%
  \BibitemOpen
  \bibfield  {author} {\bibinfo {author} {\bibfnamefont {A.}~\bibnamefont
  {Maleknejad}}\ and\ \bibinfo {author} {\bibfnamefont {M.~M.}\ \bibnamefont
  {Sheikh-Jabbari}},\ }\href {\doibase 10.1103/PhysRevD.84.043515} {\bibfield
  {journal} {\bibinfo  {journal} {Phys. Rev.}\ }\textbf {\bibinfo {volume}
  {D84}},\ \bibinfo {pages} {043515} (\bibinfo {year} {2011})},\ \Eprint
  {http://arxiv.org/abs/1102.1932} {arXiv:1102.1932 [hep-ph]} \BibitemShut
  {NoStop}%
%%CITATION = ARXIV:1102.1932;%%
\bibitem [{\citenamefont {Adshead}\ and\ \citenamefont
  {Wyman}(2012{\natexlab{a}})}]{Adshead:2012kp}%
  \BibitemOpen
  \bibfield  {author} {\bibinfo {author} {\bibfnamefont {P.}~\bibnamefont
  {Adshead}}\ and\ \bibinfo {author} {\bibfnamefont {M.}~\bibnamefont
  {Wyman}},\ }\href {\doibase 10.1103/PhysRevLett.108.261302} {\bibfield
  {journal} {\bibinfo  {journal} {Phys. Rev. Lett.}\ }\textbf {\bibinfo
  {volume} {108}},\ \bibinfo {pages} {261302} (\bibinfo {year}
  {2012}{\natexlab{a}})},\ \Eprint {http://arxiv.org/abs/1202.2366}
  {arXiv:1202.2366 [hep-th]} \BibitemShut {NoStop}%
%%CITATION = ARXIV:1202.2366;%%
\bibitem [{\citenamefont {Maleknejad}\ \emph {et~al.}(2012)\citenamefont
  {Maleknejad}, \citenamefont {Sheikh-Jabbari},\ and\ \citenamefont
  {Soda}}]{Maleknejad:2011jr}%
  \BibitemOpen
  \bibfield  {author} {\bibinfo {author} {\bibfnamefont {A.}~\bibnamefont
  {Maleknejad}}, \bibinfo {author} {\bibfnamefont {M.~M.}\ \bibnamefont
  {Sheikh-Jabbari}}, \ and\ \bibinfo {author} {\bibfnamefont {J.}~\bibnamefont
  {Soda}},\ }\href {\doibase 10.1088/1475-7516/2012/01/016} {\bibfield
  {journal} {\bibinfo  {journal} {JCAP}\ }\textbf {\bibinfo {volume} {1201}},\
  \bibinfo {pages} {016} (\bibinfo {year} {2012})},\ \Eprint
  {http://arxiv.org/abs/1109.5573} {arXiv:1109.5573 [hep-th]} \BibitemShut
  {NoStop}%
%%CITATION = ARXIV:1109.5573;%%
\bibitem [{\citenamefont {Namba}\ \emph {et~al.}(2013)\citenamefont {Namba},
  \citenamefont {Dimastrogiovanni},\ and\ \citenamefont
  {Peloso}}]{Namba:2013kia}%
  \BibitemOpen
  \bibfield  {author} {\bibinfo {author} {\bibfnamefont {R.}~\bibnamefont
  {Namba}}, \bibinfo {author} {\bibfnamefont {E.}~\bibnamefont
  {Dimastrogiovanni}}, \ and\ \bibinfo {author} {\bibfnamefont
  {M.}~\bibnamefont {Peloso}},\ }\href {\doibase 10.1088/1475-7516/2013/11/045}
  {\bibfield  {journal} {\bibinfo  {journal} {JCAP}\ }\textbf {\bibinfo
  {volume} {1311}},\ \bibinfo {pages} {045} (\bibinfo {year} {2013})},\ \Eprint
  {http://arxiv.org/abs/1308.1366} {arXiv:1308.1366 [astro-ph.CO]} \BibitemShut
  {NoStop}%
%%CITATION = ARXIV:1308.1366;%%
\bibitem [{\citenamefont {Adshead}\ \emph {et~al.}(2013)\citenamefont
  {Adshead}, \citenamefont {Martinec},\ and\ \citenamefont
  {Wyman}}]{Adshead:2013nka}%
  \BibitemOpen
  \bibfield  {author} {\bibinfo {author} {\bibfnamefont {P.}~\bibnamefont
  {Adshead}}, \bibinfo {author} {\bibfnamefont {E.}~\bibnamefont {Martinec}}, \
  and\ \bibinfo {author} {\bibfnamefont {M.}~\bibnamefont {Wyman}},\ }\href
  {\doibase 10.1007/JHEP09(2013)087} {\bibfield  {journal} {\bibinfo  {journal}
  {JHEP}\ }\textbf {\bibinfo {volume} {1309}},\ \bibinfo {pages} {087}
  (\bibinfo {year} {2013})},\ \Eprint {http://arxiv.org/abs/1305.2930}
  {arXiv:1305.2930 [hep-th]} \BibitemShut {NoStop}%
%%CITATION = ARXIV:1305.2930;%%
\bibitem [{\citenamefont {Boulware}(1970)}]{Boulware:1970zc}%
  \BibitemOpen
  \bibfield  {author} {\bibinfo {author} {\bibfnamefont {D.~G.}\ \bibnamefont
  {Boulware}},\ }\href {\doibase 10.1016/0003-4916(70)90008-4} {\bibfield
  {journal} {\bibinfo  {journal} {Annals Phys.}\ }\textbf {\bibinfo {volume}
  {56}},\ \bibinfo {pages} {140} (\bibinfo {year} {1970})}\BibitemShut
  {NoStop}%
%%CITATION = APNYA,56,140;%%
\bibitem [{\citenamefont {Shizuya}(1975)}]{Shizuya:1975ek}%
  \BibitemOpen
  \bibfield  {author} {\bibinfo {author} {\bibfnamefont {K.-i.}\ \bibnamefont
  {Shizuya}},\ }\href {\doibase 10.1016/0550-3213(75)90492-7} {\bibfield
  {journal} {\bibinfo  {journal} {Nucl. Phys.}\ }\textbf {\bibinfo {volume}
  {B94}},\ \bibinfo {pages} {260} (\bibinfo {year} {1975})}\BibitemShut
  {NoStop}%
%%CITATION = NUPHA,B94,260;%%
\bibitem [{\citenamefont {Grosse-Knetter}(1993)}]{GrosseKnetter:1993nu}%
  \BibitemOpen
  \bibfield  {author} {\bibinfo {author} {\bibfnamefont {C.}~\bibnamefont
  {Grosse-Knetter}},\ }\href {\doibase 10.1103/PhysRevD.48.2854} {\bibfield
  {journal} {\bibinfo  {journal} {Phys. Rev.}\ }\textbf {\bibinfo {volume}
  {D48}},\ \bibinfo {pages} {2854} (\bibinfo {year} {1993})},\ \Eprint
  {http://arxiv.org/abs/hep-ph/9304310} {arXiv:hep-ph/9304310 [hep-ph]}
  \BibitemShut {NoStop}%
%%CITATION = HEP-PH/9304310;%%
\bibitem [{\citenamefont {Banerjee}\ \emph {et~al.}(1995)\citenamefont
  {Banerjee}, \citenamefont {Banerjee},\ and\ \citenamefont
  {Ghosh}}]{Banerjee:1994pp}%
  \BibitemOpen
  \bibfield  {author} {\bibinfo {author} {\bibfnamefont {N.}~\bibnamefont
  {Banerjee}}, \bibinfo {author} {\bibfnamefont {R.}~\bibnamefont {Banerjee}},
  \ and\ \bibinfo {author} {\bibfnamefont {S.}~\bibnamefont {Ghosh}},\ }\href
  {\doibase 10.1006/aphy.1995.1062} {\bibfield  {journal} {\bibinfo  {journal}
  {Annals Phys.}\ }\textbf {\bibinfo {volume} {241}},\ \bibinfo {pages} {237}
  (\bibinfo {year} {1995})},\ \Eprint {http://arxiv.org/abs/hep-th/9403069}
  {arXiv:hep-th/9403069 [hep-th]} \BibitemShut {NoStop}%
%%CITATION = HEP-TH/9403069;%%
\bibitem [{\citenamefont {Su}(2002)}]{Su:2002qj}%
  \BibitemOpen
  \bibfield  {author} {\bibinfo {author} {\bibfnamefont {J.-C.}\ \bibnamefont
  {Su}},\ }\href@noop {} {\bibfield  {journal} {\bibinfo  {journal} {Nuovo
  Cim.}\ }\textbf {\bibinfo {volume} {B117}},\ \bibinfo {pages} {203} (\bibinfo
  {year} {2002})},\ \Eprint {http://arxiv.org/abs/hep-th/0506101}
  {arXiv:hep-th/0506101 [hep-th]} \BibitemShut {NoStop}%
%%CITATION = HEP-TH/0506101;%%
\bibitem [{\citenamefont {Senjanovic}(1976)}]{Senjanovic:1976br}%
  \BibitemOpen
  \bibfield  {author} {\bibinfo {author} {\bibfnamefont {P.}~\bibnamefont
  {Senjanovic}},\ }\href {\doibase 10.1016/0003-4916(76)90062-2} {\bibfield
  {journal} {\bibinfo  {journal} {Annals Phys.}\ }\textbf {\bibinfo {volume}
  {100}},\ \bibinfo {pages} {227} (\bibinfo {year} {1976})},\ \bibinfo {note}
  {[Erratum: Annals Phys.209,248(1991)]}\BibitemShut {NoStop}%
%%CITATION = APNYA,100,227;%%
\bibitem [{\citenamefont {Banerjee}\ and\ \citenamefont
  {Barcelos-Neto}(1997)}]{Banerjee:1997sf}%
  \BibitemOpen
  \bibfield  {author} {\bibinfo {author} {\bibfnamefont {R.}~\bibnamefont
  {Banerjee}}\ and\ \bibinfo {author} {\bibfnamefont {J.}~\bibnamefont
  {Barcelos-Neto}},\ }\href {\doibase 10.1016/S0550-3213(97)00296-4} {\bibfield
   {journal} {\bibinfo  {journal} {Nucl. Phys.}\ }\textbf {\bibinfo {volume}
  {B499}},\ \bibinfo {pages} {453} (\bibinfo {year} {1997})},\ \Eprint
  {http://arxiv.org/abs/hep-th/9701080} {arXiv:hep-th/9701080 [hep-th]}
  \BibitemShut {NoStop}%
%%CITATION = HEP-TH/9701080;%%
\bibitem [{\citenamefont {Slansky}(1981)}]{Slansky:1981yr}%
  \BibitemOpen
  \bibfield  {author} {\bibinfo {author} {\bibfnamefont {R.}~\bibnamefont
  {Slansky}},\ }\href {\doibase 10.1016/0370-1573(81)90092-2} {\bibfield
  {journal} {\bibinfo  {journal} {Phys.Rept.}\ }\textbf {\bibinfo {volume}
  {79}},\ \bibinfo {pages} {1} (\bibinfo {year} {1981})}\BibitemShut {NoStop}%
%%CITATION = PRPLC,79,1;%%
\bibitem [{\citenamefont {Fuchs}\ and\ \citenamefont
  {Schweigert}(2003)}]{Fuchs:1997jv}%
  \BibitemOpen
  \bibfield  {author} {\bibinfo {author} {\bibfnamefont {J.}~\bibnamefont
  {Fuchs}}\ and\ \bibinfo {author} {\bibfnamefont {C.}~\bibnamefont
  {Schweigert}},\ }\href
  {http://www.cambridge.org/uk/catalogue/catalogue.asp?isbn=0521582733} {\emph
  {\bibinfo {title} {{Symmetries, Lie algebras and representations: A graduate
  course for physicists}}}}\ (\bibinfo  {publisher} {Cambridge University
  Press},\ \bibinfo {year} {2003})\BibitemShut {NoStop}%
%%CITATION = INSPIRE-460269;%%
\bibitem [{\citenamefont {Padilla}\ \emph
  {et~al.}(2011{\natexlab{b}})\citenamefont {Padilla}, \citenamefont {Saffin},\
  and\ \citenamefont {Zhou}}]{Padilla:2010ir}%
  \BibitemOpen
  \bibfield  {author} {\bibinfo {author} {\bibfnamefont {A.}~\bibnamefont
  {Padilla}}, \bibinfo {author} {\bibfnamefont {P.~M.}\ \bibnamefont {Saffin}},
  \ and\ \bibinfo {author} {\bibfnamefont {S.-Y.}\ \bibnamefont {Zhou}},\
  }\href {\doibase 10.1103/PhysRevD.83.045009} {\bibfield  {journal} {\bibinfo
  {journal} {Phys. Rev.}\ }\textbf {\bibinfo {volume} {D83}},\ \bibinfo {pages}
  {045009} (\bibinfo {year} {2011}{\natexlab{b}})},\ \Eprint
  {http://arxiv.org/abs/1008.0745} {arXiv:1008.0745 [hep-th]} \BibitemShut
  {NoStop}%
%%CITATION = ARXIV:1008.0745;%%
\bibitem [{\citenamefont {Metha}\ \emph {et~al.}(1983)\citenamefont {Metha},
  \citenamefont {Normand},\ and\ \citenamefont {Gupta}}]{Metha:1983mng}%
  \BibitemOpen
  \bibfield  {author} {\bibinfo {author} {\bibfnamefont {M.~L.}\ \bibnamefont
  {Metha}}, \bibinfo {author} {\bibfnamefont {J.~M.}\ \bibnamefont {Normand}},
  \ and\ \bibinfo {author} {\bibfnamefont {V.}~\bibnamefont {Gupta}},\
  }\href@noop {} {\bibfield  {journal} {\bibinfo  {journal} {Commun. Math.
  Phys.}\ }\textbf {\bibinfo {volume} {90}},\ \bibinfo {pages} {69} (\bibinfo
  {year} {1983})}\BibitemShut {NoStop}%
\bibitem [{\citenamefont {Maeda}\ and\ \citenamefont
  {Yamamoto}(2013{\natexlab{a}})}]{Maeda:2012eg}%
  \BibitemOpen
  \bibfield  {author} {\bibinfo {author} {\bibfnamefont {K.-i.}\ \bibnamefont
  {Maeda}}\ and\ \bibinfo {author} {\bibfnamefont {K.}~\bibnamefont
  {Yamamoto}},\ }\href {\doibase 10.1103/PhysRevD.87.023528} {\bibfield
  {journal} {\bibinfo  {journal} {Phys. Rev.}\ }\textbf {\bibinfo {volume}
  {D87}},\ \bibinfo {pages} {023528} (\bibinfo {year} {2013}{\natexlab{a}})},\
  \Eprint {http://arxiv.org/abs/1210.4054} {arXiv:1210.4054 [astro-ph.CO]}
  \BibitemShut {NoStop}%
%%CITATION = ARXIV:1210.4054;%%
\bibitem [{\citenamefont {Adshead}\ and\ \citenamefont
  {Wyman}(2012{\natexlab{b}})}]{Adshead:2012qe}%
  \BibitemOpen
  \bibfield  {author} {\bibinfo {author} {\bibfnamefont {P.}~\bibnamefont
  {Adshead}}\ and\ \bibinfo {author} {\bibfnamefont {M.}~\bibnamefont
  {Wyman}},\ }\href {\doibase 10.1103/PhysRevD.86.043530} {\bibfield  {journal}
  {\bibinfo  {journal} {Phys. Rev.}\ }\textbf {\bibinfo {volume} {D86}},\
  \bibinfo {pages} {043530} (\bibinfo {year} {2012}{\natexlab{b}})},\ \Eprint
  {http://arxiv.org/abs/1203.2264} {arXiv:1203.2264 [hep-th]} \BibitemShut
  {NoStop}%
%%CITATION = ARXIV:1203.2264;%%
\bibitem [{\citenamefont {Nieto}\ and\ \citenamefont
  {Rodriguez}(2016)}]{Nieto:2016gnp}%
  \BibitemOpen
  \bibfield  {author} {\bibinfo {author} {\bibfnamefont {C.~M.}\ \bibnamefont
  {Nieto}}\ and\ \bibinfo {author} {\bibfnamefont {Y.}~\bibnamefont
  {Rodriguez}},\ }\href {\doibase 10.1142/S0217732316400058} {\bibfield
  {journal} {\bibinfo  {journal} {Mod. Phys. Lett.}\ }\textbf {\bibinfo
  {volume} {A31}},\ \bibinfo {pages} {1640005} (\bibinfo {year} {2016})},\
  \Eprint {http://arxiv.org/abs/1602.07197} {arXiv:1602.07197 [gr-qc]}
  \BibitemShut {NoStop}%
%%CITATION = ARXIV:1602.07197;%%
\bibitem [{\citenamefont {Davydov}\ and\ \citenamefont
  {Galtsov}(2016)}]{Davydov:2015epx}%
  \BibitemOpen
  \bibfield  {author} {\bibinfo {author} {\bibfnamefont {E.}~\bibnamefont
  {Davydov}}\ and\ \bibinfo {author} {\bibfnamefont {D.}~\bibnamefont
  {Galtsov}},\ }\href {\doibase 10.1016/j.physletb.2015.12.070} {\bibfield
  {journal} {\bibinfo  {journal} {Phys. Lett.}\ }\textbf {\bibinfo {volume}
  {B753}},\ \bibinfo {pages} {622} (\bibinfo {year} {2016})},\ \Eprint
  {http://arxiv.org/abs/1512.02164} {arXiv:1512.02164 [hep-th]} \BibitemShut
  {NoStop}%
%%CITATION = ARXIV:1512.02164;%%
\bibitem [{\citenamefont {Alexander}\ \emph {et~al.}(2015)\citenamefont
  {Alexander}, \citenamefont {Jyoti}, \citenamefont {Kosowsky},\ and\
  \citenamefont {Marciano}}]{Alexander:2014uza}%
  \BibitemOpen
  \bibfield  {author} {\bibinfo {author} {\bibfnamefont {S.}~\bibnamefont
  {Alexander}}, \bibinfo {author} {\bibfnamefont {D.}~\bibnamefont {Jyoti}},
  \bibinfo {author} {\bibfnamefont {A.}~\bibnamefont {Kosowsky}}, \ and\
  \bibinfo {author} {\bibfnamefont {A.}~\bibnamefont {Marciano}},\ }\href
  {\doibase 10.1088/1475-7516/2015/05/005} {\bibfield  {journal} {\bibinfo
  {journal} {JCAP}\ }\textbf {\bibinfo {volume} {1505}},\ \bibinfo {pages}
  {005} (\bibinfo {year} {2015})},\ \Eprint {http://arxiv.org/abs/1408.4118}
  {arXiv:1408.4118 [hep-th]} \BibitemShut {NoStop}%
%%CITATION = ARXIV:1408.4118;%%
\bibitem [{\citenamefont {Sharif}\ and\ \citenamefont
  {Saleem}(2015)}]{Sharif:2014aaa}%
  \BibitemOpen
  \bibfield  {author} {\bibinfo {author} {\bibfnamefont {M.}~\bibnamefont
  {Sharif}}\ and\ \bibinfo {author} {\bibfnamefont {R.}~\bibnamefont
  {Saleem}},\ }\href {\doibase 10.1016/j.astropartphys.2014.06.011} {\bibfield
  {journal} {\bibinfo  {journal} {Astropart. Phys.}\ }\textbf {\bibinfo
  {volume} {62}},\ \bibinfo {pages} {100} (\bibinfo {year} {2015})}\BibitemShut
  {NoStop}%
%%CITATION = APHYE,62,100;%%
\bibitem [{\citenamefont {Darabi}\ and\ \citenamefont
  {Parsiya}(2014)}]{Darabi:2014pua}%
  \BibitemOpen
  \bibfield  {author} {\bibinfo {author} {\bibfnamefont {F.}~\bibnamefont
  {Darabi}}\ and\ \bibinfo {author} {\bibfnamefont {A.}~\bibnamefont
  {Parsiya}},\ }\href {\doibase 10.1142/S0217732314501612} {\bibfield
  {journal} {\bibinfo  {journal} {Mod. Phys. Lett.}\ }\textbf {\bibinfo
  {volume} {A29}},\ \bibinfo {pages} {1450161} (\bibinfo {year} {2014})},\
  \Eprint {http://arxiv.org/abs/1401.1087} {arXiv:1401.1087 [hep-th]}
  \BibitemShut {NoStop}%
%%CITATION = ARXIV:1401.1087;%%
\bibitem [{\citenamefont {Maleknejad}\ and\ \citenamefont
  {Erfani}(2014)}]{Maleknejad:2013npa}%
  \BibitemOpen
  \bibfield  {author} {\bibinfo {author} {\bibfnamefont {A.}~\bibnamefont
  {Maleknejad}}\ and\ \bibinfo {author} {\bibfnamefont {E.}~\bibnamefont
  {Erfani}},\ }\href {\doibase 10.1088/1475-7516/2014/03/016} {\bibfield
  {journal} {\bibinfo  {journal} {JCAP}\ }\textbf {\bibinfo {volume} {1403}},\
  \bibinfo {pages} {016} (\bibinfo {year} {2014})},\ \Eprint
  {http://arxiv.org/abs/1311.3361} {arXiv:1311.3361 [hep-th]} \BibitemShut
  {NoStop}%
%%CITATION = ARXIV:1311.3361;%%
\bibitem [{\citenamefont {Maeda}\ and\ \citenamefont
  {Yamamoto}(2013{\natexlab{b}})}]{Maeda:2013daa}%
  \BibitemOpen
  \bibfield  {author} {\bibinfo {author} {\bibfnamefont {K.-i.}\ \bibnamefont
  {Maeda}}\ and\ \bibinfo {author} {\bibfnamefont {K.}~\bibnamefont
  {Yamamoto}},\ }\href {\doibase 10.1088/1475-7516/2013/12/018} {\bibfield
  {journal} {\bibinfo  {journal} {JCAP}\ }\textbf {\bibinfo {volume} {1312}},\
  \bibinfo {pages} {018} (\bibinfo {year} {2013}{\natexlab{b}})},\ \Eprint
  {http://arxiv.org/abs/1310.6916} {arXiv:1310.6916 [gr-qc]} \BibitemShut
  {NoStop}%
%%CITATION = ARXIV:1310.6916;%%
\bibitem [{\citenamefont {Setare}\ and\ \citenamefont
  {Kamali}(2014)}]{Setare:2013dra}%
  \BibitemOpen
  \bibfield  {author} {\bibinfo {author} {\bibfnamefont {M.~R.}\ \bibnamefont
  {Setare}}\ and\ \bibinfo {author} {\bibfnamefont {V.}~\bibnamefont
  {Kamali}},\ }\href {\doibase 10.1007/s10714-013-1642-6} {\bibfield  {journal}
  {\bibinfo  {journal} {Gen. Rel. Grav.}\ }\textbf {\bibinfo {volume} {46}},\
  \bibinfo {pages} {1642} (\bibinfo {year} {2014})},\ \Eprint
  {http://arxiv.org/abs/1308.5674} {arXiv:1308.5674 [gr-qc]} \BibitemShut
  {NoStop}%
%%CITATION = ARXIV:1308.5674;%%
\bibitem [{\citenamefont {Maleknejad}\ \emph {et~al.}(2013)\citenamefont
  {Maleknejad}, \citenamefont {Sheikh-Jabbari},\ and\ \citenamefont
  {Soda}}]{Maleknejad:2012fw}%
  \BibitemOpen
  \bibfield  {author} {\bibinfo {author} {\bibfnamefont {A.}~\bibnamefont
  {Maleknejad}}, \bibinfo {author} {\bibfnamefont {M.~M.}\ \bibnamefont
  {Sheikh-Jabbari}}, \ and\ \bibinfo {author} {\bibfnamefont {J.}~\bibnamefont
  {Soda}},\ }\href {\doibase 10.1016/j.physrep.2013.03.003} {\bibfield
  {journal} {\bibinfo  {journal} {Phys. Rept.}\ }\textbf {\bibinfo {volume}
  {528}},\ \bibinfo {pages} {161} (\bibinfo {year} {2013})},\ \Eprint
  {http://arxiv.org/abs/1212.2921} {arXiv:1212.2921 [hep-th]} \BibitemShut
  {NoStop}%
%%CITATION = ARXIV:1212.2921;%%
\bibitem [{\citenamefont {Cembranos}\ \emph {et~al.}(2013)\citenamefont
  {Cembranos}, \citenamefont {Maroto},\ and\ \citenamefont
  {Jareño}}]{Cembranos:2012ng}%
  \BibitemOpen
  \bibfield  {author} {\bibinfo {author} {\bibfnamefont {J.~A.~R.}\
  \bibnamefont {Cembranos}}, \bibinfo {author} {\bibfnamefont {A.~L.}\
  \bibnamefont {Maroto}}, \ and\ \bibinfo {author} {\bibfnamefont {S.~J.~N.}\
  \bibnamefont {Jareño}},\ }\href {\doibase 10.1103/PhysRevD.87.043523}
  {\bibfield  {journal} {\bibinfo  {journal} {Phys. Rev.}\ }\textbf {\bibinfo
  {volume} {D87}},\ \bibinfo {pages} {043523} (\bibinfo {year} {2013})},\
  \Eprint {http://arxiv.org/abs/1212.3201} {arXiv:1212.3201 [astro-ph.CO]}
  \BibitemShut {NoStop}%
%%CITATION = ARXIV:1212.3201;%%
\bibitem [{\citenamefont {Dimastrogiovanni}\ and\ \citenamefont
  {Peloso}(2013)}]{Dimastrogiovanni:2012ew}%
  \BibitemOpen
  \bibfield  {author} {\bibinfo {author} {\bibfnamefont {E.}~\bibnamefont
  {Dimastrogiovanni}}\ and\ \bibinfo {author} {\bibfnamefont {M.}~\bibnamefont
  {Peloso}},\ }\href {\doibase 10.1103/PhysRevD.87.103501} {\bibfield
  {journal} {\bibinfo  {journal} {Phys. Rev.}\ }\textbf {\bibinfo {volume}
  {D87}},\ \bibinfo {pages} {103501} (\bibinfo {year} {2013})},\ \Eprint
  {http://arxiv.org/abs/1212.5184} {arXiv:1212.5184 [astro-ph.CO]} \BibitemShut
  {NoStop}%
%%CITATION = ARXIV:1212.5184;%%
\bibitem [{\citenamefont {Ghalee}(2012)}]{Ghalee:2012gg}%
  \BibitemOpen
  \bibfield  {author} {\bibinfo {author} {\bibfnamefont {A.}~\bibnamefont
  {Ghalee}},\ }\href {\doibase 10.1016/j.physletb.2012.09.059} {\bibfield
  {journal} {\bibinfo  {journal} {Phys. Lett.}\ }\textbf {\bibinfo {volume}
  {B717}},\ \bibinfo {pages} {307} (\bibinfo {year} {2012})},\ \Eprint
  {http://arxiv.org/abs/1206.1650} {arXiv:1206.1650 [gr-qc]} \BibitemShut
  {NoStop}%
%%CITATION = ARXIV:1206.1650;%%
\bibitem [{\citenamefont {Li}(2015)}]{Li:2015vwa}%
  \BibitemOpen
  \bibfield  {author} {\bibinfo {author} {\bibfnamefont {W.}~\bibnamefont
  {Li}},\ }\href@noop {} {\  (\bibinfo {year} {2015})},\ \Eprint
  {http://arxiv.org/abs/1508.03247} {arXiv:1508.03247 [gr-qc]} \BibitemShut
  {NoStop}%
%%CITATION = ARXIV:1508.03247;%%
\bibitem [{\citenamefont {de~Rham}\ and\ \citenamefont
  {Heisenberg}(2011)}]{deRham:2011by}%
  \BibitemOpen
  \bibfield  {author} {\bibinfo {author} {\bibfnamefont {C.}~\bibnamefont
  {de~Rham}}\ and\ \bibinfo {author} {\bibfnamefont {L.}~\bibnamefont
  {Heisenberg}},\ }\href {\doibase 10.1103/PhysRevD.84.043503} {\bibfield
  {journal} {\bibinfo  {journal} {Phys. Rev.}\ }\textbf {\bibinfo {volume}
  {D84}},\ \bibinfo {pages} {043503} (\bibinfo {year} {2011})},\ \Eprint
  {http://arxiv.org/abs/1106.3312} {arXiv:1106.3312 [hep-th]} \BibitemShut
  {NoStop}%
%%CITATION = ARXIV:1106.3312;%%
\bibitem [{\citenamefont {Beltr\'an~Jiménez}\ \emph
  {et~al.}(2013)\citenamefont {Beltr\'an~Jiménez}, \citenamefont {Durrer},
  \citenamefont {Heisenberg},\ and\ \citenamefont
  {Thorsrud}}]{Jimenez:2013qsa}%
  \BibitemOpen
  \bibfield  {author} {\bibinfo {author} {\bibfnamefont {J.}~\bibnamefont
  {Beltr\'an~Jiménez}}, \bibinfo {author} {\bibfnamefont {R.}~\bibnamefont
  {Durrer}}, \bibinfo {author} {\bibfnamefont {L.}~\bibnamefont {Heisenberg}},
  \ and\ \bibinfo {author} {\bibfnamefont {M.}~\bibnamefont {Thorsrud}},\
  }\href {\doibase 10.1088/1475-7516/2013/10/064} {\bibfield  {journal}
  {\bibinfo  {journal} {JCAP}\ }\textbf {\bibinfo {volume} {1310}},\ \bibinfo
  {pages} {064} (\bibinfo {year} {2013})},\ \Eprint
  {http://arxiv.org/abs/1308.1867} {arXiv:1308.1867 [hep-th]} \BibitemShut
  {NoStop}%
%%CITATION = ARXIV:1308.1867;%%
\bibitem [{\citenamefont {Hull}\ \emph {et~al.}(2016)\citenamefont {Hull},
  \citenamefont {Koyama},\ and\ \citenamefont {Tasinato}}]{Hull:2015uwa}%
  \BibitemOpen
  \bibfield  {author} {\bibinfo {author} {\bibfnamefont {M.}~\bibnamefont
  {Hull}}, \bibinfo {author} {\bibfnamefont {K.}~\bibnamefont {Koyama}}, \ and\
  \bibinfo {author} {\bibfnamefont {G.}~\bibnamefont {Tasinato}},\ }\href
  {\doibase 10.1103/PhysRevD.93.064012} {\bibfield  {journal} {\bibinfo
  {journal} {Phys. Rev.}\ }\textbf {\bibinfo {volume} {D93}},\ \bibinfo {pages}
  {064012} (\bibinfo {year} {2016})},\ \Eprint
  {http://arxiv.org/abs/1510.07029} {arXiv:1510.07029 [hep-th]} \BibitemShut
  {NoStop}%
%%CITATION = ARXIV:1510.07029;%%
\bibitem [{\citenamefont {Deffayet}\ \emph {et~al.}(2010)\citenamefont
  {Deffayet}, \citenamefont {Deser},\ and\ \citenamefont
  {Esposito-Farese}}]{Deffayet:2010zh}%
  \BibitemOpen
  \bibfield  {author} {\bibinfo {author} {\bibfnamefont {C.}~\bibnamefont
  {Deffayet}}, \bibinfo {author} {\bibfnamefont {S.}~\bibnamefont {Deser}}, \
  and\ \bibinfo {author} {\bibfnamefont {G.}~\bibnamefont {Esposito-Farese}},\
  }\href {\doibase 10.1103/PhysRevD.82.061501} {\bibfield  {journal} {\bibinfo
  {journal} {Phys. Rev.}\ }\textbf {\bibinfo {volume} {D82}},\ \bibinfo {pages}
  {061501} (\bibinfo {year} {2010})},\ \Eprint {http://arxiv.org/abs/1007.5278}
  {arXiv:1007.5278 [gr-qc]} \BibitemShut {NoStop}%
%%CITATION = ARXIV:1007.5278;%%
\bibitem [{\citenamefont {Padilla}\ \emph {et~al.}(2010)\citenamefont
  {Padilla}, \citenamefont {Saffin},\ and\ \citenamefont
  {Zhou}}]{Padilla:2010de}%
  \BibitemOpen
  \bibfield  {author} {\bibinfo {author} {\bibfnamefont {A.}~\bibnamefont
  {Padilla}}, \bibinfo {author} {\bibfnamefont {P.~M.}\ \bibnamefont {Saffin}},
  \ and\ \bibinfo {author} {\bibfnamefont {S.-Y.}\ \bibnamefont {Zhou}},\
  }\href {\doibase 10.1007/JHEP12(2010)031} {\bibfield  {journal} {\bibinfo
  {journal} {JHEP}\ }\textbf {\bibinfo {volume} {1012}},\ \bibinfo {pages}
  {031} (\bibinfo {year} {2010})},\ \Eprint {http://arxiv.org/abs/1007.5424}
  {arXiv:1007.5424 [hep-th]} \BibitemShut {NoStop}%
%%CITATION = ARXIV:1007.5424;%%
\bibitem [{\citenamefont {Hinterbichler}\ \emph {et~al.}(2010)\citenamefont
  {Hinterbichler}, \citenamefont {Trodden},\ and\ \citenamefont
  {Wesley}}]{Hinterbichler:2010xn}%
  \BibitemOpen
  \bibfield  {author} {\bibinfo {author} {\bibfnamefont {K.}~\bibnamefont
  {Hinterbichler}}, \bibinfo {author} {\bibfnamefont {M.}~\bibnamefont
  {Trodden}}, \ and\ \bibinfo {author} {\bibfnamefont {D.}~\bibnamefont
  {Wesley}},\ }\href {\doibase 10.1103/PhysRevD.82.124018} {\bibfield
  {journal} {\bibinfo  {journal} {Phys. Rev.}\ }\textbf {\bibinfo {volume}
  {D82}},\ \bibinfo {pages} {124018} (\bibinfo {year} {2010})},\ \Eprint
  {http://arxiv.org/abs/1008.1305} {arXiv:1008.1305 [hep-th]} \BibitemShut
  {NoStop}%
%%CITATION = ARXIV:1008.1305;%%
\bibitem [{\citenamefont {Padilla}\ and\ \citenamefont
  {Sivanesan}(2013)}]{Padilla:2012dx}%
  \BibitemOpen
  \bibfield  {author} {\bibinfo {author} {\bibfnamefont {A.}~\bibnamefont
  {Padilla}}\ and\ \bibinfo {author} {\bibfnamefont {V.}~\bibnamefont
  {Sivanesan}},\ }\href {\doibase 10.1007/JHEP04(2013)032} {\bibfield
  {journal} {\bibinfo  {journal} {JHEP}\ }\textbf {\bibinfo {volume} {1304}},\
  \bibinfo {pages} {032} (\bibinfo {year} {2013})},\ \Eprint
  {http://arxiv.org/abs/1210.4026} {arXiv:1210.4026 [gr-qc]} \BibitemShut
  {NoStop}%
%%CITATION = ARXIV:1210.4026;%%
\bibitem [{\citenamefont {Sivanesan}(2014)}]{Sivanesan:2013tba}%
  \BibitemOpen
  \bibfield  {author} {\bibinfo {author} {\bibfnamefont {V.}~\bibnamefont
  {Sivanesan}},\ }\href {\doibase 10.1103/PhysRevD.90.104006} {\bibfield
  {journal} {\bibinfo  {journal} {Phys. Rev.}\ }\textbf {\bibinfo {volume}
  {D90}},\ \bibinfo {pages} {104006} (\bibinfo {year} {2014})},\ \Eprint
  {http://arxiv.org/abs/1307.8081} {arXiv:1307.8081 [gr-qc]} \BibitemShut
  {NoStop}%
%%CITATION = ARXIV:1307.8081;%%
\bibitem [{\citenamefont {Ramond}(2010)}]{Ramond:2010zz}%
  \BibitemOpen
  \bibfield  {author} {\bibinfo {author} {\bibfnamefont {P.}~\bibnamefont
  {Ramond}},\ }\href {http://www.cambridge.org/de/knowledge/isbn/item2710157}
  {\emph {\bibinfo {title} {{Group theory: A physicist's survey}}}}\ (\bibinfo
  {publisher} {Cambridge University Press},\ \bibinfo {year}
  {2010})\BibitemShut {NoStop}%
%%CITATION = INSPIRE-1108114;%%
\bibitem [{\citenamefont {Feger}\ and\ \citenamefont
  {Kephart}(2015)}]{Feger:2012bs}%
  \BibitemOpen
  \bibfield  {author} {\bibinfo {author} {\bibfnamefont {R.}~\bibnamefont
  {Feger}}\ and\ \bibinfo {author} {\bibfnamefont {T.~W.}\ \bibnamefont
  {Kephart}},\ }\href {\doibase 10.1016/j.cpc.2014.12.023} {\bibfield
  {journal} {\bibinfo  {journal} {Comput. Phys. Commun.}\ }\textbf {\bibinfo
  {volume} {192}},\ \bibinfo {pages} {166} (\bibinfo {year} {2015})},\ \Eprint
  {http://arxiv.org/abs/1206.6379} {arXiv:1206.6379 [math-ph]} \BibitemShut
  {NoStop}%
%%CITATION = ARXIV:1206.6379;%%
\bibitem [{\citenamefont {Fleury}\ \emph {et~al.}(2014)\citenamefont {Fleury},
  \citenamefont {Almeida}, \citenamefont {Pitrou},\ and\ \citenamefont
  {Uzan}}]{Fleury:2014qfa}%
  \BibitemOpen
  \bibfield  {author} {\bibinfo {author} {\bibfnamefont {P.}~\bibnamefont
  {Fleury}}, \bibinfo {author} {\bibfnamefont {J.~P.~B.}\ \bibnamefont
  {Almeida}}, \bibinfo {author} {\bibfnamefont {C.}~\bibnamefont {Pitrou}}, \
  and\ \bibinfo {author} {\bibfnamefont {J.-P.}\ \bibnamefont {Uzan}},\ }\href
  {\doibase 10.1088/1475-7516/2014/11/043} {\bibfield  {journal} {\bibinfo
  {journal} {JCAP}\ }\textbf {\bibinfo {volume} {1411}},\ \bibinfo {pages}
  {043} (\bibinfo {year} {2014})},\ \Eprint {http://arxiv.org/abs/1406.6254}
  {arXiv:1406.6254 [hep-th]} \BibitemShut {NoStop}%
%%CITATION = ARXIV:1406.6254;%%
\end{thebibliography}%

\end{document}